\documentclass[aps,amsmath,amssymb,prb,preprint]{revtex4-1}
\usepackage{amsmath, latexsym, color, graphicx, tabularx, booktabs, enumerate}
\usepackage{multirow}
\usepackage[normalem]{ulem} 
\usepackage{caption}
\usepackage{epstopdf}
\usepackage{mathtools}
\usepackage{hyperref}
\usepackage{cleveref}
\usepackage{adjustbox}
\usepackage[english]{babel}
\begin{document}
Defect-induced states, defect-induced phase transition and excitonic states in bent transition metal dichalcogenide (TMD) nanoribbons: density functional vs. many body theory
\author{Santosh Neupane}
\author{Hong Tang}
\author{Adrienn Ruzsinszky}

\affiliation{Department of Physics, Temple University, Philadelphia, Pennsylvania 19122, United States}

\begin{abstract}
Two-dimensional (2D) transition metal dichalcogenide (TMD) materials have versatile electronic and optical properties.TMD nanoribbons exhibit interesting properties due to reduced dimensionality, quantum confinement, and edge states, which make them suitable for various electronic and optoelectronic applications. In a previous work conducted by our group, we demonstrated that the edge bands evolved with bending can tune the optical properties for various widths of TMD nanoribbons. Defects are commonly present in 2D TMD materials, and can dramatically change the material properties. In this following work, we investigate the interaction between the edge  and the defect states in tungsten disulfide (WS$_{2}$) nanoribbons with line defects under different bending conditions, using density functional theory (DFT). To gain understanding about the limits of density functional approximations, we compare results on band gaps and energies of defect states with quasiparticle GW. We reveal interesting semiconductor-metal phase transitions, suggesting potential applications in nano-electronics or molecular electronics. We also calculate the optical absorption of the bent and defective nanoribbons  with the many-body GW-BSE (Bethe-Salpeter equation) approach, revealing a tunable optical spectrum and diverse exciton states in the defective TMD nanoribbons.
\end{abstract}
\maketitle

\section{Introduction}
Condensed matter physics \cite{cm1,cm2,cm3,cm4,cm5,cm6,cm7} has undergone significant advancements since the discovery of single-layer graphene in 2004 \cite{graphene}, with the emergence of a new area of theoretical and experimental study focused on two-dimensional (2D) materials \cite{2d1,2d2,2d3,2d4,2d5,2d6}. This research has revealed novel physical properties in 2D materials, and has driven the search for new 2D semiconducting materials, leading to breakthroughs in nanoelectronics \cite{ne1,ne2,ne3}, sensing \cite{s1,s2}, energy storage \cite{es1,es2}, energy conversion \cite{ec1,ec2}, photonics \cite{ph1,ph2}, optoelectronics \cite{2d6}, magnetoresistance \cite{ms}, and valleytronics \cite{val1,val2}. However, like any other materials, defects are inevitable in 2D materials, and their presence can significantly affect their properties \cite{defects1-2dmaterials, defects2, defects3, defects4}. Structural defects, such as vacancies, dislocations, and grain boundaries, can alter their electronic and mechanical properties \cite{defects-efects, defects-effects2, defects-effects3, defects-efffects4-ws2}. Thus, understanding the role of defects in 2D materials is crucial for optimizing their properties and developing new applications. 

Among 2D materials, recent experiments have shown promising breakthroughs in monolayer transition metal dichalcogenides (TMDs) \cite{tmd1,tmd2,tmd3,tmd4,tmd5}. Monolayer tungsten disulfide (WS$_{2}$) emits strong photoluminescence (PL) with a high quantum yield \cite{pl1,pl2}, making it a promising material for 2D optoelectronic devices \cite{ws2app1,ws2app2,ws2app3}. However, the lack of true ohmic contact between the 2D semiconductor and metallic electrodes remains a significant hurdle for device design and fabrication, due to the issue of Schottky contact. Recent research has proposed the use of heterophase homojunctions, which involve co-existing semiconducting and metallic phases built from the same 2D layered materials, as a solution to this problem. Such homojunctions have been shown to enable nearly perfect ohmic connections between semiconducting 2D layers and electrodes by introducing conducting phase patches or regions through locally induced defects and/or strains via laser beam \cite{Cho_2015} or ion beam \cite{Wang_2019} irradiations. 

Moreover, recent studies have highlighted the crucial role played by defects in the exciton dynamics of monolayer TMDs. Zhou \textit{et al.} \cite{dark-excitons} and Attaccalite \textit{et al.} \cite{Attaccale} have demonstrated the intricate interplay between defects, electronic structure, and optical properties in TMDs. Their findings have provided novel insights into the mechanisms governing exciton generation, recombination, and transport in TMDs. They have shown that defects can significantly affect the exciton lifetime, the radiative and non-radiative decay rates, and the exciton diffusion length in TMDs. These results have important implications for the development of TMD-based optoelectronic devices and suggest new avenues for future research in this field.

Nanoribbons, which are quasi-one-dimensional structures with more spatial confinement effects and rich edge states than their 2D counterparts, offer promising potential for designing nanoscale devices \cite{nr1,nr2,nr3}. By applying uniaxial strains or bending, one can create local strains on the nanoribbons and effectively modify the edge states and band structures, leading to controllable electronic and optical properties \cite{bending2us,bending1bn}. In our previous work \cite{spin-anisotropy}, we have found that the incorporation of bending and doping in WS$_{2}$, WSe$_{2}$, and WTe$_{2}$ nanoribbons can lead to enhanced spin-orbit coupling (SOC) effects and controlled magnetism. This results in spatially varying spin polarization, which has a significant impact on the spin configuration of exciton states. 

The focus of this work is two-fold: (1) investigating the potential of armchair WS$_{2}$ nanoribbons with line defects under different bending radii as a means to realize the semiconductor-metal phase connection in WS$_{2}$ homojunctions, (2) understanding the optical response of bent nanoribbons with defects. Meanwhile we are assessing advanced density functional approximations for the electronic structure including defect states.

We examine the correlation between defect and edge states for armchair WS$_{2}$ nanoribbons with defects under various bending curvatures using the Perdew-Burke-Ernzerhof (PBE) \cite{PBE}, r$^{2}$SCAN meta-GGA \cite{r2scan}, the modified TASK (mTASK) \cite{mTASk} functional approximations, the higher-level screened hybrid functional HSE06 \cite{HSE06} and the GW method \cite{LH65, LH69, HMLS85, AG98}. We also investigate optical absorption and excitonic states using the many-body perturbation GW and BSE (Bethe-Salpeter equation) methods. We confirm that the appropriate choice of the defect sites can induce magnetic properties on the WS$_{2}$ nanoribbons.  Our findings suggest that defects in WS$_{2}$ nanoribbons are crucial for inducing magnetic properties. Specifically, we have observed that the appropriate choice of defect sites can result in net magnetization. In contrast, our previous work showed that net magnetization could be achieved only with electron or hole doping in pristine WS$_{2}$ nanoribbons \cite{spin-anisotropy}. Therefore, we conclude that defects play a significant role in the magnetic behavior of WS$_{2}$ nanoribbons. Our results show that a semiconductor-metal phase transition in WS$_{2}$ nanoribbons can be driven by line defects under various bending curvatures with the appropriate choice of the defect site. These results broaden the device design strategies for WS$_{2}$ homojunctions, and have potential applications in phase-change electrical devices. Our work demonstrates that the optical absorption in A13WS$_{2}$ nanoribbons is robust against central sulfur defects, while the spatial extension of excitons can be controlled through bending. Furthermore, intrinsic exciton lifetimes can be modified and controlled through defect introduction and bending, which has important implications for exciton-based quantum controls and probing exciton dynamics.

\section{Computational details}
Density functional theory (DFT) \cite{HK64, KS65} calculations were performed using the Vienna Ab initio Software Package (VASP)\cite{VASP} with projector augmented-wave pseudopotentials \cite{pp1,pp2}. In order to avoid the interactions between the nanoribbon and its periodic images, a vacuum layer of more than 12 {\AA} is added along the direction of nanoribbon's width and inserted along the direction perpendicular to the 2D surface of the nanoribbon. The energy cutoff is 580 eV. The k-point mesh of 1 × 1 × 24 was used for all nanoribbons. The pre-relaxed WS$_{2}$ monolayer, which was relaxed with Perdew-Burke-Ernzerhof (PBE) functional \cite{PBE}, with the in-plane lattice constants of a = b = 3.321 {\AA} was used to build the nanoribbons. Then a full structural relaxation for all the nanoribbons was done with PBE with all forces less than 0.01 eV/{\AA}. During the relaxation, the x and y coordinates of the two outermost metal atoms on the two edge sides were fixed, while their coordinates along the ribbon axis direction (z direction) and all the coordinates of other atoms were allowed to relax. The supercell vectors a and b along the x and y axes were also fixed. 

The $G_0W_0$ \cite{LH65, LH69, HMLS85, AG98} and $G_0W_0$+BSE \cite{Onida02, ASR98, BLS98, RML98, HSL80} calculations were carried out using BerkeleyGW code \cite{BGW} by pairing with Quantum ESPRESSO \cite{QE}. The wavefunction energy cutoff is 70 Ry ($\approx$950 eV). The energy cutoff for the epsilon matrix is 18 Ry ($\approx$240 eV). The k-point mesh of 1 × 1 × 42 and both valence and conduction bands of 8 were set for optical absorption calculations. The band number for summation is 1100. The correction of the exact static remainder and the wire Coulomb truncation for 1D systems were also used.

\section{RESULTS AND DISCUSSIONS}
\subsection{Defect-induced phase transition}
The nanoribbon with armchair edges is cut from a WS$_{2}$ monolayer of the 2H phase, in which the tungsten (W) and chalcogen sulphur (S) atoms form a trigonal prism, as illustrated in Figures 1a-c. The edge atoms are passivated with hydrogen (H) atoms. The vacuum layer is aligned along the y axis, the width directions are along the x axis, and the periodic (or ribbon length) direction of the nanoribbons is along the z axis. The supercell vectors a, b, and c are aligned with the x, y, and z axes, respectively. As illustrated in Figure 1a, we employ AnWS$_{2}$ to refer to the WS$_{2}$ nanoribbons, where n denotes the number of W atoms in the supercell, and n=13 indicates 13 W atoms of the nanoribbon without defects in the supercell. Figure 1b and 1c depict the relaxed structures of A13WS$_{2}$ with W and S defects under various bending curvature radii R, respectively. The calculated band structures of the flat and bent A13WS$_{2}$ nanoribbons without defects and with S and W defects using PBE functional \cite{PBE} are shown in Figure 2a-i. Here, S defect in the nanoribbon refers to the A13WS$_{2}$ nanoribbon with a line of sulfur vacancies that is created by removing the central S atom in the lower S atom layer in the supercell and hence forming a S vacancy line parallel to the nanoribbon’s length direction.  The W vacancy refers to the line of tungsten vacancies that is created in a similar manner as sulfur vacancies.

The fundamental band gap of the WS$_2$ nanoribbon is indirect or direct, depending on the density functional approximation or the choice of bending curvatures and defects in the nanoribbons. By definition it is the difference between the valence band maximum (VBM) and conduction band minimum (CBM) (Table I).
The edge band gap (EG), specific to nanoribbons as quasi-one-dimensional systems, is defined as the difference between CBM and VBM of these localized in-gap states, while the non-edge band gap (NEG) is defined as the difference between the bottom of the conduction band continuum (CBC) and the top of the valence band continuum (VBC), both at the $\Gamma$ point, (Figure 2e). In addition to the edge bands near the Fermi level as in-gap states, there are four defect-induced bands above the upper edge bands in the conduction band region, clearly seen in the Figures 2d, 2e, and 2f. The defect gap (DG) at $\Gamma$  is defined as the difference between the minimum of the upper defect bands at $\Gamma$ and the maximum of the lower defect bands at $\Gamma$ as shown in Figure 2e. Here we report a similar variation in the NEG and EG of the pristine A13WS$_{2}$ nanoribbons with bending curvature as discussed recently by Tang \textit{et al.} \cite{spin-anisotropy} for A13WSe$_{2}$. In case of S defect, the conduction defect bands move close to the edge bands with increasing bending curvatures. For the largest bending curvature $\kappa$ = 0.142/Å ( R = 7 Å ), the lower defect bands merge into the upper edge bands. We find a semiconducting behavior in the pristine and S-defective A13WS$_{2}$ nanoribbons with various bending radii. In contrast, the A13WS$_{2}$ nanoribbons with central W vacancy exhibit metallic nature. Inspired by the potential of defects in semiconductor-metal phase change in  A13WS$_{2}$ nanoribbons, we have extented our study to relatively smaller and larger sized  (A7WS$_{2}$ and A25WS$_{2}$) nanoribbons with different bending curvatures. We show that a very controlled semiconductor-metal phase transition can be achieved in AnWS$_{2}$ (n = 7, 13, 25) nanoribbons with the combination of bending and defects as displayed in Figure 3. This can be proposed as an alternative way to realize a heterophase homojunction that shows ohmic contact in TMDs transistors via laser irradiation \cite{laser-driven}. This finding in WS$_{2}$ nanoribbons has the  potential to impact various fields and create new opportunities for technological advancement.

The density of states (DOS) analysis for A13WS$_{2}$ without defect, with S  and  W defects is shown in the top, middle and bottom panels of Figure 4, respectively. The total four W atoms located at the edges, two W atoms at each edge, are counted as the edge W atoms. The three W atoms near the position of the S defect and four W atoms near the  position of W defect of the nanoribbon are counted as the near defect W atoms, as shown in Supplemental Information (SI) Figure S1. In the pristine nanoribbons, the edge W atoms dominate the DOS around the Fermi level, as shown in the top panels of Figure 4, and the edge bands show up in the range of 0-0.5 and -0.5-0 eV. Similar to the findings for A13WSe$_{2}$ by Tang \textit{et al.} \cite{spin-anisotropy}, the W atoms close to the mid-ribbon area contribute more to the top of the valence band (approximately -0.5 to 0 eV) with increasing bending curvature.

In the nanoribbons with S defect, the edge W atoms again dominate the DOS around the Fermi level, as shown in the middle panels of Figure 4. In the flat nanoribbon, the near defect W atoms dominate the DOS at about 0.5 to 1 eV from the conduction bands. Those bands corresponding to the DOS at about -0.5 to -1 eV is contributed by all W atoms.

In the nanoribbons with W defect, the DOS analysis shows that the bands crossing the Fermi level  mainly consist of  the edge W atoms and near defect W atoms in flat and bent (R = 12 Å) nanoribbons, while mainly contributed by near defect W atoms in bent (R = 7 Å) nanoribbon, as shown in the bottom panels of Figure 4.

In our previous work, we demonstrate that only doping induces net magnetization. Doping makes our nanoribbons electrically charged, but the nanoribbons remain still electrically neutral with S line defects. Doping can introduce magnetic impurities into the material, which can lead to the formation of local magnetic moments and induce net magnetization. For example, doping a TMDs nanoribbon with transition metal impurities can introduce local magnetic moments that contribute to the overall magnetization of the material. Defects, on the other hand, can induce net magnetization through the modification of the electronic structure of the material. For example, defects can introduce localized states within the band gap of the material, which can lead to the formation of magnetic polarons or the localization of magnetic moments in the vicinity of the defects. Such magnetic moments are great testbeds for density functional approximations. We investigate the magnetization of pristine and defective A13WS$_{2}$ nanoribbons under three different bending curvatures: flat, R = 12 Å, and R = 7 Å. Our results, presented in Table II using the PBE functional, indicate that the pristine and S-defective nanoribbons have either no net or negligible magnetization, which is consistent with the results of our previous work \cite{spin-anisotropy}. However, we observe a net magnetization in the nanoribbons in the presence of a W vacancy.  Specifically, we find that for the flat nanoribbon, there is net magnetization along the Y-axis only, which is perpendicular to the plane of the nanoribbon. However, when we bend the nanoribbon along its width direction (X-axis), we observe a net magnetization along the ribbon's width direction as well. This suggests that bending can have a significant impact and controlling power on the magnetic properties of defective A13WS$_{2}$ nanoribbons.

For the pristine A13WS$_{2}$ nanoribbons the site resolved SOC energy on each W atom shows a similar pattern as it was found by Tang et al. \cite{spin-anisotropy}, as shown in Figure S3. The detailed analysis for the site resolved SOC energy on each W atom for defective A13WS$_{2}$ nanoribbons is presented in the SI.

Next, we analyze the fundamental band gaps and defect levels with semilocal density functional approximations, and compare the energy levels to the ones of higher-level hybrid functional HSE06 and one-shot GW.
We present the band gaps between VBM and CBM for the pristine A13WS$_{2}$ nanoribbons with various bending radii and with central S vacancy in Table I.

Meta-generalized gradient approximations (meta-GGAs) of the third-rung of Jacob's ladder \cite{Jacob-ladder} such as SCAN, r$^{2}$SCAN in the generalized Kohn-Sham (gKS) approximation \cite{gKS} are known to slightly open band gaps compared to PBE-GGA \cite{scan-bandgap,r2scan-bandgap}. The r$^{2}$SCAN results in larger band gaps in A13WS$_{2}$ nanoribbons than PBE. The TASK and mTASK meta-GGA's achieve even more nonlocality at the price of losing accuracy for ground state properties \cite{task,mTASk}. Within this work we focus on mTASK that proved a more suitable choice for the fundamental band gaps of materials with low dimension. The mTASK improves the band gaps in comparison to both PBE and r$^{2}$SCAN, capturing 55-59\% of the GW value for the flat pristine and S-defective nanoribbon. The accuracy decreases with bending, specifically when S vacancy is present, and the decrease in accuracy is more conspicious with mTASK than with r$^{2}$SCAN. r$^{2}$SCAN also makes the band gaps correctly direct as HSE06 and GW do. mTASK, though improves band gaps in general, but band dispersion was found unreliable \cite{task}. The GW method with the nonlocal self-energy correction produces significantly larger band gaps than density functional approximations, highlighting the importance of the non-local self energy in these quasi-one-dimensional materials. 

The computationally more feasible semilocal DFT is useful to establish the trend between in-gap edge, and defect states in bent and defected A13WS$_{2}$ nanoribbons, as shown in Table III and Figure 5. Our results reveal that the NEG band gap remains almost the same for the flat and bent (R = 12 Å) nanoribbons using PBE, while it slightly increases using r$^{2}$SCAN, mTASK and HSE06 . This result emphasizes the predictive power of r$^{2}$SCAN, mTASK and HSE06 compared to PBE in simulations of more localized defect states. However, increasing the bending curvature leads to a decrease in the NEG band gap using PBE, r$^{2}$SCAN, mTASk and HSE06 density functionals. Using the GW method, we observe a more enhanced decrease in the NEG band gap with increasing bending curvature.
Edge and defect states are more localized than bulk-like band states. Despite that they are localized states, density functional approximations display a different trend in accuracy. While there is no remarkable discrepancy between the accuracy of semilocal density functional approximations and HSE06 or GW for defect band gaps, the accuracy of the edge states varies more significantly from switching from PBE to meta-GGA and non-local approximations.

\begin{table}
\caption{Comparison of the actual band gaps between VBM and CBM using PBE, r$^{2}$SCAN, and mTASK density functionals with the higher-level hybrid functional HSE06 or with the G0W0 approximation within many-body perturbation theory for bent and defective A13WS$_{2}$ nanoribbons.}
\centering
\begin{tabular}{llllllll}
\hline \hline
\multicolumn{1}{l}{Bending} & \multicolumn{1}{l}{Defect} & \multicolumn{1}{l}{PBE (eV)} & \multicolumn{1}{l}{r$^{2}$SCAN (eV)}  &  \multicolumn{1}{l} {mTASK (eV)} & \multicolumn{1}{l} {HSE06 (eV)} & \multicolumn{1}{l}{GW (eV)} \\
\hline
R = $\infty$ (flat) &   None       & 0.47 (indirect)     & 0.73 (direct)      & 1.00 (direct)  & 1.25 (direct)  & 1.82   (direct)          \\
                      & S  & 0.49 (indirect)    & 0.75  (direct)     & 1.08 (direct)  & 1.27  (direct)  & 1.83  (direct)          \\
    R = 12 \AA        & None       & 0.42 (indirect)    & 0.67 (direct)      & 0.84 (direct)   & 1.13  (direct) & 1.80  (direct)          \\
                     & S   & 0.53 (indirect)    & 0.79  (direct)     & 0.96 (indirect)   & 1.28 (direct)  & 1.95   (direct)         \\
R = 7 \AA            & None       & 0.35 (indirect)   & 0.54  (direct)     & 0.77  (direct)    & 1.00 (direct)  & 1.67   (direct)        \\
                     & S  & 0.45 (indirect)   & 0.53  (indirect)     & 0.53 (indirect)  & 1.00  (direct)  & 1.66  (direct)           \\
\hline \hline  
\end{tabular}
\end{table}

\begin{table}[!ht]
\caption{Calculated magnetization (in unit Bohr magneton $\mu$${_B}$ per supercell) for bent and defective A13WS$_{2}$ nanoribbons with the PBE approximation.}
\centering
\setlength\tabcolsep{2pt}
\begin{tabularx}{\linewidth}{l l *{12}{X}}
\hline\hline
Defect & \multicolumn{3}{c}{Flat} & \multicolumn{3}{c}{R12} & \multicolumn{3}{c}{R7} \\ \hline
& ~~X-axis & ~~Y-axis & Z-axis & X-axis & Y-axis & Z-axis & X-axis & Y-axis & Z-axis \\ \hline
None & ~~0 & ~~0 & 0 & -0.004 & 0 & 0 & -0.004 & 0 & 0\\
S & ~~-0.007 & ~~0 & 0 & -0.003 & 0 & 0 & 0 & 0 & 0 \\
W & ~~0 & ~~0.275 & 0 & -0.022 & 0.116 & 0.002 & -0.064 & 0.106 & 0  \\
\hline\hline
\end{tabularx}
\end{table}

\begin{table}[!ht]
\caption{Comparison of NEG, EG and DG for PBE, r$^{2}$SCAN, and mTASK density functionals with the higher-level hybrid functional HSE06 or with the G0W0 approximation within many-body perturbation theory for bent and defective A13WS$_{2}$ nanoribbons. The fundamental band gaps are in eV.}
\centering
\setlength\tabcolsep{2pt}
\begin{tabularx}{\linewidth}{l l *{22}{X}}
\hline\hline
& Defect & \multicolumn{3}{c}{PBE} & \multicolumn{3}{c}{r$^{2}$SCAN} & \multicolumn{3}{c}{mTASK} & \multicolumn{3}{c}{HSE06} & \multicolumn{3}{c}{GW} \\  \cline{3-5} \cline{6-8} \cline{9-11} \cline{12-14} \cline{15-17}
&& NEG & ~EG & DG & NEG & ~EG & DG & NEG & ~EG & DG & NEG & ~EG & DG & NEG & ~EG & DG \\ \hline
Flat & None & 1.90 & 0.48 & N/A & 2.02 & 0.73 & N/A & 2.18 & 1.00 & N/A & 2.60 & 1.25 & N/A & 3.43 & 1.82 & N/A \\
& S & 1.97 & 0.50 & 0.21 & 2.05 & 0.75 & 0.24 & 2.26 & 1.08 & 0.18 & 2.61 & 1.27 & 0.20 & 3.52 & 1.83 & 0.23 \\
R12 & None & 1.92 & 0.42 & N/A & 2.10 & 0.67 & N/A & 2.33 & 0.84 & N/A & 2.70 & 1.13 & N/A & 3.37 & 1.80 & N/A \\
& S & 1.98 & 0.54 & 0.63 & 2.14 & 0.79 & 0.64 & 2.36 & 0.96 & 0.58 & 2.78 & 1.28 & 0.68 & 3.47 & 1.95 & 0.67 \\
R7 & None & 1.21 & 0.38 & N/A & 1.33 & 0.54 & N/A & 1.47 & 0.77 & N/A & 1.81 & 1.00 & N/A & 2.59 & 1.67 & N/A \\
& S & 1.60 & 0.46 & 0.77 & 1.73 & 0.53 & 0.57 & 1.87 & 0.54 & 0.48 & 2.33 & 0.99 & 0.63 & 3.07 & 1.66 & 0.74 \\ \hline\hline
\end{tabularx}
\end{table}

\begin{figure}[h!]
    \centering
    \begin{adjustbox}{right=18.1cm}
    \includegraphics[scale=0.75]{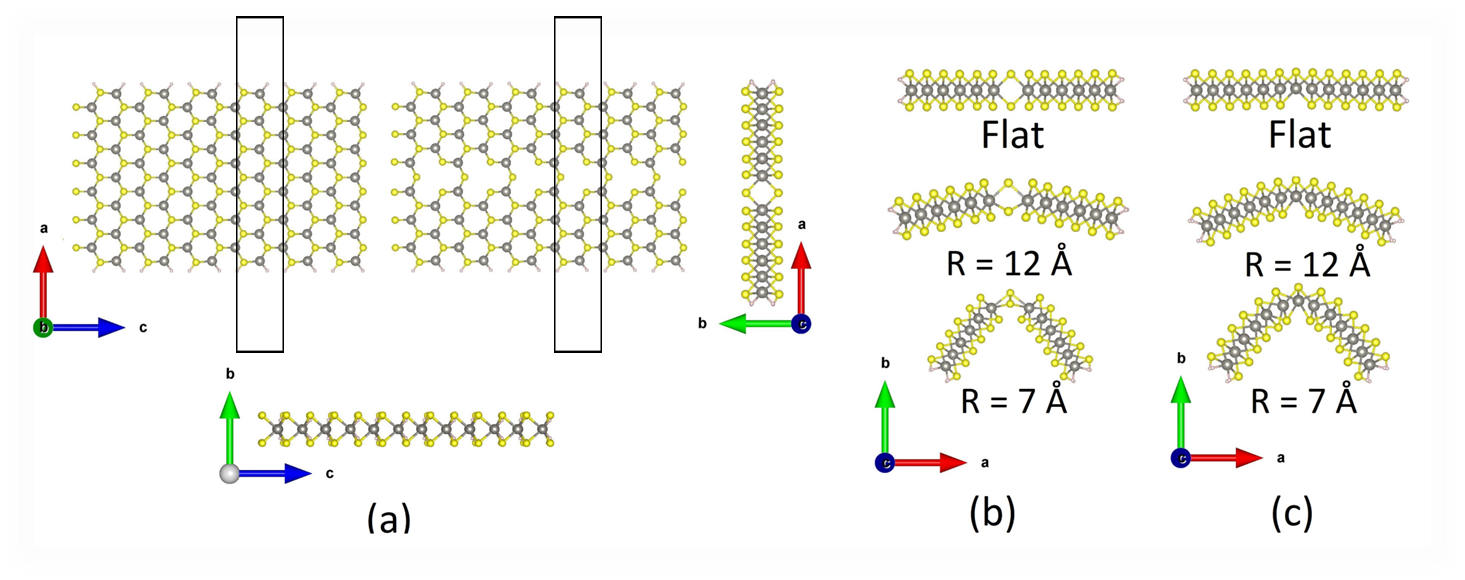}
    \end{adjustbox}
    \caption{The structures of the armchair WS$_{2}$ nanoribbons and their relaxed structures with S defect and W defect under different bending curvature radii R. The flat nanoribbon AnWS$_{2}$ with n=13 is shown in (a). n in AnWS$_{2}$ represents the number of W atoms in the pristine nanoribbon in the periodic supercell, whose unit vectors a, b, and c are aligned with axes x, y, and z, respectively. Blue balls represent W atoms, yellow balls are for S atoms and small white balls are H atoms. The side views of the A13WS$_{2}$ nanoribbon are also shown in (a). Panels (b) and (c) show the relaxed structures of the A13WS$_{2}$ nanoribbon under flat, R = 12 Å, and R = 7 Å with W defect and S defect, respectively.}
    \label{fig:fig1}
\end{figure} 

\begin{figure}[h!]
    \begin{adjustbox}{right=16.3cm}
    \includegraphics[scale=0.16]{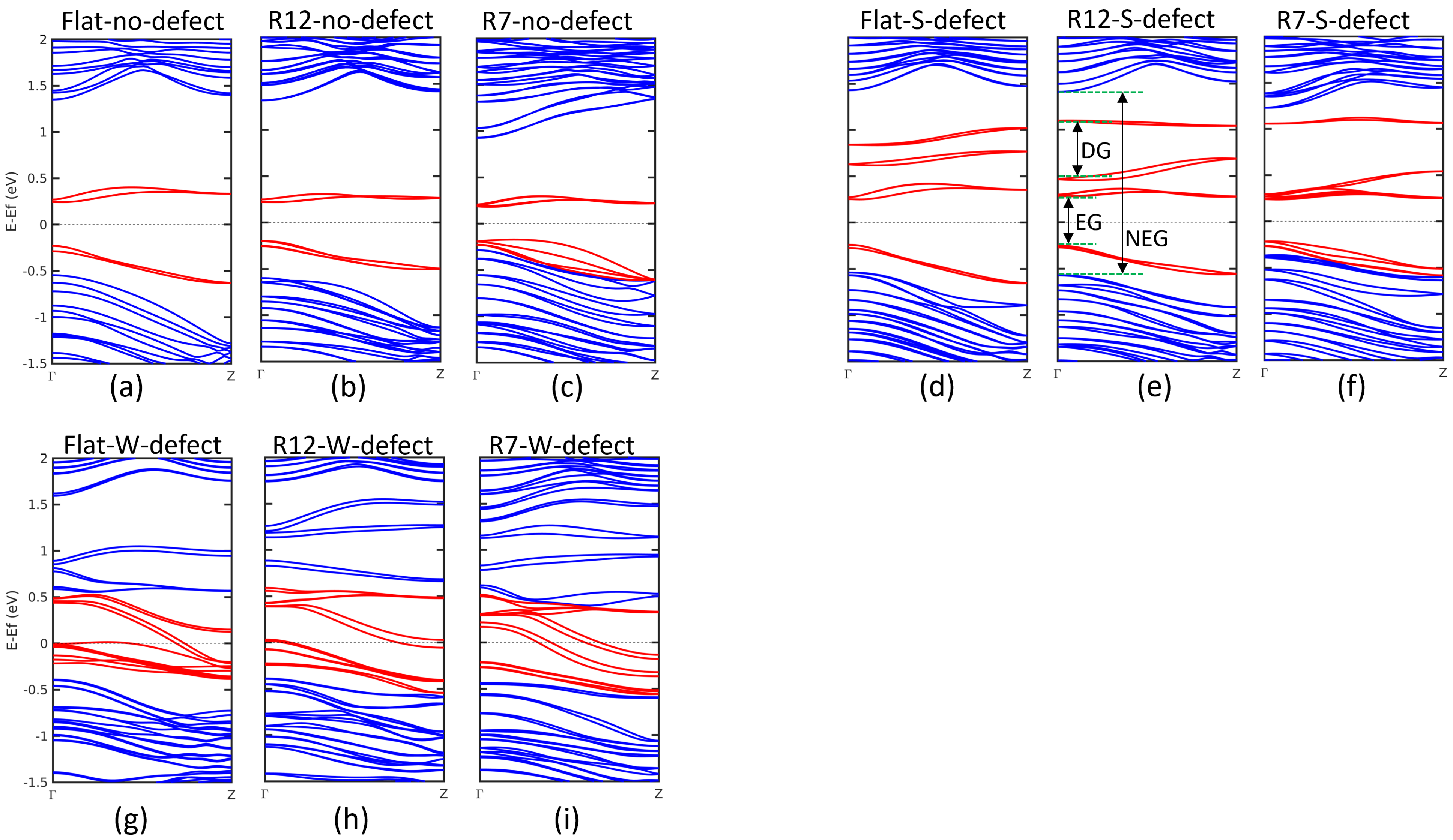}
    \end{adjustbox}
    \caption{The band structures of A13WS$_{2}$ nanoribbons without defect, S defect and W defect under different bending curvature radii R. Panels (a), (b) and (c) are the calculated band structures of the A13WS$_{2}$ nanoribbons without defects for various bending curvature radii of  R = $\infty$ (flat), R = 12 Å and R = 7 Å, respectively. Panels (d), (e), and (f) are the calculated band structures of the A13WS$_{2}$ nanoribbons with S defect for various bending curvature radii of  R = $\infty$ (flat), R = 12 Å and R = 7 Å, respectively. Panels (g), (h), and (i) are the calculated band structures of the A13WS$_{2}$ nanoribbons with W defect for various bending curvature radii of  R = $\infty$ (flat), R = 12 Å and R = 7 Å, respectively. The edge bands around the Fermi level and defect bands above the upper edge bands are shown in red and other bands are in blue.}
    \label{fig:fig2}
\end{figure} 

\begin{figure}[h!]
    \includegraphics[scale=0.45]{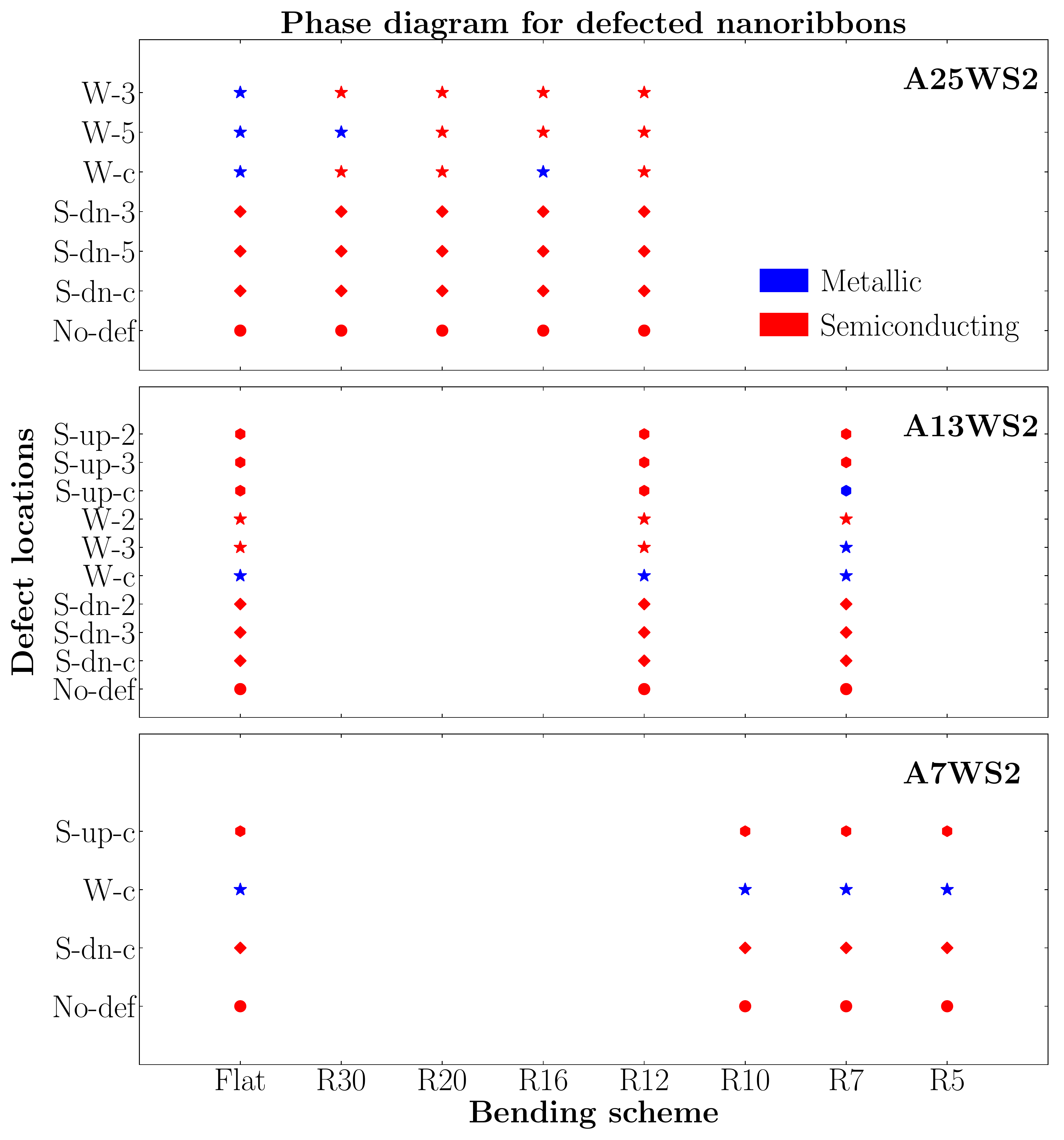}
    \caption{The phase diagram for different sizes of WS$_{2}$ nanoribbon with various bending radii. The defect locations labeled in y-axis for different sizes of the nanoribbons are shown in the Figure S2. No-def in y-axis refers to the nanoribbons without defects.}
    \label{fig:fig2}
\end{figure} 

\begin{figure}[h!]
    \includegraphics[scale=0.20]{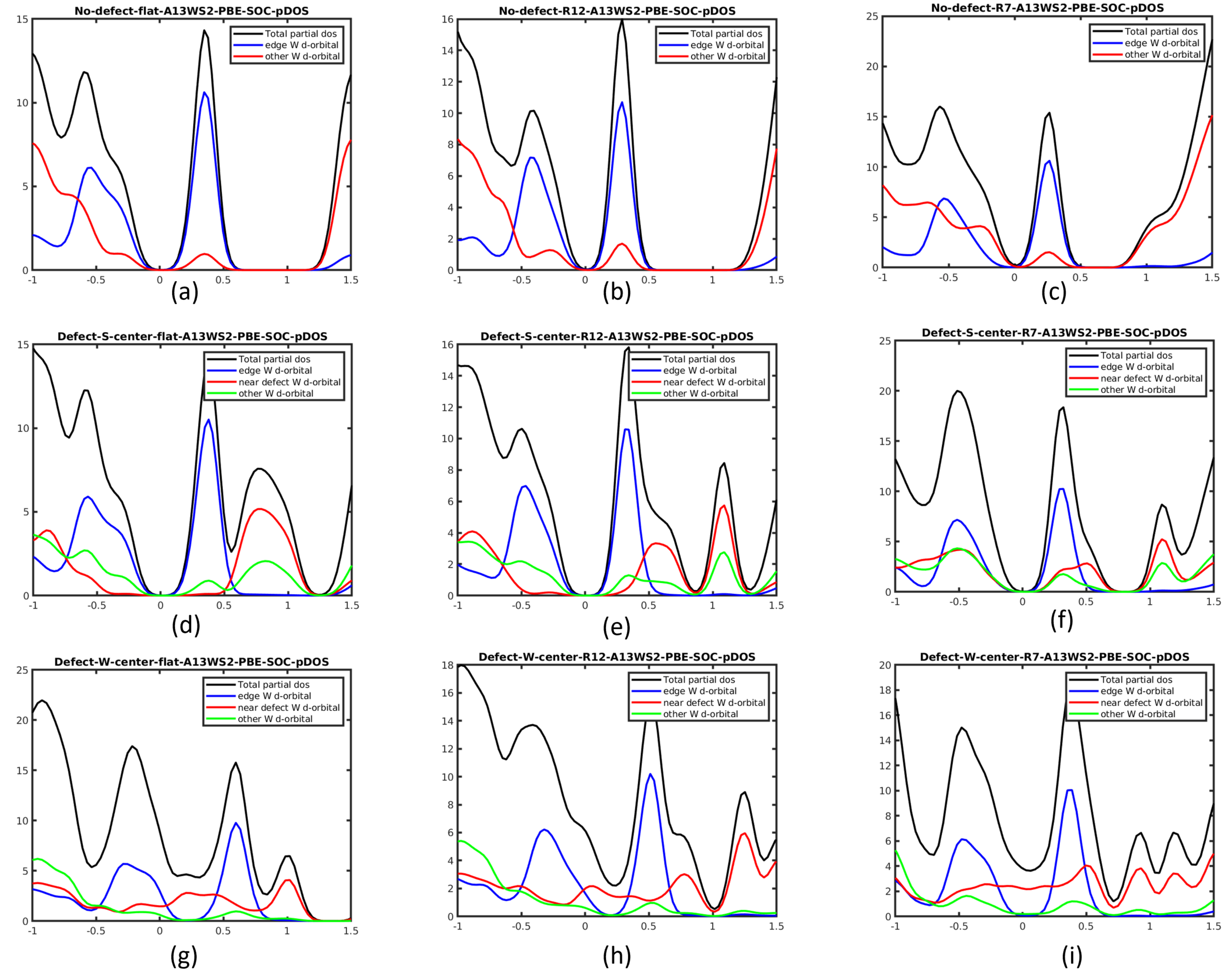}
    \caption{Panels (a), (b) and (c) represent the analysis of the density of states of A13WS$_{2}$ nanoribbon without defect under different bending radii. Panels (d), (e) and (f) represent the analysis of the density of states of A13WS$_{2}$ nanoribbon with S defect under different bending radii. See the title on each plot for the bending radius. Panels (g), (h) and (i) represent the analysis of the density of states of A13WS$_{2}$ nanoribbon with W defect under different bending radii. See the title on each plot for the bending radius. They are calculated with PBE and the SOC effect included.}
    \label{fig:fig3}
\end{figure}

\begin{figure}[h!]
\centering
\includegraphics[scale=0.19]{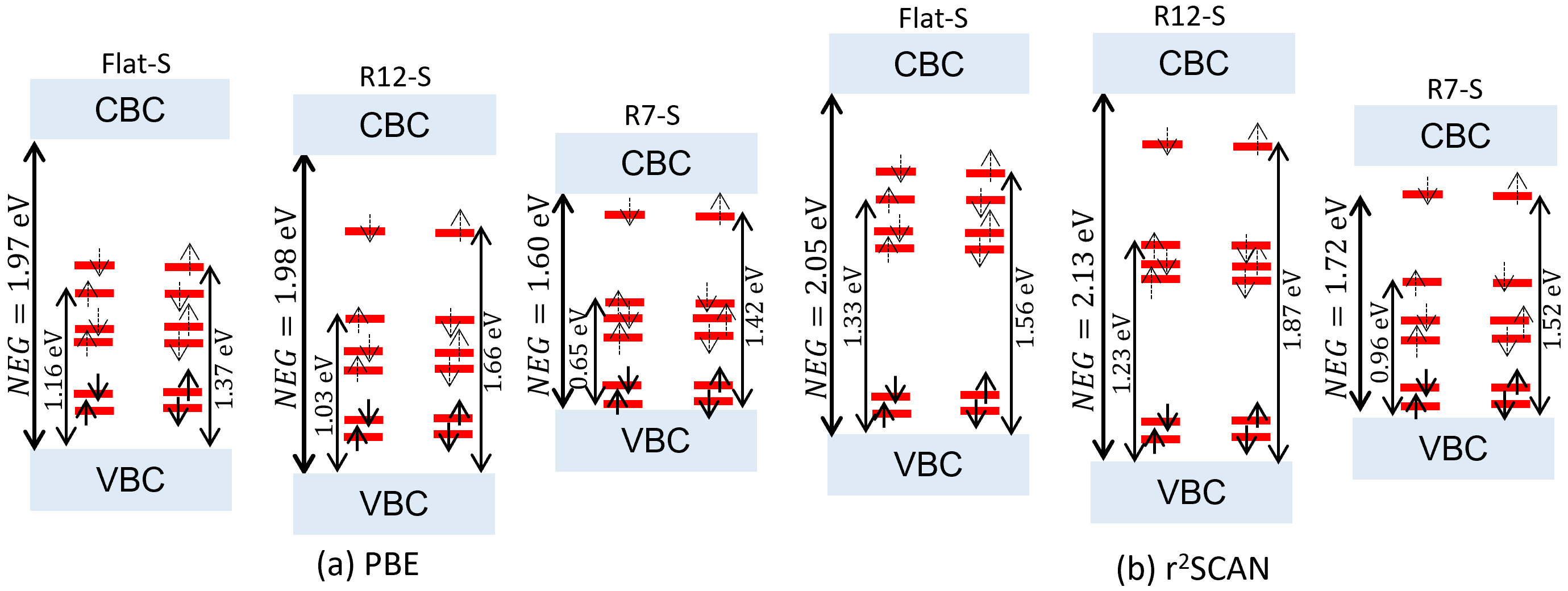}\vspace{1\baselineskip}
\includegraphics[scale=0.19]{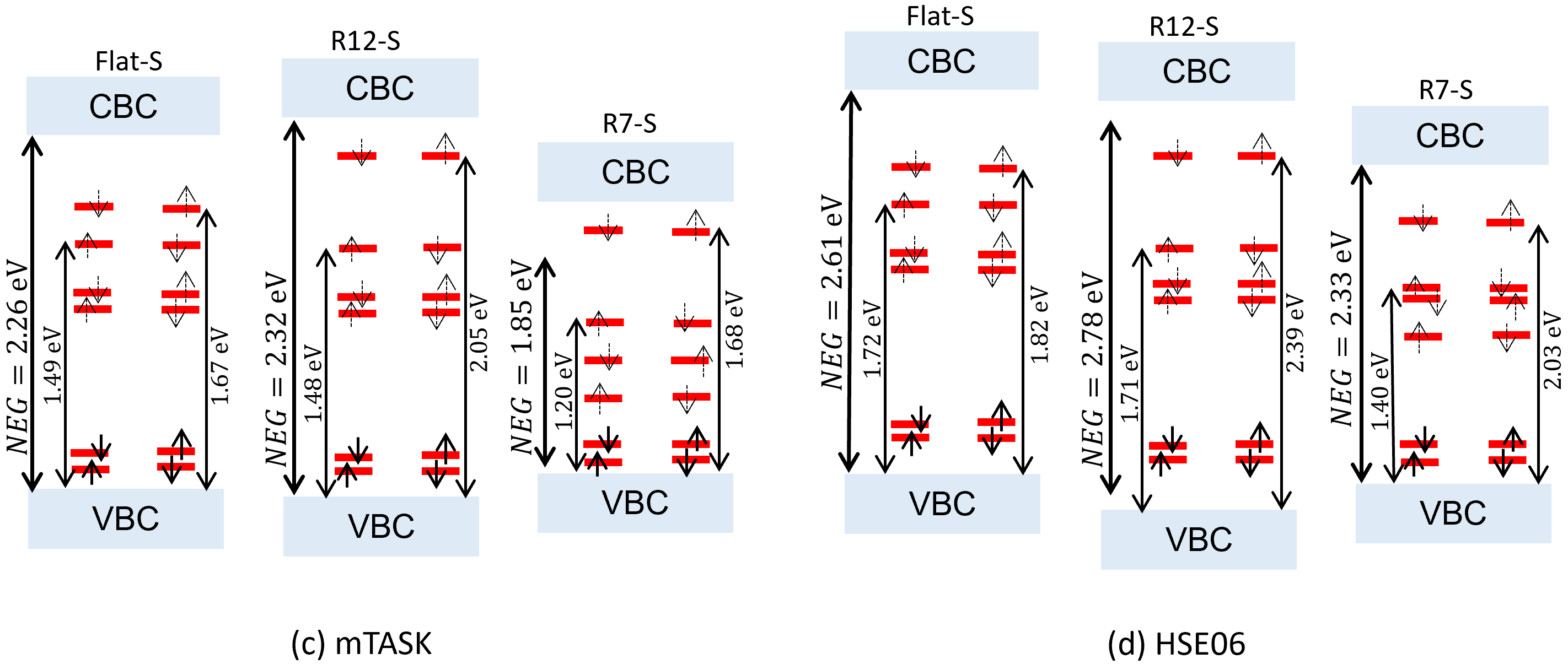}\vspace{1\baselineskip}
\begin{adjustbox}{left=14.0cm}
\includegraphics[scale=0.19]{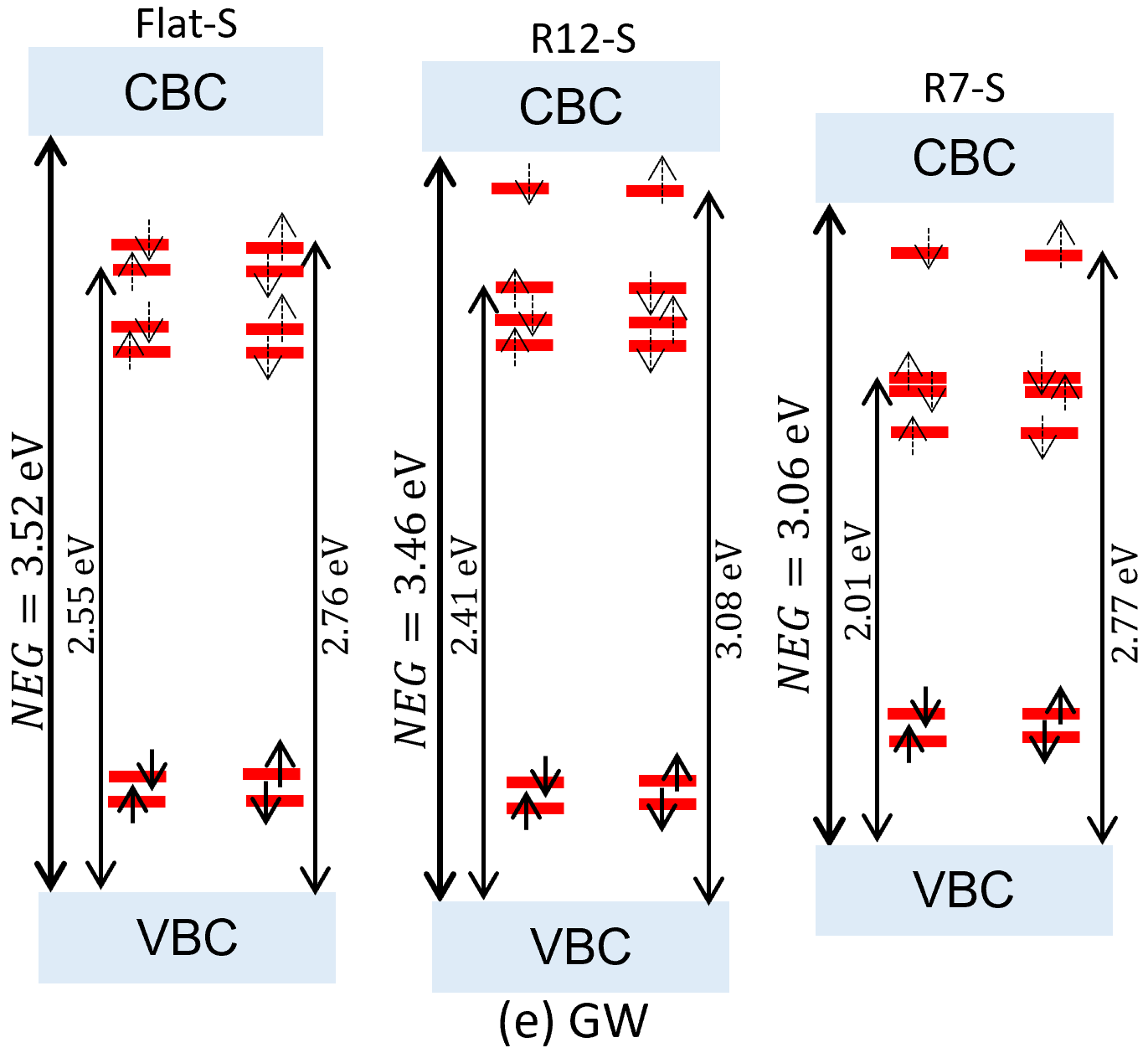}
\end{adjustbox}
\caption{Schematic diagram for quasiparticle energies of edge states and defect states at the $\Gamma$ point of A13WS$_{2}$ using PBE, r2SCAN and mTASK density functionals, the higher-level hybrid functional HSE06, and the GW method. The bottom four levels correspond to the lower edge bands occupied with electrons with spin up or down, while the other remaining levels correspond to the unoccupied upper edge bands and defect bands with spin up or down.}
\label{fig:fig4}
\end{figure}

\subsection{Optical absorption and exciton states}
\subsubsection{A13WS$_{2}$ nanoribbon without defects}

Panels (a)-(h) of Figure 6 represent different physical parameters of the band structures for the flat and bent (with bending curvature radius R = 7 Å) A13WS$_{2}$ nanoribbon without defects. In the case of the flat nanoribbon, the site-resolved DOS (density of states) plot (panel (b)), shows that the relatively separated eight bands around the Fermi level  mainly arise from the edge W atoms (blue dots). Among the eight bands, there are four conduction bands forming the upper edge bands and four valence bands forming the lower edge bands. The spin-polarization of those edge bands calculated from PBE (panel (c)) and $G_0W_0$ (panel (d)) are similar, although $G_0W_0$ produces a much larger band gap (1.82 eV) than that (0.47 eV) of  PBE, as a manifestation of the significant self-energy correction from the many-body effects. In the bent nanoribbon at R = 7 \(\text{\AA}\), the four lower edge bands are merged into the valence band continuum, as shown in the DOS plot (panel (f)). The PBE band gap of the bent nanoribbon is reduced to 0.35 eV from the 0.47 eV of the flat nanoribbon, and the $G_0W_0$ gap of the bent nanoribbon is reduced to 1.67 eV from the 1.82 eV of the flat one. 

This rich interplay of bending and screening contributes to the optical response of nanoribbons, that reflects the alteration of screening with increased bending.
Panels (i)-(l) of Figure 6 show the calculated optical absorption and exciton spectra for the flat and bent nanoribbons, showing strong excitonic effects. For the flat nanoribbon, the absorption mainly peaks around 0.65 eV and 1.00 eV. The bent nanoribbon shows two broad absorption peaks, which slightly extend to the low energy direction by about 0.05 eV, compared with those of the flat case, due to the slightly decreased fundamental band gap. A few small peaks are detected in the bent nanoribbon at around 1.5 to 1.8 eV, which are not present in the flat nanoribbon. Bending induces more bright exciton states within the range of 0.5-1.0 eV. 

Both the flat and bent nanoribbons show several dark exciton states, and the lowest exciton state is a dark exciton, with mainly  transitions from the lower edge to the upper edge bands (as shown in Figure 7(e)). The dark exciton at 0.45 eV exhibits similar physics (Figure S5). The bright excitons makes peak A at 0.64 eV in the flat nanoribbon (Figure 6(i), and its spatial extension is shown in the Supplementary Material, (Figures S6 and S7). Another  bright exciton with high oscillator strength appears at peak B at 0.99 eV (Figure S8).

In the bent nanoribbon with R = 7 Å, the dark excitons are shifted to 0.41 eV and 0.48, and they become localized at the edge (Figure S9 and S10). Bright excitons shift similarly by bending,  peak A to 0.63 eV, and peak B to 0.95 eV, mostly preserving the nature of their transition states, (Figures S11 and S12).                                        
 
\subsubsection{A13WS$_{2}$ nanoribbon with a line of S vacancy defects}                                               
We also calculated the optical absorption and exciton states for the A13WS$_{2}$ nanoribbon with a line of sulfur vacancy defects. In the flat defective nanoribbon, the relaxed structure almost remains flat and only slightly deformed around the middle of the ribbon. The PBE band structure and the DOS resolved band structure are shown in Figure 8(a) and (b), respectively. The four upper edge and the four lower edge bands remain around the Fermi level. The defect-derived four conduction bands are located higher than the upper edge bands. The defect-derived valence bands are mixed with bulk bands in the valence band continuum (VBC), while the top of VBC is mainly bulk atom derived. The fundamental band gap is 0.49 eV from PBE and 1.83 eV from $G_0W_0$.

For the same defective A13WS$_{2}$ nanoribbon under bending at R = 7 Å, the PBE band structure and the DOS resolved band structure are shown in Figure 8 (e) and (f), respectively. The conduction bands C1 and C2 are almost degenerate in energy. The DOS of C1 and C2 shows a mixture of edge (blue spots) and defect states (green spots) in Figure 8(f), while the valence band V1 mainly consists of the bulk states (red spots) of the nanoribbon. The spin-polarization resolved band structures of the same bent nanoribbon calculated from both PBE and $G_0W_0$ methods (Figure 8(g) and (h)), indicate that C1 is spin-up, while C2 is spin-down polarized within PBE, while the polarization is switched in $G_0W_0$. The fundamental gap increases from 0.45 eV in PBE to 1.66 eV in $G_0W_0$.

The optical absorption and exciton spectra are shown in Figure 8 (i) and (j) respectively for the flat and defective nanoribbon and in Figure 8(k) and (l) for the bent defective nanoribbon with R = 7 Å. The optical absorption of the defective nanoribbon (Figure 8 (i) ) remains similar to that of the pristine nanoribbon (Figure 6(i)), especially regarding the positions of peaks A and B and the overall shapes of the peaks, as indicators that line defects behave similarly as edge state, and they do not significantly alter the optical response. Since the S vacancy line almost has no coupling with the edge bands of the flat ribbon, the positions of the upper and lower edge bands in the $G_0W_0$ band structures do not differ in the pristine and defective flat nanoribbon, leaving the optical absorption spectra with only some minor changes with excitonic peaks around 1.5 eV in the defective nanoribbon.

In the flat defective nanoribbon, the exciton states with the lowest energy at 0.46 and 0.47 eV is still a dark exciton, (Figure S13, S14). There are two bright excitons with almost degenerate energy at 0.65 eV, and with mixed spin configurations (Figures S15 and S16). Peak B consists of several bright excitons and there are two energy-degenerate ones at 0.997 eV. The wavefunctions of the two bright excitons show slightly nodal features both in the real space and momentum space and they occupy different edge (Figures S17 and S18). There is also a bright exciton at 1.01 eV. The valence bands V1-V8 have major contributions from the edge and bulk atoms and a minor contribution from the defect, (Figure S19).

In the bent defective nanoribbon, the optical absorption spectrum shows a similar shape and peak positions with those of the pristine nanoribbon with the same bending curvature radius R = 7 Å, especially within the energy range of 0.5-1.0 eV. Similar to the pristine bent nanoribbon, the lower four edge bands merge into the valence band continuum under bending, and the V1 and V2 are bulk-derived. The four upper edge bands are mixed to some extent with the two lower defect-derived conduction bands. These common features make the optical absorption spectra of the pristine and defective bent nanoribbons share similar shapes within 0.5-1.0 eV.

Due to the mixture of the upper edge bands and the defect derived conduction bands, the exciton states formed in the defective bent nanoribbon show some different features. The exciton with the lowest energy at 0.43 eV is a dark exciton, which is due to the transitions from V1-V8 (including the bulk and lower edge bands) to C1-C6 (including the four upper edge bands and two defect derived conduction bands). The wavefunction of this exciton in real space is mainly located on one of the edges and also extended slightly along the central defect S line, confirming the relevant transitions to defect derived conduction bands (see supplemental Figure S20). The dark exciton at 0.51 eV is very similar to the one at 0.43 eV, and it is located on another edge and extended along the central defect S line (see supplemental Figure S21). Peak A consists of several bright excitons. The one at 0.57 eV is mainly due to the transition from V1 to C2 around $\Gamma$. Since V1 and C2 are both up-spin polarized, this bright exciton is mainly of like-spin configuration. Also, because of the admixing of edge and defect contributions in V1, the wavefunction of this exciton is largely extended along the central defect S line (see supplemental Figure S22). The wavefunctions of the two bright excitons at 0.64 eV and 0.66 eV also show the similar spatial extension along the central defect line (see supplemental Figures S23 and S24), since their transitioned bands have defect derived features. The two excitons involve more band transitions and have a mixed spin configuration. The excitons in peak B show spatially more extended and nodal features, as can be seen from those of excitons at 0.93 eV and 0.95 eV (see supplemental Figures S25 and S26), which are related to transitions from V1-V8 to C1-C6. Generally, they have a rich and mixed spin configuration. 

The optical absorption shows a robust feature against the central sulfur defect line, while the exciton’s extension can be controlled across the nanoribbon with appropriately applied bending. This may provide a fine-tuning knob for the exciton dynamics in the nanoribbons through the defect line and bending engineering. For example, one can switch the exciton between the edge and the center of the nanoribbon by bending, and selectively switch the interaction of excitons with other attached species.\\~\\

Unlike the optical absorption spectra, exciton lifetimes are strongly impacted by line defects and bending. We calculated the intrinsic exciton lifetimes for the pristine and defective A13WS$_{2}$ nanoribbons under various bending curvatures, and the results are presented in Figure 9. Typically, low-energy dark excitons have intrinsic lifetimes ranging from 0.1 to around 300 $\mu$s, much smaller than those observed in monolayer WS$_{2}$ which is 3.7 ms \cite{radlife}. The further reduced dimensionality of the quasi-1D nanoribbon system increases electron-electron interactions and the strength of the exciton transition dipole, and hence reduces the intrinsic lifetime. The flat and non-defective nanoribbon has a larger lifetime (\raisebox{0.5ex}{\texttildelow} 300 $\mu$s) for the lower energy dark excitons in comparison to the flat and defective nanoribbon which has the lifetime of \raisebox{0.5ex}{\texttildelow} 2 $\mu$s, consistent with the usual experimental results of faster recombination of excitons in defected monolayer systems \cite{exciton-dynamics}. For the bent nanoribbons with or without defects, the intrinsic lifetimes (\raisebox{0.5ex}{\texttildelow} 0.1 $\mu$s) for the low energy dark excitons are further reduced. This may be due to the merging of the lower edge states into valence band continuum at large bending, which results in a large mixing of edge and bulk states for the hole wavefunction, increasing the overlapping of wavefunctions of the electron and hole in the excitons, hence increasing the recombination rate of electron and hole and decreasing the lifetime of excitons. our results show that bending and defects can manifestly modify and control the lifetimes of excitons in the WS$_{2}$ nanoribbons. These relatively large lifetimes (\raisebox{0.5ex}{\texttildelow} 0.1 - 300 $\mu$s) and tunability are instrumental in realizing exciton-based quantum controls and probing exciton dynamics.

\begin{figure}[h!]
    \includegraphics[scale=0.09]{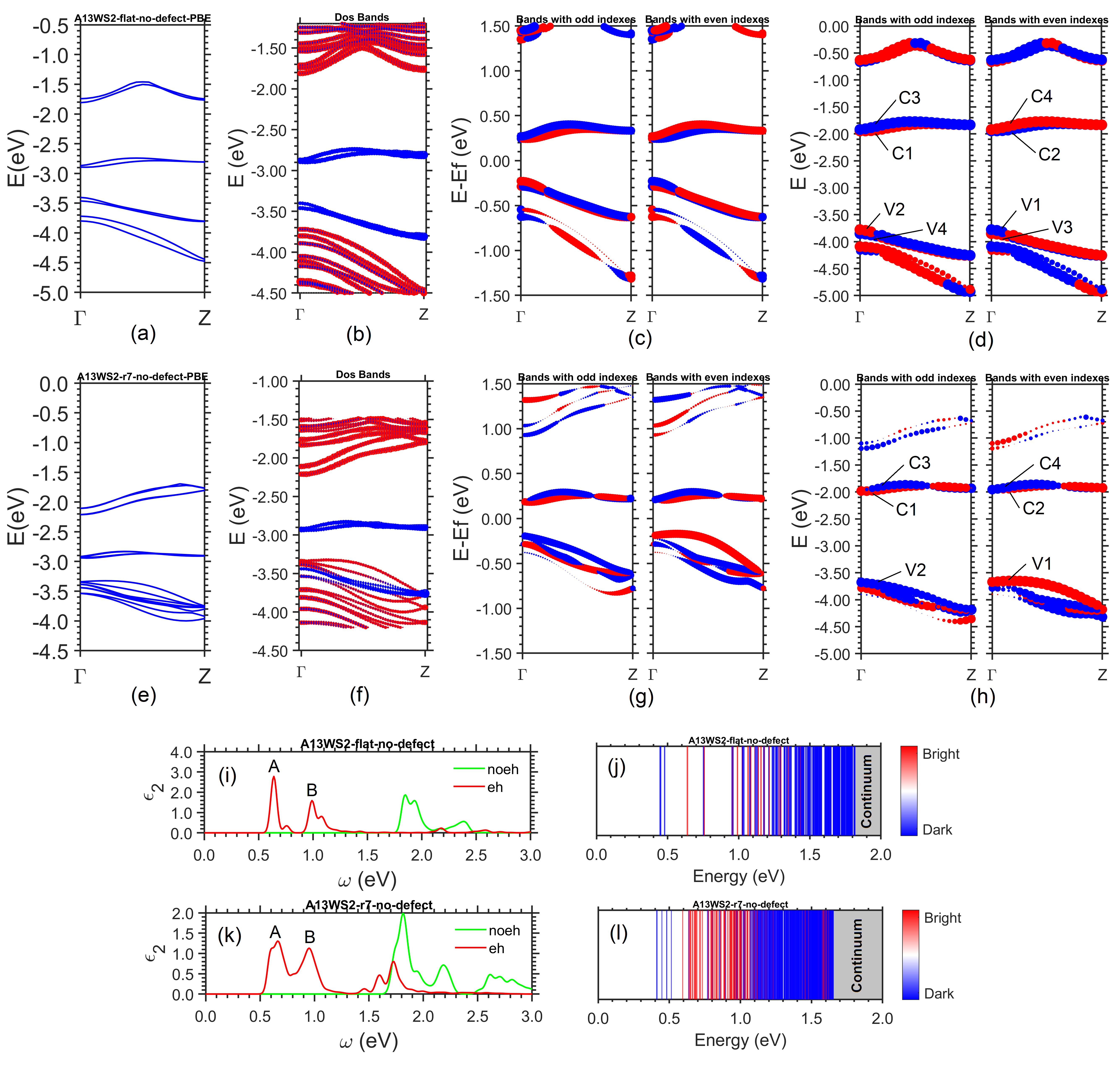}
    \caption{The various band structures of the pristine A13WS$_{2}$ nanoribbon and its optical absorption and exciton spectra. Panels (a), (b), (c) and (d) are the PBE+SOC band structure, the site-DOS-resolved band structure, the PBE+SOC spin-polarization-resolved band structure, and the $G_0W_0$ spin-polarization-resolved band structure for the flat nanoribbon, respectively.  Panels (e), (f), (g) and (h) are the similarly organized band structures for the bent nanoribbon at R = 7 Å. The blue dots in panels (b) and (f) represent the contribution by edge W atoms, while red ones by other atoms in the nanoribbon. In panels (c), (d), (g) and (h), the odd number indexed bands are shown on the left sub-panel and the even number indexed ones on the right sub-panel for clarity, and the spin-up and spin-down polarizations are represented by red dots and blue dots respectively. Band labels V1, V2, … and C1, C2, … are the other way to count the bands away from the Fermi level. Panels (i) and (j) show the optical absorption and exciton spectra, respectively, for the flat nanoribbon without defects. The former is plotted as the imaginary part of the dielectric function as a function of photon energy with the red curve representing the GW+BSE result with electron-hole (eh) interactions and the green one for that without eh (noeh) interactions, both with constant broadening of 28 meV. Panels (k)and (l) are similarly plotted and arranged for the bent (with R = 7 Å) nanoribbon without defects. Bright (dark) exciton states are represented by red (blue) lines in panels (j) and (l).}
    \label{fig:fig6}
\end{figure} 

\begin{figure}[h!]
    \includegraphics[scale=0.19]{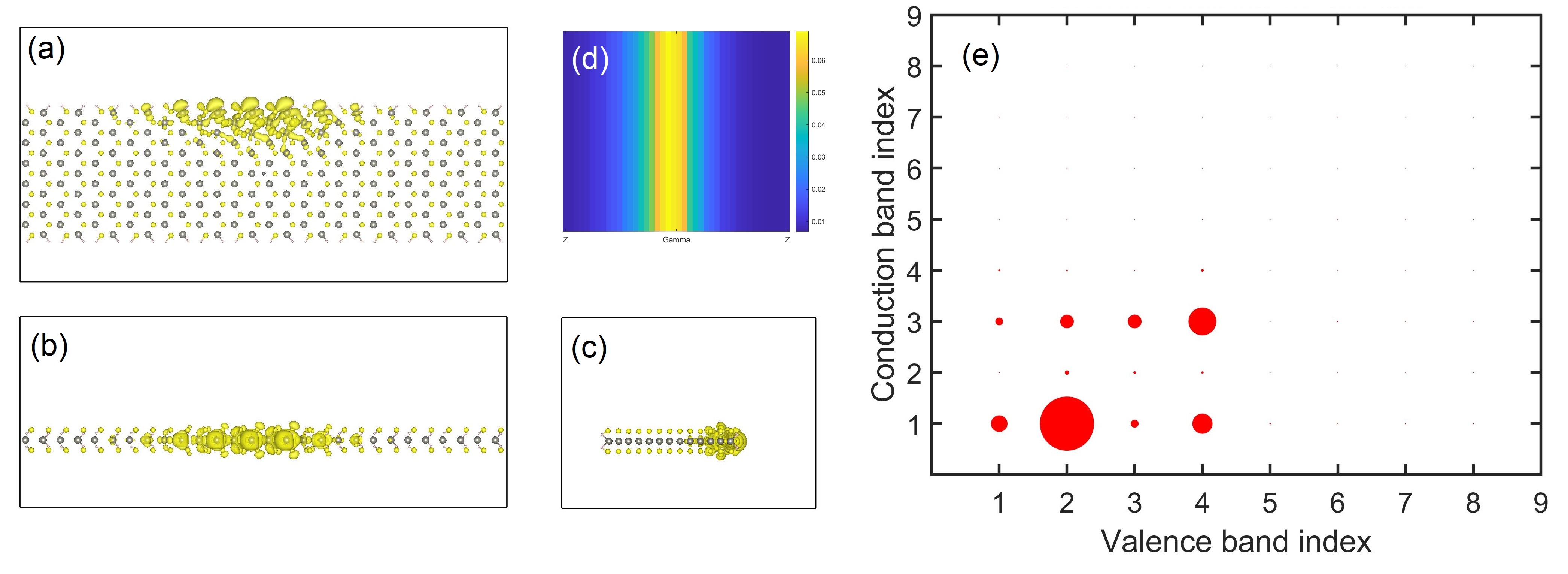}
    \caption{The dark exciton at 0.44 eV of the pristine flat A13WS$_{2}$ nanoribbon. Panels (a), (b) and (c) represent the three views of the isosurface contour of the modulus squared of the exciton wavefunction in real space, where the hole (black spot in (a)) is located at the center of the ribbon and near a W atom. The profile of the modulus squared exciton wavefunction in k space is shown in panel (d). Panel (e) shows the contributing hole (valence) and electron (conduction) bands for each exciton. The valence (conduction) band index is counted downwards (upwards) from the Fermi level. The spot size in (e) is proportional to $\sum_{k}|{A_v,_c(k)|^2}$  and represents the contributing weight from the v-c pair.}
    \label{fig:fig7}
\end{figure} 

\begin{figure}[h!]
    \includegraphics[scale=0.15]{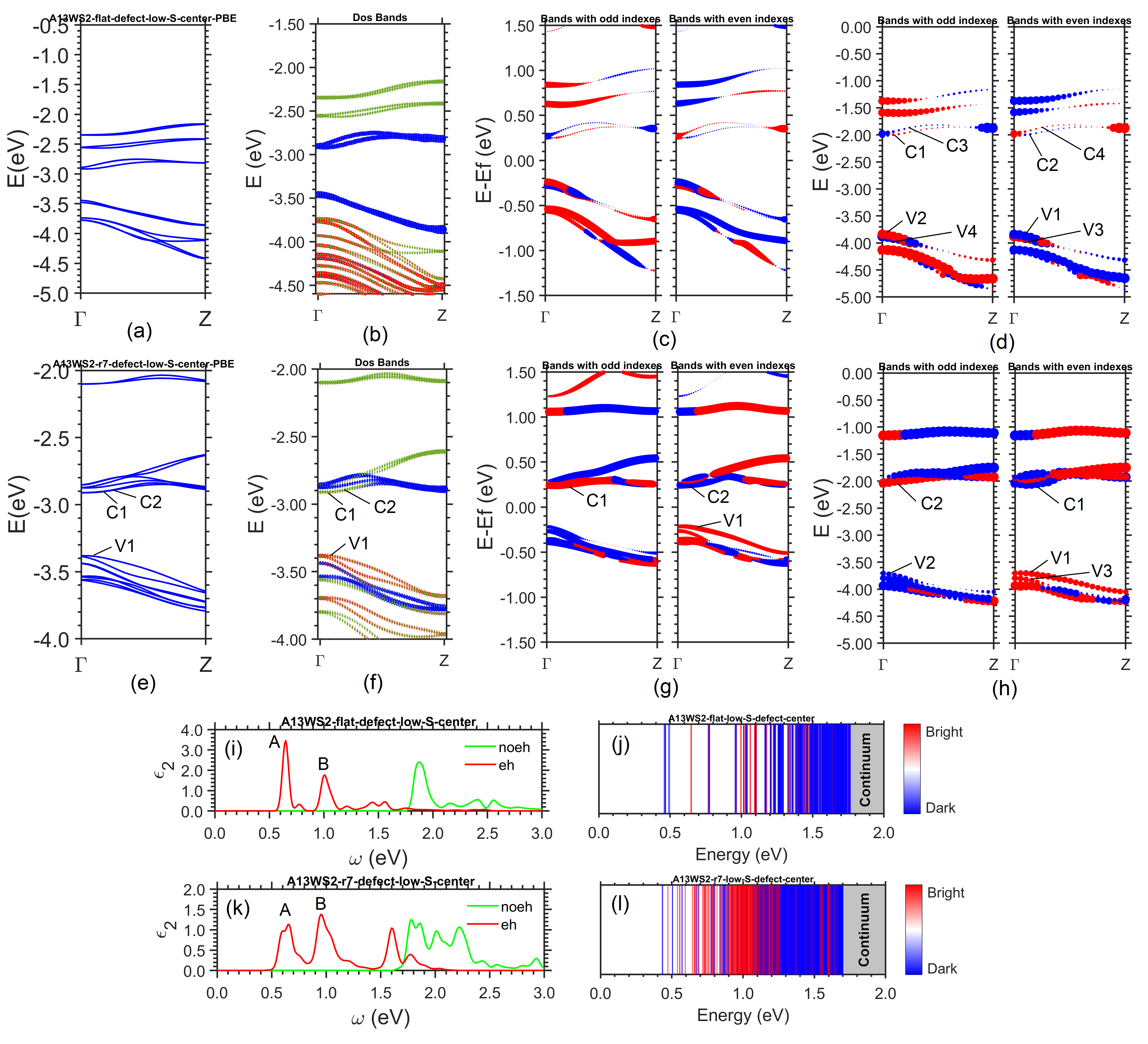}
    \caption{The various band structures of the defective A13WS$_{2}$ nanoribbon with a sulfur defect line, and its optical absorption and exciton spectra. Panels (a)-(l) are similarly obtained and arranged as in Figure 6, with panels (a)-(d), (i) and (j) for the defected flat nanoribbon and panels (e)-(h), (k) and (l) for the defective bent nanoribbon with R = 7 Å. The blue dots in panels (b) and (f) are contributed by edge W atoms, the green ones from the atoms near the defect, and the red ones from other bulk atoms in the nanoribbon.}
    \label{fig:fig8}
\end{figure} 

\begin{figure}[h!]
    \includegraphics[scale=0.6]{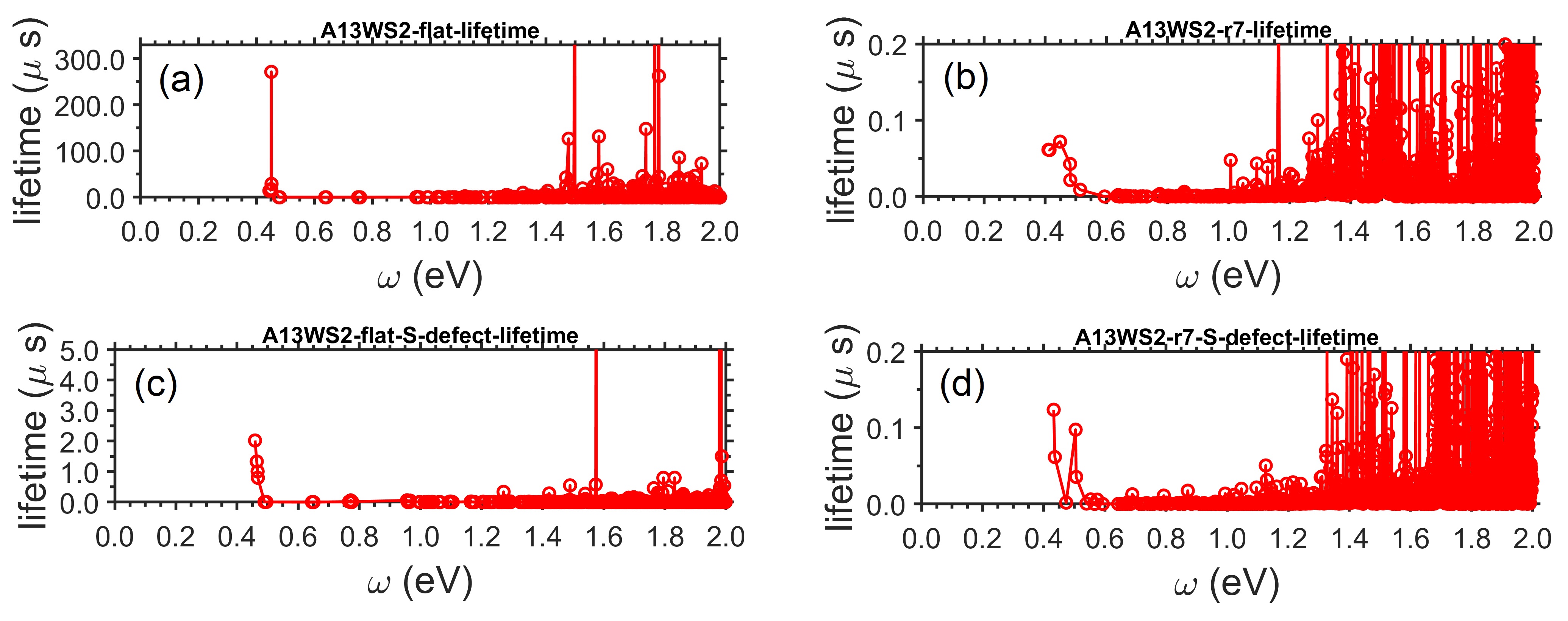}
    \caption{The calculated exciton intrinsic lifetimes of the A13WS$_{2}$ nanoribbons. (a) for flat A13WS$_{2}$ without defects, (b) for bent (with R = 7 Å) A13WS$_{2}$ without defects, (c) for the flat and S-defective A13WS$_{2}$, and (d) for the bent (with R = 7 Å) and S-defective A13WS$_{2}$.}
    \label{fig:fig9}
\end{figure}

\section{Conclusions}
In conclusion, we calculated the band structures and band gaps of armchair WS$_{2}$ nanoribbons with and without defects under various bending curvatures using density functional approximations (DFA), hybrid functional, and the high-level GW method. We also investigated the optical absorption and excitonic states of these nanoribbons using the many-body perturbation GW and BSE methods.\\
The calculations revealed that the d-orbitals of the edge W atoms are the primary source of both the upper edge bands and lower edge bands around the Fermi level. The defect bands are mainly derived from the d-orbitals of the nearby defect W atoms in the nanoribbons. The phase of the nanoribbon can be controlled by the interplay of the defect location and bending in the nanoribbon. The A13WS$_{2}$ nanoribbon with a central lower S defect line remains semiconducting with increasing bending, where the upper defect bands move downward and eventually mix with the upper edge bands at a large bending curvature. The A13WS$_{2}$ nanoribbon with a central W defect line is metallic under all bending curvatures considered, while the A13WS$_{2}$ nanoribbon with a non-central W defect line can exhibit a metal-to-semiconductor or semiconductor-to-metal phase transition with an appropriate bending curvature. Additionally, the A13WS$_{2}$ nanoribbon with a W defect line is magnetic, and the magnetization shows slight spatial variation with bending. These results suggest that properly designed defected and bent WS$_{2}$ nanoribbons can be utilized in heterophase homojunctions to avoid Schottky contact in phase-change electrical devices, as well as in magnetic nano-electronics or molecular electronics.\\
The band gaps calculated using the mTASK functional are generally larger than those from PBE and r2SCAN and are close to the values obtained from the hybrid HSE06 functional. This is due to the more enhanced derivative of exchange energy density over the kinetic energy density in the mTASK functional. The GW method yields larger values of band gaps than PBE, r2SCAN, mTASK, and HSE06, indicating the large self-energy correction due to reduced screening in the quasi one-dimensional nanoribbon system.\\
The calculated optical absorption spectra of flat WS$_{2}$ nanoribbons with and without a central sulfur defect line share a similar curve shape, emphasizing the dominance of edge-band transitions around the Fermi level. At a large bending curvature, the defect-derived upper conduction bands  move downward and mix with the edge-derived bands, while the lower edge bands merge into the valence band continuum. This again makes the overall curve shape of the optical absorption spectra for the bent WS$_{2}$ nanoribbons with and without a central Sulphur defect line similar. However, the spatial distributions of the excitons of the defected bent nanoribbon are generally more in the central area of the nanoribbon, indicating the involvement of the defect-derived bands in the optical transitions. The defective A13WS$_{2}$ nanoribbon typically exhibits a rich and mixed spin configuration due to the mixing of upper edge bands and defect-derived conduction bands. Also, the A13WS$_{2}$ nanoribbons show large calculated intrinsic lifetimes of dark excitons and a tunability in the intrinsic lifetime with bending, suggesting a useful application in exciton dynamics and exciton-based quantum technologies.\\
Overall, our study provides valuable insights into the electronic and optical properties of defected and bent WS$_{2}$ nanoribbons, offering prospects for their utilization in various technological fields.\\
  
\section{Acknowledgement}
This material is based upon work supported by the U.S. Department of Energy, Office of Science, Office of Basic Energy Sciences, under Award Number DE-SC0021263. Computational support was provided by Temple University’s HPC resources, resources of the National Energy Research Scientific Computing Center, a DOE Office of Science User Facility supported by the Office of Science of the U.S. Department of Energy under Contract No. DE-AC02-05CH11231.\\ We thank Professor John P. Perdew for his valuable comments on the manuscript.

\bibliography{lit}

\begin{thebibliography}{84}%
\makeatletter
\providecommand \@ifxundefined [1]{%
 \@ifx{#1\undefined}
}%
\providecommand \@ifnum [1]{%
 \ifnum #1\expandafter \@firstoftwo
 \else \expandafter \@secondoftwo
 \fi
}%
\providecommand \@ifx [1]{%
 \ifx #1\expandafter \@firstoftwo
 \else \expandafter \@secondoftwo
 \fi
}%
\providecommand \natexlab [1]{#1}%
\providecommand \enquote  [1]{``#1''}%
\providecommand \bibnamefont  [1]{#1}%
\providecommand \bibfnamefont [1]{#1}%
\providecommand \citenamefont [1]{#1}%
\providecommand \href@noop [0]{\@secondoftwo}%
\providecommand \href [0]{\begingroup \@sanitize@url \@href}%
\providecommand \@href[1]{\@@startlink{#1}\@@href}%
\providecommand \@@href[1]{\endgroup#1\@@endlink}%
\providecommand \@sanitize@url [0]{\catcode `\\12\catcode `\$12\catcode
  `\&12\catcode `\#12\catcode `\^12\catcode `\_12\catcode `\%12\relax}%
\providecommand \@@startlink[1]{}%
\providecommand \@@endlink[0]{}%
\providecommand \url  [0]{\begingroup\@sanitize@url \@url }%
\providecommand \@url [1]{\endgroup\@href {#1}{\urlprefix }}%
\providecommand \urlprefix  [0]{URL }%
\providecommand \Eprint [0]{\href }%
\providecommand \doibase [0]{http://dx.doi.org/}%
\providecommand \selectlanguage [0]{\@gobble}%
\providecommand \bibinfo  [0]{\@secondoftwo}%
\providecommand \bibfield  [0]{\@secondoftwo}%
\providecommand \translation [1]{[#1]}%
\providecommand \BibitemOpen [0]{}%
\providecommand \bibitemStop [0]{}%
\providecommand \bibitemNoStop [0]{.\EOS\space}%
\providecommand \EOS [0]{\spacefactor3000\relax}%
\providecommand \BibitemShut  [1]{\csname bibitem#1\endcsname}%
\let\auto@bib@innerbib\@empty
\bibitem [{\citenamefont {Ferrari}\ \emph {et~al.}(2015)\citenamefont
  {Ferrari}, \citenamefont {Bonaccorso}, \citenamefont {Fal{'}ko},
  \citenamefont {Novoselov}, \citenamefont {Roche}, \citenamefont {Bøggild},
  \citenamefont {Borini}, \citenamefont {Koppens}, \citenamefont {Palermo},
  \citenamefont {Pugno}, \citenamefont {Garrido}, \citenamefont {Sordan},
  \citenamefont {Bianco}, \citenamefont {Ballerini}, \citenamefont {Prato},
  \citenamefont {Lidorikis}, \citenamefont {Kivioja}, \citenamefont
  {Marinelli}, \citenamefont {Ryhänen}, \citenamefont {Morpurgo},
  \citenamefont {Coleman}, \citenamefont {Nicolosi}, \citenamefont {Colombo},
  \citenamefont {Fert}, \citenamefont {Garcia-Hernandez}, \citenamefont
  {Bachtold}, \citenamefont {Schneider}, \citenamefont {Guinea}, \citenamefont
  {Dekker}, \citenamefont {Barbone}, \citenamefont {Sun}, \citenamefont
  {Galiotis}, \citenamefont {Grigorenko}, \citenamefont {Konstantatos},
  \citenamefont {Kis}, \citenamefont {Katsnelson}, \citenamefont {Vandersypen},
  \citenamefont {Loiseau}, \citenamefont {Morandi}, \citenamefont {Neumaier},
  \citenamefont {Treossi}, \citenamefont {Pellegrini}, \citenamefont {Polini},
  \citenamefont {Tredicucci}, \citenamefont {Williams}, \citenamefont
  {Hee~Hong}, \citenamefont {Ahn}, \citenamefont {Min~Kim}, \citenamefont
  {Zirath}, \citenamefont {van Wees}, \citenamefont {van~der Zant},
  \citenamefont {Occhipinti}, \citenamefont {Di~Matteo}, \citenamefont
  {Kinloch}, \citenamefont {Seyller}, \citenamefont {Quesnel}, \citenamefont
  {Feng}, \citenamefont {Teo}, \citenamefont {Rupesinghe}, \citenamefont
  {Hakonen}, \citenamefont {Neil}, \citenamefont {Tannock}, \citenamefont
  {Löfwander},\ and\ \citenamefont {Kinaret}}]{cm1}%
  \BibitemOpen
  \bibfield  {author} {\bibinfo {author} {\bibfnamefont {A.~C.}\ \bibnamefont
  {Ferrari}}, \bibinfo {author} {\bibfnamefont {F.}~\bibnamefont {Bonaccorso}},
  \bibinfo {author} {\bibfnamefont {V.}~\bibnamefont {Fal{'}ko}}, \bibinfo
  {author} {\bibfnamefont {K.~S.}\ \bibnamefont {Novoselov}}, \bibinfo {author}
  {\bibfnamefont {S.}~\bibnamefont {Roche}}, \bibinfo {author} {\bibfnamefont
  {P.}~\bibnamefont {Bøggild}}, \bibinfo {author} {\bibfnamefont
  {S.}~\bibnamefont {Borini}}, \bibinfo {author} {\bibfnamefont {F.~H.~L.}\
  \bibnamefont {Koppens}}, \bibinfo {author} {\bibfnamefont {V.}~\bibnamefont
  {Palermo}}, \bibinfo {author} {\bibfnamefont {N.}~\bibnamefont {Pugno}},
  \bibinfo {author} {\bibfnamefont {J.~A.}\ \bibnamefont {Garrido}}, \bibinfo
  {author} {\bibfnamefont {R.}~\bibnamefont {Sordan}}, \bibinfo {author}
  {\bibfnamefont {A.}~\bibnamefont {Bianco}}, \bibinfo {author} {\bibfnamefont
  {L.}~\bibnamefont {Ballerini}}, \bibinfo {author} {\bibfnamefont
  {M.}~\bibnamefont {Prato}}, \bibinfo {author} {\bibfnamefont
  {E.}~\bibnamefont {Lidorikis}}, \bibinfo {author} {\bibfnamefont
  {J.}~\bibnamefont {Kivioja}}, \bibinfo {author} {\bibfnamefont
  {C.}~\bibnamefont {Marinelli}}, \bibinfo {author} {\bibfnamefont
  {T.}~\bibnamefont {Ryhänen}}, \bibinfo {author} {\bibfnamefont
  {A.}~\bibnamefont {Morpurgo}}, \bibinfo {author} {\bibfnamefont {J.~N.}\
  \bibnamefont {Coleman}}, \bibinfo {author} {\bibfnamefont {V.}~\bibnamefont
  {Nicolosi}}, \bibinfo {author} {\bibfnamefont {L.}~\bibnamefont {Colombo}},
  \bibinfo {author} {\bibfnamefont {A.}~\bibnamefont {Fert}}, \bibinfo {author}
  {\bibfnamefont {M.}~\bibnamefont {Garcia-Hernandez}}, \bibinfo {author}
  {\bibfnamefont {A.}~\bibnamefont {Bachtold}}, \bibinfo {author}
  {\bibfnamefont {G.~F.}\ \bibnamefont {Schneider}}, \bibinfo {author}
  {\bibfnamefont {F.}~\bibnamefont {Guinea}}, \bibinfo {author} {\bibfnamefont
  {C.}~\bibnamefont {Dekker}}, \bibinfo {author} {\bibfnamefont
  {M.}~\bibnamefont {Barbone}}, \bibinfo {author} {\bibfnamefont
  {Z.}~\bibnamefont {Sun}}, \bibinfo {author} {\bibfnamefont {C.}~\bibnamefont
  {Galiotis}}, \bibinfo {author} {\bibfnamefont {A.~N.}\ \bibnamefont
  {Grigorenko}}, \bibinfo {author} {\bibfnamefont {G.}~\bibnamefont
  {Konstantatos}}, \bibinfo {author} {\bibfnamefont {A.}~\bibnamefont {Kis}},
  \bibinfo {author} {\bibfnamefont {M.}~\bibnamefont {Katsnelson}}, \bibinfo
  {author} {\bibfnamefont {L.}~\bibnamefont {Vandersypen}}, \bibinfo {author}
  {\bibfnamefont {A.}~\bibnamefont {Loiseau}}, \bibinfo {author} {\bibfnamefont
  {V.}~\bibnamefont {Morandi}}, \bibinfo {author} {\bibfnamefont
  {D.}~\bibnamefont {Neumaier}}, \bibinfo {author} {\bibfnamefont
  {E.}~\bibnamefont {Treossi}}, \bibinfo {author} {\bibfnamefont
  {V.}~\bibnamefont {Pellegrini}}, \bibinfo {author} {\bibfnamefont
  {M.}~\bibnamefont {Polini}}, \bibinfo {author} {\bibfnamefont
  {A.}~\bibnamefont {Tredicucci}}, \bibinfo {author} {\bibfnamefont {G.~M.}\
  \bibnamefont {Williams}}, \bibinfo {author} {\bibfnamefont {B.}~\bibnamefont
  {Hee~Hong}}, \bibinfo {author} {\bibfnamefont {J.-H.}\ \bibnamefont {Ahn}},
  \bibinfo {author} {\bibfnamefont {J.}~\bibnamefont {Min~Kim}}, \bibinfo
  {author} {\bibfnamefont {H.}~\bibnamefont {Zirath}}, \bibinfo {author}
  {\bibfnamefont {B.~J.}\ \bibnamefont {van Wees}}, \bibinfo {author}
  {\bibfnamefont {H.}~\bibnamefont {van~der Zant}}, \bibinfo {author}
  {\bibfnamefont {L.}~\bibnamefont {Occhipinti}}, \bibinfo {author}
  {\bibfnamefont {A.}~\bibnamefont {Di~Matteo}}, \bibinfo {author}
  {\bibfnamefont {I.~A.}\ \bibnamefont {Kinloch}}, \bibinfo {author}
  {\bibfnamefont {T.}~\bibnamefont {Seyller}}, \bibinfo {author} {\bibfnamefont
  {E.}~\bibnamefont {Quesnel}}, \bibinfo {author} {\bibfnamefont
  {X.}~\bibnamefont {Feng}}, \bibinfo {author} {\bibfnamefont {K.}~\bibnamefont
  {Teo}}, \bibinfo {author} {\bibfnamefont {N.}~\bibnamefont {Rupesinghe}},
  \bibinfo {author} {\bibfnamefont {P.}~\bibnamefont {Hakonen}}, \bibinfo
  {author} {\bibfnamefont {S.~R.~T.}\ \bibnamefont {Neil}}, \bibinfo {author}
  {\bibfnamefont {Q.}~\bibnamefont {Tannock}}, \bibinfo {author} {\bibfnamefont
  {T.}~\bibnamefont {Löfwander}}, \ and\ \bibinfo {author} {\bibfnamefont
  {J.}~\bibnamefont {Kinaret}},\ }\href {\doibase 10.1039/C4NR01600A}
  {\bibfield  {journal} {\bibinfo  {journal} {Nanoscale}\ }\textbf {\bibinfo
  {volume} {7}},\ \bibinfo {pages} {4598} (\bibinfo {year} {2015})}\BibitemShut
  {NoStop}%
\bibitem [{\citenamefont {Radisavljevic}\ \emph {et~al.}(2011)\citenamefont
  {Radisavljevic}, \citenamefont {Radenovic}, \citenamefont {Brivio},
  \citenamefont {Giacometti},\ and\ \citenamefont {Kis}}]{cm2}%
  \BibitemOpen
  \bibfield  {author} {\bibinfo {author} {\bibfnamefont {B.}~\bibnamefont
  {Radisavljevic}}, \bibinfo {author} {\bibfnamefont {A.}~\bibnamefont
  {Radenovic}}, \bibinfo {author} {\bibfnamefont {J.}~\bibnamefont {Brivio}},
  \bibinfo {author} {\bibfnamefont {V.}~\bibnamefont {Giacometti}}, \ and\
  \bibinfo {author} {\bibfnamefont {A.}~\bibnamefont {Kis}},\ }\href {\doibase
  10.1038/nnano.2010.279} {\bibfield  {journal} {\bibinfo  {journal} {Nature
  Nanotechnology}\ }\textbf {\bibinfo {volume} {6}},\ \bibinfo {pages} {147}
  (\bibinfo {year} {2011})}\BibitemShut {NoStop}%
\bibitem [{\citenamefont {Coleman}\ \emph {et~al.}(2011)\citenamefont
  {Coleman}, \citenamefont {Lotya}, \citenamefont {O’Neill}, \citenamefont
  {Bergin}, \citenamefont {King}, \citenamefont {Khan}, \citenamefont {Young},
  \citenamefont {Gaucher}, \citenamefont {De}, \citenamefont {Smith},
  \citenamefont {Shvets}, \citenamefont {Arora}, \citenamefont {Stanton},
  \citenamefont {Kim}, \citenamefont {Lee}, \citenamefont {Kim}, \citenamefont
  {Duesberg}, \citenamefont {Hallam}, \citenamefont {Boland}, \citenamefont
  {Wang}, \citenamefont {Donegan}, \citenamefont {Grunlan}, \citenamefont
  {Moriarty}, \citenamefont {Shmeliov}, \citenamefont {Nicholls}, \citenamefont
  {Perkins}, \citenamefont {Grieveson}, \citenamefont {Theuwissen},
  \citenamefont {McComb}, \citenamefont {Nellist},\ and\ \citenamefont
  {Nicolosi}}]{cm3}%
  \BibitemOpen
  \bibfield  {author} {\bibinfo {author} {\bibfnamefont {J.~N.}\ \bibnamefont
  {Coleman}}, \bibinfo {author} {\bibfnamefont {M.}~\bibnamefont {Lotya}},
  \bibinfo {author} {\bibfnamefont {A.}~\bibnamefont {O’Neill}}, \bibinfo
  {author} {\bibfnamefont {S.~D.}\ \bibnamefont {Bergin}}, \bibinfo {author}
  {\bibfnamefont {P.~J.}\ \bibnamefont {King}}, \bibinfo {author}
  {\bibfnamefont {U.}~\bibnamefont {Khan}}, \bibinfo {author} {\bibfnamefont
  {K.}~\bibnamefont {Young}}, \bibinfo {author} {\bibfnamefont
  {A.}~\bibnamefont {Gaucher}}, \bibinfo {author} {\bibfnamefont
  {S.}~\bibnamefont {De}}, \bibinfo {author} {\bibfnamefont {R.~J.}\
  \bibnamefont {Smith}}, \bibinfo {author} {\bibfnamefont {I.~V.}\ \bibnamefont
  {Shvets}}, \bibinfo {author} {\bibfnamefont {S.~K.}\ \bibnamefont {Arora}},
  \bibinfo {author} {\bibfnamefont {G.}~\bibnamefont {Stanton}}, \bibinfo
  {author} {\bibfnamefont {H.-Y.}\ \bibnamefont {Kim}}, \bibinfo {author}
  {\bibfnamefont {K.}~\bibnamefont {Lee}}, \bibinfo {author} {\bibfnamefont
  {G.~T.}\ \bibnamefont {Kim}}, \bibinfo {author} {\bibfnamefont {G.~S.}\
  \bibnamefont {Duesberg}}, \bibinfo {author} {\bibfnamefont {T.}~\bibnamefont
  {Hallam}}, \bibinfo {author} {\bibfnamefont {J.~J.}\ \bibnamefont {Boland}},
  \bibinfo {author} {\bibfnamefont {J.~J.}\ \bibnamefont {Wang}}, \bibinfo
  {author} {\bibfnamefont {J.~F.}\ \bibnamefont {Donegan}}, \bibinfo {author}
  {\bibfnamefont {J.~C.}\ \bibnamefont {Grunlan}}, \bibinfo {author}
  {\bibfnamefont {G.}~\bibnamefont {Moriarty}}, \bibinfo {author}
  {\bibfnamefont {A.}~\bibnamefont {Shmeliov}}, \bibinfo {author}
  {\bibfnamefont {R.~J.}\ \bibnamefont {Nicholls}}, \bibinfo {author}
  {\bibfnamefont {J.~M.}\ \bibnamefont {Perkins}}, \bibinfo {author}
  {\bibfnamefont {E.~M.}\ \bibnamefont {Grieveson}}, \bibinfo {author}
  {\bibfnamefont {K.}~\bibnamefont {Theuwissen}}, \bibinfo {author}
  {\bibfnamefont {D.~W.}\ \bibnamefont {McComb}}, \bibinfo {author}
  {\bibfnamefont {P.~D.}\ \bibnamefont {Nellist}}, \ and\ \bibinfo {author}
  {\bibfnamefont {V.}~\bibnamefont {Nicolosi}},\ }\href {\doibase
  10.1126/science.1194975} {\bibfield  {journal} {\bibinfo  {journal}
  {Science}\ }\textbf {\bibinfo {volume} {331}},\ \bibinfo {pages} {568}
  (\bibinfo {year} {2011})},\ \Eprint
  {http://arxiv.org/abs/https://www.science.org/doi/pdf/10.1126/science.1194975}
  {https://www.science.org/doi/pdf/10.1126/science.1194975} \BibitemShut
  {NoStop}%
\bibitem [{\citenamefont {Ataca}\ \emph {et~al.}(2012)\citenamefont {Ataca},
  \citenamefont {Şahin},\ and\ \citenamefont {Ciraci}}]{cm4}%
  \BibitemOpen
  \bibfield  {author} {\bibinfo {author} {\bibfnamefont {C.}~\bibnamefont
  {Ataca}}, \bibinfo {author} {\bibfnamefont {H.}~\bibnamefont {Şahin}}, \
  and\ \bibinfo {author} {\bibfnamefont {S.}~\bibnamefont {Ciraci}},\ }\href
  {\doibase 10.1021/jp212558p} {\bibfield  {journal} {\bibinfo  {journal} {The
  Journal of Physical Chemistry C}\ }\textbf {\bibinfo {volume} {116}},\
  \bibinfo {pages} {8983} (\bibinfo {year} {2012})},\ \Eprint
  {http://arxiv.org/abs/https://doi.org/10.1021/jp212558p}
  {https://doi.org/10.1021/jp212558p} \BibitemShut {NoStop}%
\bibitem [{\citenamefont {Leb\`egue}\ \emph {et~al.}(2013)\citenamefont
  {Leb\`egue}, \citenamefont {Bj\"orkman}, \citenamefont {Klintenberg},
  \citenamefont {Nieminen},\ and\ \citenamefont {Eriksson}}]{cm5}%
  \BibitemOpen
  \bibfield  {author} {\bibinfo {author} {\bibfnamefont {S.}~\bibnamefont
  {Leb\`egue}}, \bibinfo {author} {\bibfnamefont {T.}~\bibnamefont
  {Bj\"orkman}}, \bibinfo {author} {\bibfnamefont {M.}~\bibnamefont
  {Klintenberg}}, \bibinfo {author} {\bibfnamefont {R.~M.}\ \bibnamefont
  {Nieminen}}, \ and\ \bibinfo {author} {\bibfnamefont {O.}~\bibnamefont
  {Eriksson}},\ }\href {\doibase 10.1103/PhysRevX.3.031002} {\bibfield
  {journal} {\bibinfo  {journal} {Phys. Rev. X}\ }\textbf {\bibinfo {volume}
  {3}},\ \bibinfo {pages} {031002} (\bibinfo {year} {2013})}\BibitemShut
  {NoStop}%
\bibitem [{\citenamefont {Terrones}\ \emph {et~al.}(2013)\citenamefont
  {Terrones}, \citenamefont {L{\'o}pez-Ur{\'i}as},\ and\ \citenamefont
  {Terrones}}]{cm6}%
  \BibitemOpen
  \bibfield  {author} {\bibinfo {author} {\bibfnamefont {H.}~\bibnamefont
  {Terrones}}, \bibinfo {author} {\bibfnamefont {F.}~\bibnamefont
  {L{\'o}pez-Ur{\'i}as}}, \ and\ \bibinfo {author} {\bibfnamefont
  {M.}~\bibnamefont {Terrones}},\ }\href {\doibase 10.1038/srep01549}
  {\bibfield  {journal} {\bibinfo  {journal} {Scientific Reports}\ }\textbf
  {\bibinfo {volume} {3}},\ \bibinfo {pages} {1549} (\bibinfo {year}
  {2013})}\BibitemShut {NoStop}%
\bibitem [{\citenamefont {Rasmussen}\ and\ \citenamefont
  {Thygesen}(2015)}]{cm7}%
  \BibitemOpen
  \bibfield  {author} {\bibinfo {author} {\bibfnamefont {F.~A.}\ \bibnamefont
  {Rasmussen}}\ and\ \bibinfo {author} {\bibfnamefont {K.~S.}\ \bibnamefont
  {Thygesen}},\ }\href {\doibase 10.1021/acs.jpcc.5b02950} {\bibfield
  {journal} {\bibinfo  {journal} {The Journal of Physical Chemistry C}\
  }\textbf {\bibinfo {volume} {119}},\ \bibinfo {pages} {13169} (\bibinfo
  {year} {2015})},\ \Eprint
  {http://arxiv.org/abs/https://doi.org/10.1021/acs.jpcc.5b02950}
  {https://doi.org/10.1021/acs.jpcc.5b02950} \BibitemShut {NoStop}%
\bibitem [{\citenamefont {Novoselov}\ \emph {et~al.}(2004)\citenamefont
  {Novoselov}, \citenamefont {Geim}, \citenamefont {Morozov}, \citenamefont
  {Jiang}, \citenamefont {Zhang}, \citenamefont {Dubonos}, \citenamefont
  {Grigorieva},\ and\ \citenamefont {Firsov}}]{graphene}%
  \BibitemOpen
  \bibfield  {author} {\bibinfo {author} {\bibfnamefont {K.~S.}\ \bibnamefont
  {Novoselov}}, \bibinfo {author} {\bibfnamefont {A.~K.}\ \bibnamefont {Geim}},
  \bibinfo {author} {\bibfnamefont {S.~V.}\ \bibnamefont {Morozov}}, \bibinfo
  {author} {\bibfnamefont {D.}~\bibnamefont {Jiang}}, \bibinfo {author}
  {\bibfnamefont {Y.}~\bibnamefont {Zhang}}, \bibinfo {author} {\bibfnamefont
  {S.~V.}\ \bibnamefont {Dubonos}}, \bibinfo {author} {\bibfnamefont {I.~V.}\
  \bibnamefont {Grigorieva}}, \ and\ \bibinfo {author} {\bibfnamefont {A.~A.}\
  \bibnamefont {Firsov}},\ }\href {\doibase 10.1126/science.1102896} {\bibfield
   {journal} {\bibinfo  {journal} {Science}\ }\textbf {\bibinfo {volume}
  {306}},\ \bibinfo {pages} {666} (\bibinfo {year} {2004})},\ \Eprint
  {http://arxiv.org/abs/https://www.science.org/doi/pdf/10.1126/science.1102896}
  {https://www.science.org/doi/pdf/10.1126/science.1102896} \BibitemShut
  {NoStop}%
\bibitem [{\citenamefont {Katsnelson}\ \emph {et~al.}(2006)\citenamefont
  {Katsnelson}, \citenamefont {Novoselov},\ and\ \citenamefont {Geim}}]{2d1}%
  \BibitemOpen
  \bibfield  {author} {\bibinfo {author} {\bibfnamefont {M.~I.}\ \bibnamefont
  {Katsnelson}}, \bibinfo {author} {\bibfnamefont {K.~S.}\ \bibnamefont
  {Novoselov}}, \ and\ \bibinfo {author} {\bibfnamefont {A.~K.}\ \bibnamefont
  {Geim}},\ }\href
  {http://libproxy.temple.edu/login?url=https://www.proquest.com/scholarly-journals/chiral-tunnelling-klein-paradox-graphene/docview/194685949/se-2}
  {\bibfield  {journal} {\bibinfo  {journal} {Nature Physics}\ }\textbf
  {\bibinfo {volume} {2}},\ \bibinfo {pages} {620} (\bibinfo {year} {Sep
  2006})},\ \bibinfo {note} {copyright - Copyright Nature Publishing Group Sep
  2006}\BibitemShut {NoStop}%
\bibitem [{\citenamefont {Novoselov}\ \emph
  {et~al.}(2005{\natexlab{a}})\citenamefont {Novoselov}, \citenamefont {Jiang},
  \citenamefont {Schedin}, \citenamefont {Booth}, \citenamefont {Khotkevich},
  \citenamefont {Morozov},\ and\ \citenamefont {Geim}}]{2d2}%
  \BibitemOpen
  \bibfield  {author} {\bibinfo {author} {\bibfnamefont {K.~S.}\ \bibnamefont
  {Novoselov}}, \bibinfo {author} {\bibfnamefont {D.}~\bibnamefont {Jiang}},
  \bibinfo {author} {\bibfnamefont {F.}~\bibnamefont {Schedin}}, \bibinfo
  {author} {\bibfnamefont {T.~J.}\ \bibnamefont {Booth}}, \bibinfo {author}
  {\bibfnamefont {V.~V.}\ \bibnamefont {Khotkevich}}, \bibinfo {author}
  {\bibfnamefont {S.~V.}\ \bibnamefont {Morozov}}, \ and\ \bibinfo {author}
  {\bibfnamefont {A.~K.}\ \bibnamefont {Geim}},\ }\href {\doibase
  10.1073/pnas.0502848102} {\bibfield  {journal} {\bibinfo  {journal}
  {Proceedings of the National Academy of Sciences}\ }\textbf {\bibinfo
  {volume} {102}},\ \bibinfo {pages} {10451} (\bibinfo {year}
  {2005}{\natexlab{a}})},\ \Eprint
  {http://arxiv.org/abs/https://www.pnas.org/doi/pdf/10.1073/pnas.0502848102}
  {https://www.pnas.org/doi/pdf/10.1073/pnas.0502848102} \BibitemShut {NoStop}%
\bibitem [{\citenamefont {Novoselov}\ \emph
  {et~al.}(2005{\natexlab{b}})\citenamefont {Novoselov}, \citenamefont {Geim},
  \citenamefont {Morozov}, \citenamefont {Jiang}, \citenamefont {Katsnelson},
  \citenamefont {Grigorieva}, \citenamefont {Dubonos},\ and\ \citenamefont
  {Firsov}}]{2d3}%
  \BibitemOpen
  \bibfield  {author} {\bibinfo {author} {\bibfnamefont {K.~S.}\ \bibnamefont
  {Novoselov}}, \bibinfo {author} {\bibfnamefont {A.~K.}\ \bibnamefont {Geim}},
  \bibinfo {author} {\bibfnamefont {S.~V.}\ \bibnamefont {Morozov}}, \bibinfo
  {author} {\bibfnamefont {D.}~\bibnamefont {Jiang}}, \bibinfo {author}
  {\bibfnamefont {M.~I.}\ \bibnamefont {Katsnelson}}, \bibinfo {author}
  {\bibfnamefont {I.~V.}\ \bibnamefont {Grigorieva}}, \bibinfo {author}
  {\bibfnamefont {S.~V.}\ \bibnamefont {Dubonos}}, \ and\ \bibinfo {author}
  {\bibfnamefont {A.~A.}\ \bibnamefont {Firsov}},\ }\href {\doibase
  10.1038/nature04233} {\bibfield  {journal} {\bibinfo  {journal} {Nature}\
  }\textbf {\bibinfo {volume} {438}},\ \bibinfo {pages} {197} (\bibinfo {year}
  {2005}{\natexlab{b}})}\BibitemShut {NoStop}%
\bibitem [{\citenamefont {Zhang}\ \emph {et~al.}(2005)\citenamefont {Zhang},
  \citenamefont {Tan}, \citenamefont {Stormer},\ and\ \citenamefont
  {Kim}}]{2d4}%
  \BibitemOpen
  \bibfield  {author} {\bibinfo {author} {\bibfnamefont {Y.}~\bibnamefont
  {Zhang}}, \bibinfo {author} {\bibfnamefont {Y.-W.}\ \bibnamefont {Tan}},
  \bibinfo {author} {\bibfnamefont {H.~L.}\ \bibnamefont {Stormer}}, \ and\
  \bibinfo {author} {\bibfnamefont {P.}~\bibnamefont {Kim}},\ }\href {\doibase
  10.1038/nature04235} {\bibfield  {journal} {\bibinfo  {journal} {Nature}\
  }\textbf {\bibinfo {volume} {438}},\ \bibinfo {pages} {201} (\bibinfo {year}
  {2005})}\BibitemShut {NoStop}%
\bibitem [{\citenamefont {Chhowalla}\ \emph {et~al.}(2013)\citenamefont
  {Chhowalla}, \citenamefont {Shin}, \citenamefont {Eda}, \citenamefont {Li},
  \citenamefont {Loh},\ and\ \citenamefont {Zhang}}]{2d5}%
  \BibitemOpen
  \bibfield  {author} {\bibinfo {author} {\bibfnamefont {M.}~\bibnamefont
  {Chhowalla}}, \bibinfo {author} {\bibfnamefont {H.~S.}\ \bibnamefont {Shin}},
  \bibinfo {author} {\bibfnamefont {G.}~\bibnamefont {Eda}}, \bibinfo {author}
  {\bibfnamefont {L.-J.}\ \bibnamefont {Li}}, \bibinfo {author} {\bibfnamefont
  {K.~P.}\ \bibnamefont {Loh}}, \ and\ \bibinfo {author} {\bibfnamefont
  {H.}~\bibnamefont {Zhang}},\ }\href {\doibase 10.1038/nchem.1589} {\bibfield
  {journal} {\bibinfo  {journal} {Nature Chemistry}\ }\textbf {\bibinfo
  {volume} {5}},\ \bibinfo {pages} {263} (\bibinfo {year} {2013})}\BibitemShut
  {NoStop}%
\bibitem [{\citenamefont {Wang}\ \emph
  {et~al.}(2012{\natexlab{a}})\citenamefont {Wang}, \citenamefont
  {Kalantar-Zadeh}, \citenamefont {Kis}, \citenamefont {Coleman},\ and\
  \citenamefont {Strano}}]{2d6}%
  \BibitemOpen
  \bibfield  {author} {\bibinfo {author} {\bibfnamefont {Q.~H.}\ \bibnamefont
  {Wang}}, \bibinfo {author} {\bibfnamefont {K.}~\bibnamefont
  {Kalantar-Zadeh}}, \bibinfo {author} {\bibfnamefont {A.}~\bibnamefont {Kis}},
  \bibinfo {author} {\bibfnamefont {J.~N.}\ \bibnamefont {Coleman}}, \ and\
  \bibinfo {author} {\bibfnamefont {M.~S.}\ \bibnamefont {Strano}},\ }\href
  {\doibase 10.1038/nnano.2012.193} {\bibfield  {journal} {\bibinfo  {journal}
  {Nature Nanotechnology}\ }\textbf {\bibinfo {volume} {7}},\ \bibinfo {pages}
  {699} (\bibinfo {year} {2012}{\natexlab{a}})}\BibitemShut {NoStop}%
\bibitem [{\citenamefont {Lu}\ and\ \citenamefont {Lieber}(2007)}]{ne1}%
  \BibitemOpen
  \bibfield  {author} {\bibinfo {author} {\bibfnamefont {W.}~\bibnamefont
  {Lu}}\ and\ \bibinfo {author} {\bibfnamefont {C.~M.}\ \bibnamefont
  {Lieber}},\ }\href {\doibase 10.1038/nmat2028} {\bibfield  {journal}
  {\bibinfo  {journal} {Nature Materials}\ }\textbf {\bibinfo {volume} {6}},\
  \bibinfo {pages} {841} (\bibinfo {year} {2007})}\BibitemShut {NoStop}%
\bibitem [{\citenamefont {Qian}\ \emph {et~al.}(2014)\citenamefont {Qian},
  \citenamefont {Liu}, \citenamefont {Fu},\ and\ \citenamefont {Li}}]{ne2}%
  \BibitemOpen
  \bibfield  {author} {\bibinfo {author} {\bibfnamefont {X.}~\bibnamefont
  {Qian}}, \bibinfo {author} {\bibfnamefont {J.}~\bibnamefont {Liu}}, \bibinfo
  {author} {\bibfnamefont {L.}~\bibnamefont {Fu}}, \ and\ \bibinfo {author}
  {\bibfnamefont {J.}~\bibnamefont {Li}},\ }\href {\doibase
  10.1126/science.1256815} {\bibfield  {journal} {\bibinfo  {journal}
  {Science}\ }\textbf {\bibinfo {volume} {346}},\ \bibinfo {pages} {1344}
  (\bibinfo {year} {2014})},\ \Eprint
  {http://arxiv.org/abs/https://www.science.org/doi/pdf/10.1126/science.1256815}
  {https://www.science.org/doi/pdf/10.1126/science.1256815} \BibitemShut
  {NoStop}%
\bibitem [{\citenamefont {Fang}\ \emph {et~al.}(2012)\citenamefont {Fang},
  \citenamefont {Chuang}, \citenamefont {Chang}, \citenamefont {Takei},
  \citenamefont {Takahashi},\ and\ \citenamefont {Javey}}]{ne3}%
  \BibitemOpen
  \bibfield  {author} {\bibinfo {author} {\bibfnamefont {H.}~\bibnamefont
  {Fang}}, \bibinfo {author} {\bibfnamefont {S.}~\bibnamefont {Chuang}},
  \bibinfo {author} {\bibfnamefont {T.~C.}\ \bibnamefont {Chang}}, \bibinfo
  {author} {\bibfnamefont {K.}~\bibnamefont {Takei}}, \bibinfo {author}
  {\bibfnamefont {T.}~\bibnamefont {Takahashi}}, \ and\ \bibinfo {author}
  {\bibfnamefont {A.}~\bibnamefont {Javey}},\ }\href {\doibase
  10.1021/nl301702r} {\bibfield  {journal} {\bibinfo  {journal} {Nano Letters}\
  }\textbf {\bibinfo {volume} {12}},\ \bibinfo {pages} {3788} (\bibinfo {year}
  {2012})},\ \bibinfo {note} {pMID: 22697053},\ \Eprint
  {http://arxiv.org/abs/https://doi.org/10.1021/nl301702r}
  {https://doi.org/10.1021/nl301702r} \BibitemShut {NoStop}%
\bibitem [{\citenamefont {Schwierz}(2010)}]{s1}%
  \BibitemOpen
  \bibfield  {author} {\bibinfo {author} {\bibfnamefont {F.}~\bibnamefont
  {Schwierz}},\ }\href {\doibase 10.1038/nnano.2010.89} {\bibfield  {journal}
  {\bibinfo  {journal} {Nature Nanotechnology}\ }\textbf {\bibinfo {volume}
  {5}},\ \bibinfo {pages} {487} (\bibinfo {year} {2010})}\BibitemShut {NoStop}%
\bibitem [{\citenamefont {Fowler}\ \emph {et~al.}(2009)\citenamefont {Fowler},
  \citenamefont {Allen}, \citenamefont {Tung}, \citenamefont {Yang},
  \citenamefont {Kaner},\ and\ \citenamefont {Weiller}}]{s2}%
  \BibitemOpen
  \bibfield  {author} {\bibinfo {author} {\bibfnamefont {J.~D.}\ \bibnamefont
  {Fowler}}, \bibinfo {author} {\bibfnamefont {M.~J.}\ \bibnamefont {Allen}},
  \bibinfo {author} {\bibfnamefont {V.~C.}\ \bibnamefont {Tung}}, \bibinfo
  {author} {\bibfnamefont {Y.}~\bibnamefont {Yang}}, \bibinfo {author}
  {\bibfnamefont {R.~B.}\ \bibnamefont {Kaner}}, \ and\ \bibinfo {author}
  {\bibfnamefont {B.~H.}\ \bibnamefont {Weiller}},\ }\href {\doibase
  10.1021/nn800593m} {\bibfield  {journal} {\bibinfo  {journal} {ACS Nano}\
  }\textbf {\bibinfo {volume} {3}},\ \bibinfo {pages} {301} (\bibinfo {year}
  {2009})},\ \bibinfo {note} {pMID: 19236064},\ \Eprint
  {http://arxiv.org/abs/https://doi.org/10.1021/nn800593m}
  {https://doi.org/10.1021/nn800593m} \BibitemShut {NoStop}%
\bibitem [{\citenamefont {Chang}\ and\ \citenamefont {Chen}(2011)}]{es1}%
  \BibitemOpen
  \bibfield  {author} {\bibinfo {author} {\bibfnamefont {K.}~\bibnamefont
  {Chang}}\ and\ \bibinfo {author} {\bibfnamefont {W.}~\bibnamefont {Chen}},\
  }\href {\doibase 10.1021/nn200659w} {\bibfield  {journal} {\bibinfo
  {journal} {ACS Nano}\ }\textbf {\bibinfo {volume} {5}},\ \bibinfo {pages}
  {4720} (\bibinfo {year} {2011})},\ \bibinfo {note} {pMID: 21574610},\ \Eprint
  {http://arxiv.org/abs/https://doi.org/10.1021/nn200659w}
  {https://doi.org/10.1021/nn200659w} \BibitemShut {NoStop}%
\bibitem [{\citenamefont {Bhandavat}\ \emph {et~al.}(2012)\citenamefont
  {Bhandavat}, \citenamefont {David},\ and\ \citenamefont {Singh}}]{es2}%
  \BibitemOpen
  \bibfield  {author} {\bibinfo {author} {\bibfnamefont {R.}~\bibnamefont
  {Bhandavat}}, \bibinfo {author} {\bibfnamefont {L.}~\bibnamefont {David}}, \
  and\ \bibinfo {author} {\bibfnamefont {G.}~\bibnamefont {Singh}},\ }\href
  {\doibase 10.1021/jz300480w} {\bibfield  {journal} {\bibinfo  {journal} {The
  Journal of Physical Chemistry Letters}\ }\textbf {\bibinfo {volume} {3}},\
  \bibinfo {pages} {1523} (\bibinfo {year} {2012})},\ \bibinfo {note} {pMID:
  26285632},\ \Eprint {http://arxiv.org/abs/https://doi.org/10.1021/jz300480w}
  {https://doi.org/10.1021/jz300480w} \BibitemShut {NoStop}%
\bibitem [{\citenamefont {Beal}\ \emph {et~al.}(1975)\citenamefont {Beal},
  \citenamefont {Hughes},\ and\ \citenamefont {Liang}}]{ec1}%
  \BibitemOpen
  \bibfield  {author} {\bibinfo {author} {\bibfnamefont {A.~R.}\ \bibnamefont
  {Beal}}, \bibinfo {author} {\bibfnamefont {H.~P.}\ \bibnamefont {Hughes}}, \
  and\ \bibinfo {author} {\bibfnamefont {W.~Y.}\ \bibnamefont {Liang}},\
  }\href@noop {} {\bibfield  {journal} {\bibinfo  {journal} {Journal of Physics
  C: Solid State Physics}\ }\textbf {\bibinfo {volume} {8}},\ \bibinfo {pages}
  {4236} (\bibinfo {year} {1975})}\BibitemShut {NoStop}%
\bibitem [{\citenamefont {Chandra}\ \emph {et~al.}(1984)\citenamefont
  {Chandra}, \citenamefont {Singh}, \citenamefont {Srivastava},\ and\
  \citenamefont {Sahu}}]{ec2}%
  \BibitemOpen
  \bibfield  {author} {\bibinfo {author} {\bibfnamefont {S.}~\bibnamefont
  {Chandra}}, \bibinfo {author} {\bibfnamefont {D.~P.}\ \bibnamefont {Singh}},
  \bibinfo {author} {\bibfnamefont {P.~C.}\ \bibnamefont {Srivastava}}, \ and\
  \bibinfo {author} {\bibfnamefont {S.~N.}\ \bibnamefont {Sahu}},\ }\href
  {\doibase 10.1088/0022-3727/17/10/023} {\bibfield  {journal} {\bibinfo
  {journal} {Journal of Physics D: Applied Physics}\ }\textbf {\bibinfo
  {volume} {17}},\ \bibinfo {pages} {2125} (\bibinfo {year}
  {1984})}\BibitemShut {NoStop}%
\bibitem [{\citenamefont {Gokus}\ \emph {et~al.}(2009)\citenamefont {Gokus},
  \citenamefont {Nair}, \citenamefont {Bonetti}, \citenamefont {Böhmler},
  \citenamefont {Lombardo}, \citenamefont {Novoselov}, \citenamefont {Geim},
  \citenamefont {Ferrari},\ and\ \citenamefont {Hartschuh}}]{ph1}%
  \BibitemOpen
  \bibfield  {author} {\bibinfo {author} {\bibfnamefont {T.}~\bibnamefont
  {Gokus}}, \bibinfo {author} {\bibfnamefont {R.~R.}\ \bibnamefont {Nair}},
  \bibinfo {author} {\bibfnamefont {A.}~\bibnamefont {Bonetti}}, \bibinfo
  {author} {\bibfnamefont {M.}~\bibnamefont {Böhmler}}, \bibinfo {author}
  {\bibfnamefont {A.}~\bibnamefont {Lombardo}}, \bibinfo {author}
  {\bibfnamefont {K.~S.}\ \bibnamefont {Novoselov}}, \bibinfo {author}
  {\bibfnamefont {A.~K.}\ \bibnamefont {Geim}}, \bibinfo {author}
  {\bibfnamefont {A.~C.}\ \bibnamefont {Ferrari}}, \ and\ \bibinfo {author}
  {\bibfnamefont {A.}~\bibnamefont {Hartschuh}},\ }\href {\doibase
  10.1021/nn9012753} {\bibfield  {journal} {\bibinfo  {journal} {ACS Nano}\
  }\textbf {\bibinfo {volume} {3}},\ \bibinfo {pages} {3963} (\bibinfo {year}
  {2009})},\ \bibinfo {note} {pMID: 19925014},\ \Eprint
  {http://arxiv.org/abs/https://doi.org/10.1021/nn9012753}
  {https://doi.org/10.1021/nn9012753} \BibitemShut {NoStop}%
\bibitem [{\citenamefont {Eda}\ \emph {et~al.}(2010)\citenamefont {Eda},
  \citenamefont {Lin}, \citenamefont {Mattevi}, \citenamefont {Yamaguchi},
  \citenamefont {Chen}, \citenamefont {Chen}, \citenamefont {Chen},\ and\
  \citenamefont {Chhowalla}}]{ph2}%
  \BibitemOpen
  \bibfield  {author} {\bibinfo {author} {\bibfnamefont {G.}~\bibnamefont
  {Eda}}, \bibinfo {author} {\bibfnamefont {Y.-Y.}\ \bibnamefont {Lin}},
  \bibinfo {author} {\bibfnamefont {C.}~\bibnamefont {Mattevi}}, \bibinfo
  {author} {\bibfnamefont {H.}~\bibnamefont {Yamaguchi}}, \bibinfo {author}
  {\bibfnamefont {H.-A.}\ \bibnamefont {Chen}}, \bibinfo {author}
  {\bibfnamefont {I.-S.}\ \bibnamefont {Chen}}, \bibinfo {author}
  {\bibfnamefont {C.-W.}\ \bibnamefont {Chen}}, \ and\ \bibinfo {author}
  {\bibfnamefont {M.}~\bibnamefont {Chhowalla}},\ }\href {\doibase
  10.1002/adma.200901996} {\bibfield  {journal} {\bibinfo  {journal} {Advanced
  Materials}\ }\textbf {\bibinfo {volume} {22}},\ \bibinfo {pages} {505}
  (\bibinfo {year} {2010})}\BibitemShut {NoStop}%
\bibitem [{\citenamefont {Ali}\ \emph {et~al.}(2014)\citenamefont {Ali},
  \citenamefont {Xiong}, \citenamefont {Flynn}, \citenamefont {Tao},
  \citenamefont {Gibson}, \citenamefont {Schoop}, \citenamefont {Liang},
  \citenamefont {Haldolaarachchige}, \citenamefont {Hirschberger},
  \citenamefont {Ong},\ and\ \citenamefont {Cava}}]{ms}%
  \BibitemOpen
  \bibfield  {author} {\bibinfo {author} {\bibfnamefont {M.~N.}\ \bibnamefont
  {Ali}}, \bibinfo {author} {\bibfnamefont {J.}~\bibnamefont {Xiong}}, \bibinfo
  {author} {\bibfnamefont {S.}~\bibnamefont {Flynn}}, \bibinfo {author}
  {\bibfnamefont {J.}~\bibnamefont {Tao}}, \bibinfo {author} {\bibfnamefont
  {Q.~D.}\ \bibnamefont {Gibson}}, \bibinfo {author} {\bibfnamefont {L.~M.}\
  \bibnamefont {Schoop}}, \bibinfo {author} {\bibfnamefont {T.}~\bibnamefont
  {Liang}}, \bibinfo {author} {\bibfnamefont {N.}~\bibnamefont
  {Haldolaarachchige}}, \bibinfo {author} {\bibfnamefont {M.}~\bibnamefont
  {Hirschberger}}, \bibinfo {author} {\bibfnamefont {N.~P.}\ \bibnamefont
  {Ong}}, \ and\ \bibinfo {author} {\bibfnamefont {R.~J.}\ \bibnamefont
  {Cava}},\ }\href {\doibase 10.1038/nature13763} {\bibfield  {journal}
  {\bibinfo  {journal} {Nature}\ }\textbf {\bibinfo {volume} {514}},\ \bibinfo
  {pages} {205} (\bibinfo {year} {2014})}\BibitemShut {NoStop}%
\bibitem [{\citenamefont {Mak}\ \emph {et~al.}(2012)\citenamefont {Mak},
  \citenamefont {He}, \citenamefont {Shan},\ and\ \citenamefont
  {Heinz}}]{val1}%
  \BibitemOpen
  \bibfield  {author} {\bibinfo {author} {\bibfnamefont {K.~F.}\ \bibnamefont
  {Mak}}, \bibinfo {author} {\bibfnamefont {K.}~\bibnamefont {He}}, \bibinfo
  {author} {\bibfnamefont {J.}~\bibnamefont {Shan}}, \ and\ \bibinfo {author}
  {\bibfnamefont {T.~F.}\ \bibnamefont {Heinz}},\ }\href {\doibase
  10.1038/nnano.2012.96} {\bibfield  {journal} {\bibinfo  {journal} {Nature
  Nanotechnology}\ }\textbf {\bibinfo {volume} {7}},\ \bibinfo {pages} {494}
  (\bibinfo {year} {2012})}\BibitemShut {NoStop}%
\bibitem [{\citenamefont {Zeng}\ \emph {et~al.}(2012)\citenamefont {Zeng},
  \citenamefont {Dai}, \citenamefont {Yao}, \citenamefont {Xiao},\ and\
  \citenamefont {Cui}}]{val2}%
  \BibitemOpen
  \bibfield  {author} {\bibinfo {author} {\bibfnamefont {H.}~\bibnamefont
  {Zeng}}, \bibinfo {author} {\bibfnamefont {J.}~\bibnamefont {Dai}}, \bibinfo
  {author} {\bibfnamefont {W.}~\bibnamefont {Yao}}, \bibinfo {author}
  {\bibfnamefont {D.}~\bibnamefont {Xiao}}, \ and\ \bibinfo {author}
  {\bibfnamefont {X.}~\bibnamefont {Cui}},\ }\href {\doibase
  10.1038/nnano.2012.95} {\bibfield  {journal} {\bibinfo  {journal} {Nature
  Nanotechnology}\ }\textbf {\bibinfo {volume} {7}},\ \bibinfo {pages} {490}
  (\bibinfo {year} {2012})}\BibitemShut {NoStop}%
\bibitem [{\citenamefont {Lin}\ \emph {et~al.}(2016)\citenamefont {Lin},
  \citenamefont {McCreary}, \citenamefont {Briggs}, \citenamefont
  {Subramanian}, \citenamefont {Zhang}, \citenamefont {Sun}, \citenamefont
  {Li}, \citenamefont {Borys}, \citenamefont {Yuan}, \citenamefont
  {Fullerton-Shirey}, \citenamefont {Chernikov}, \citenamefont {Zhao},
  \citenamefont {McDonnell}, \citenamefont {Lindenberg}, \citenamefont {Xiao},
  \citenamefont {LeRoy}, \citenamefont {Drndić}, \citenamefont {Hwang},
  \citenamefont {Park}, \citenamefont {Chhowalla}, \citenamefont {Schaak},
  \citenamefont {Javey}, \citenamefont {Hersam}, \citenamefont {Robinson},\
  and\ \citenamefont {Terrones}}]{defects1-2dmaterials}%
  \BibitemOpen
  \bibfield  {author} {\bibinfo {author} {\bibfnamefont {Z.}~\bibnamefont
  {Lin}}, \bibinfo {author} {\bibfnamefont {A.}~\bibnamefont {McCreary}},
  \bibinfo {author} {\bibfnamefont {N.}~\bibnamefont {Briggs}}, \bibinfo
  {author} {\bibfnamefont {S.}~\bibnamefont {Subramanian}}, \bibinfo {author}
  {\bibfnamefont {K.}~\bibnamefont {Zhang}}, \bibinfo {author} {\bibfnamefont
  {Y.}~\bibnamefont {Sun}}, \bibinfo {author} {\bibfnamefont {X.}~\bibnamefont
  {Li}}, \bibinfo {author} {\bibfnamefont {N.~J.}\ \bibnamefont {Borys}},
  \bibinfo {author} {\bibfnamefont {H.}~\bibnamefont {Yuan}}, \bibinfo {author}
  {\bibfnamefont {S.~K.}\ \bibnamefont {Fullerton-Shirey}}, \bibinfo {author}
  {\bibfnamefont {A.}~\bibnamefont {Chernikov}}, \bibinfo {author}
  {\bibfnamefont {H.}~\bibnamefont {Zhao}}, \bibinfo {author} {\bibfnamefont
  {S.}~\bibnamefont {McDonnell}}, \bibinfo {author} {\bibfnamefont {A.~M.}\
  \bibnamefont {Lindenberg}}, \bibinfo {author} {\bibfnamefont
  {K.}~\bibnamefont {Xiao}}, \bibinfo {author} {\bibfnamefont {B.~J.}\
  \bibnamefont {LeRoy}}, \bibinfo {author} {\bibfnamefont {M.}~\bibnamefont
  {Drndić}}, \bibinfo {author} {\bibfnamefont {J.~C.~M.}\ \bibnamefont
  {Hwang}}, \bibinfo {author} {\bibfnamefont {J.}~\bibnamefont {Park}},
  \bibinfo {author} {\bibfnamefont {M.}~\bibnamefont {Chhowalla}}, \bibinfo
  {author} {\bibfnamefont {R.~E.}\ \bibnamefont {Schaak}}, \bibinfo {author}
  {\bibfnamefont {A.}~\bibnamefont {Javey}}, \bibinfo {author} {\bibfnamefont
  {M.~C.}\ \bibnamefont {Hersam}}, \bibinfo {author} {\bibfnamefont
  {J.}~\bibnamefont {Robinson}}, \ and\ \bibinfo {author} {\bibfnamefont
  {M.}~\bibnamefont {Terrones}},\ }\href {\doibase
  10.1088/2053-1583/3/4/042001} {\bibfield  {journal} {\bibinfo  {journal} {2D
  Materials}\ }\textbf {\bibinfo {volume} {3}},\ \bibinfo {pages} {042001}
  (\bibinfo {year} {2016})}\BibitemShut {NoStop}%
\bibitem [{\citenamefont {Wu}\ and\ \citenamefont {Ni}(2017)}]{defects2}%
  \BibitemOpen
  \bibfield  {author} {\bibinfo {author} {\bibfnamefont {Z.}~\bibnamefont
  {Wu}}\ and\ \bibinfo {author} {\bibfnamefont {Z.}~\bibnamefont {Ni}},\ }\href
  {\doibase doi:10.1515/nanoph-2016-0151} {\bibfield  {journal} {\bibinfo
  {journal} {Nanophotonics}\ }\textbf {\bibinfo {volume} {6}},\ \bibinfo
  {pages} {1219} (\bibinfo {year} {2017})}\BibitemShut {NoStop}%
\bibitem [{\citenamefont {Wang}\ \emph {et~al.}(2018)\citenamefont {Wang},
  \citenamefont {Robertson},\ and\ \citenamefont {Warner}}]{defects3}%
  \BibitemOpen
  \bibfield  {author} {\bibinfo {author} {\bibfnamefont {S.}~\bibnamefont
  {Wang}}, \bibinfo {author} {\bibfnamefont {A.}~\bibnamefont {Robertson}}, \
  and\ \bibinfo {author} {\bibfnamefont {J.~H.}\ \bibnamefont {Warner}},\
  }\href {\doibase 10.1039/C8CS00236C} {\bibfield  {journal} {\bibinfo
  {journal} {Chem. Soc. Rev.}\ }\textbf {\bibinfo {volume} {47}},\ \bibinfo
  {pages} {6764} (\bibinfo {year} {2018})}\BibitemShut {NoStop}%
\bibitem [{\citenamefont {Rhodes}\ \emph {et~al.}(2019)\citenamefont {Rhodes},
  \citenamefont {Chae}, \citenamefont {Ribeiro-Palau},\ and\ \citenamefont
  {Hone}}]{defects4}%
  \BibitemOpen
  \bibfield  {author} {\bibinfo {author} {\bibfnamefont {D.}~\bibnamefont
  {Rhodes}}, \bibinfo {author} {\bibfnamefont {S.~H.}\ \bibnamefont {Chae}},
  \bibinfo {author} {\bibfnamefont {R.}~\bibnamefont {Ribeiro-Palau}}, \ and\
  \bibinfo {author} {\bibfnamefont {J.}~\bibnamefont {Hone}},\ }\href {\doibase
  10.1038/s41563-019-0366-8} {\bibfield  {journal} {\bibinfo  {journal} {Nature
  Materials}\ }\textbf {\bibinfo {volume} {18}},\ \bibinfo {pages} {541}
  (\bibinfo {year} {2019})}\BibitemShut {NoStop}%
\bibitem [{\citenamefont {Chow}\ \emph {et~al.}(2015)\citenamefont {Chow},
  \citenamefont {Jacobs-Gedrim}, \citenamefont {Gao}, \citenamefont {Lu},
  \citenamefont {Yu}, \citenamefont {Terrones},\ and\ \citenamefont
  {Koratkar}}]{defects-efects}%
  \BibitemOpen
  \bibfield  {author} {\bibinfo {author} {\bibfnamefont {P.~K.}\ \bibnamefont
  {Chow}}, \bibinfo {author} {\bibfnamefont {R.~B.}\ \bibnamefont
  {Jacobs-Gedrim}}, \bibinfo {author} {\bibfnamefont {J.}~\bibnamefont {Gao}},
  \bibinfo {author} {\bibfnamefont {T.-M.}\ \bibnamefont {Lu}}, \bibinfo
  {author} {\bibfnamefont {B.}~\bibnamefont {Yu}}, \bibinfo {author}
  {\bibfnamefont {H.}~\bibnamefont {Terrones}}, \ and\ \bibinfo {author}
  {\bibfnamefont {N.}~\bibnamefont {Koratkar}},\ }\href {\doibase
  10.1021/nn5073495} {\bibfield  {journal} {\bibinfo  {journal} {ACS Nano}\
  }\textbf {\bibinfo {volume} {9}},\ \bibinfo {pages} {1520} (\bibinfo {year}
  {2015})},\ \bibinfo {note} {pMID: 25603228},\ \Eprint
  {http://arxiv.org/abs/https://doi.org/10.1021/nn5073495}
  {https://doi.org/10.1021/nn5073495} \BibitemShut {NoStop}%
\bibitem [{\citenamefont {Hong}\ \emph {et~al.}(2015)\citenamefont {Hong},
  \citenamefont {Hu}, \citenamefont {Probert}, \citenamefont {Li},
  \citenamefont {Lv}, \citenamefont {Yang}, \citenamefont {Gu}, \citenamefont
  {Mao}, \citenamefont {Feng}, \citenamefont {Xie}, \citenamefont {Zhang},
  \citenamefont {Wu}, \citenamefont {Zhang}, \citenamefont {Jin}, \citenamefont
  {Ji}, \citenamefont {Zhang}, \citenamefont {Yuan},\ and\ \citenamefont
  {Zhang}}]{defects-effects2}%
  \BibitemOpen
  \bibfield  {author} {\bibinfo {author} {\bibfnamefont {J.}~\bibnamefont
  {Hong}}, \bibinfo {author} {\bibfnamefont {Z.}~\bibnamefont {Hu}}, \bibinfo
  {author} {\bibfnamefont {M.}~\bibnamefont {Probert}}, \bibinfo {author}
  {\bibfnamefont {K.}~\bibnamefont {Li}}, \bibinfo {author} {\bibfnamefont
  {D.}~\bibnamefont {Lv}}, \bibinfo {author} {\bibfnamefont {X.}~\bibnamefont
  {Yang}}, \bibinfo {author} {\bibfnamefont {L.}~\bibnamefont {Gu}}, \bibinfo
  {author} {\bibfnamefont {N.}~\bibnamefont {Mao}}, \bibinfo {author}
  {\bibfnamefont {Q.}~\bibnamefont {Feng}}, \bibinfo {author} {\bibfnamefont
  {L.}~\bibnamefont {Xie}}, \bibinfo {author} {\bibfnamefont {J.}~\bibnamefont
  {Zhang}}, \bibinfo {author} {\bibfnamefont {D.}~\bibnamefont {Wu}}, \bibinfo
  {author} {\bibfnamefont {Z.}~\bibnamefont {Zhang}}, \bibinfo {author}
  {\bibfnamefont {C.}~\bibnamefont {Jin}}, \bibinfo {author} {\bibfnamefont
  {W.}~\bibnamefont {Ji}}, \bibinfo {author} {\bibfnamefont {X.}~\bibnamefont
  {Zhang}}, \bibinfo {author} {\bibfnamefont {J.}~\bibnamefont {Yuan}}, \ and\
  \bibinfo {author} {\bibfnamefont {Z.}~\bibnamefont {Zhang}},\ }\href
  {\doibase 10.1038/ncomms7293} {\bibfield  {journal} {\bibinfo  {journal}
  {Nature Communications}\ }\textbf {\bibinfo {volume} {6}},\ \bibinfo {pages}
  {6293} (\bibinfo {year} {2015})}\BibitemShut {NoStop}%
\bibitem [{\citenamefont {Wu}\ \emph {et~al.}(2016)\citenamefont {Wu},
  \citenamefont {Luo}, \citenamefont {Shen}, \citenamefont {Zhao},
  \citenamefont {Wang}, \citenamefont {Nan}, \citenamefont {Guo}, \citenamefont
  {Sun}, \citenamefont {Wang}, \citenamefont {You},\ and\ \citenamefont
  {Ni}}]{defects-effects3}%
  \BibitemOpen
  \bibfield  {author} {\bibinfo {author} {\bibfnamefont {Z.}~\bibnamefont
  {Wu}}, \bibinfo {author} {\bibfnamefont {Z.}~\bibnamefont {Luo}}, \bibinfo
  {author} {\bibfnamefont {Y.}~\bibnamefont {Shen}}, \bibinfo {author}
  {\bibfnamefont {W.}~\bibnamefont {Zhao}}, \bibinfo {author} {\bibfnamefont
  {W.}~\bibnamefont {Wang}}, \bibinfo {author} {\bibfnamefont {H.}~\bibnamefont
  {Nan}}, \bibinfo {author} {\bibfnamefont {X.}~\bibnamefont {Guo}}, \bibinfo
  {author} {\bibfnamefont {L.}~\bibnamefont {Sun}}, \bibinfo {author}
  {\bibfnamefont {X.}~\bibnamefont {Wang}}, \bibinfo {author} {\bibfnamefont
  {Y.}~\bibnamefont {You}}, \ and\ \bibinfo {author} {\bibfnamefont
  {Z.}~\bibnamefont {Ni}},\ }\href {\doibase 10.1007/s12274-016-1232-5}
  {\bibfield  {journal} {\bibinfo  {journal} {Nano Research}\ }\textbf
  {\bibinfo {volume} {9}},\ \bibinfo {pages} {3622} (\bibinfo {year}
  {2016})}\BibitemShut {NoStop}%
\bibitem [{\citenamefont {Wan}\ \emph {et~al.}(2022)\citenamefont {Wan},
  \citenamefont {Li}, \citenamefont {Yu}, \citenamefont {Huang}, \citenamefont
  {Li}, \citenamefont {Chou}, \citenamefont {Lee}, \citenamefont {Lee},
  \citenamefont {Hsu}, \citenamefont {Zhan}, \citenamefont {Aljarb},
  \citenamefont {Fu}, \citenamefont {Chiu}, \citenamefont {Wang}, \citenamefont
  {Lin}, \citenamefont {Chiu}, \citenamefont {Chang}, \citenamefont {Wang},
  \citenamefont {Shi}, \citenamefont {Lin}, \citenamefont {Cheng},
  \citenamefont {Tung},\ and\ \citenamefont {Li}}]{defects-efffects4-ws2}%
  \BibitemOpen
  \bibfield  {author} {\bibinfo {author} {\bibfnamefont {Y.}~\bibnamefont
  {Wan}}, \bibinfo {author} {\bibfnamefont {E.}~\bibnamefont {Li}}, \bibinfo
  {author} {\bibfnamefont {Z.}~\bibnamefont {Yu}}, \bibinfo {author}
  {\bibfnamefont {J.-K.}\ \bibnamefont {Huang}}, \bibinfo {author}
  {\bibfnamefont {M.-Y.}\ \bibnamefont {Li}}, \bibinfo {author} {\bibfnamefont
  {A.-S.}\ \bibnamefont {Chou}}, \bibinfo {author} {\bibfnamefont {Y.-T.}\
  \bibnamefont {Lee}}, \bibinfo {author} {\bibfnamefont {C.-J.}\ \bibnamefont
  {Lee}}, \bibinfo {author} {\bibfnamefont {H.-C.}\ \bibnamefont {Hsu}},
  \bibinfo {author} {\bibfnamefont {Q.}~\bibnamefont {Zhan}}, \bibinfo {author}
  {\bibfnamefont {A.}~\bibnamefont {Aljarb}}, \bibinfo {author} {\bibfnamefont
  {J.-H.}\ \bibnamefont {Fu}}, \bibinfo {author} {\bibfnamefont {S.-P.}\
  \bibnamefont {Chiu}}, \bibinfo {author} {\bibfnamefont {X.}~\bibnamefont
  {Wang}}, \bibinfo {author} {\bibfnamefont {J.-J.}\ \bibnamefont {Lin}},
  \bibinfo {author} {\bibfnamefont {Y.-P.}\ \bibnamefont {Chiu}}, \bibinfo
  {author} {\bibfnamefont {W.-H.}\ \bibnamefont {Chang}}, \bibinfo {author}
  {\bibfnamefont {H.}~\bibnamefont {Wang}}, \bibinfo {author} {\bibfnamefont
  {Y.}~\bibnamefont {Shi}}, \bibinfo {author} {\bibfnamefont {N.}~\bibnamefont
  {Lin}}, \bibinfo {author} {\bibfnamefont {Y.}~\bibnamefont {Cheng}}, \bibinfo
  {author} {\bibfnamefont {V.}~\bibnamefont {Tung}}, \ and\ \bibinfo {author}
  {\bibfnamefont {L.-J.}\ \bibnamefont {Li}},\ }\href {\doibase
  10.1038/s41467-022-31886-0} {\bibfield  {journal} {\bibinfo  {journal}
  {Nature Communications}\ }\textbf {\bibinfo {volume} {13}},\ \bibinfo {pages}
  {4149} (\bibinfo {year} {2022})}\BibitemShut {NoStop}%
\bibitem [{\citenamefont {Zhang}\ \emph {et~al.}(2017)\citenamefont {Zhang},
  \citenamefont {Cao}, \citenamefont {Lu}, \citenamefont {Lin}, \citenamefont
  {Zhang}, \citenamefont {Wang}, \citenamefont {Li}, \citenamefont {Hone},
  \citenamefont {Robinson}, \citenamefont {Smirnov}, \citenamefont {Louie},\
  and\ \citenamefont {Heinz}}]{tmd1}%
  \BibitemOpen
  \bibfield  {author} {\bibinfo {author} {\bibfnamefont {X.-X.}\ \bibnamefont
  {Zhang}}, \bibinfo {author} {\bibfnamefont {T.}~\bibnamefont {Cao}}, \bibinfo
  {author} {\bibfnamefont {Z.}~\bibnamefont {Lu}}, \bibinfo {author}
  {\bibfnamefont {Y.-C.}\ \bibnamefont {Lin}}, \bibinfo {author} {\bibfnamefont
  {F.}~\bibnamefont {Zhang}}, \bibinfo {author} {\bibfnamefont
  {Y.}~\bibnamefont {Wang}}, \bibinfo {author} {\bibfnamefont {Z.}~\bibnamefont
  {Li}}, \bibinfo {author} {\bibfnamefont {J.~C.}\ \bibnamefont {Hone}},
  \bibinfo {author} {\bibfnamefont {J.~A.}\ \bibnamefont {Robinson}}, \bibinfo
  {author} {\bibfnamefont {D.}~\bibnamefont {Smirnov}}, \bibinfo {author}
  {\bibfnamefont {S.~G.}\ \bibnamefont {Louie}}, \ and\ \bibinfo {author}
  {\bibfnamefont {T.~F.}\ \bibnamefont {Heinz}},\ }\href {\doibase
  10.1038/nnano.2017.105} {\bibfield  {journal} {\bibinfo  {journal} {Nature
  Nanotechnology}\ }\textbf {\bibinfo {volume} {12}},\ \bibinfo {pages} {883}
  (\bibinfo {year} {2017})}\BibitemShut {NoStop}%
\bibitem [{\citenamefont {Li}\ \emph {et~al.}(2021{\natexlab{a}})\citenamefont
  {Li}, \citenamefont {Jiang}, \citenamefont {Li}, \citenamefont {Zhang},
  \citenamefont {Kang}, \citenamefont {Zhu}, \citenamefont {Watanabe},
  \citenamefont {Taniguchi}, \citenamefont {Chowdhury}, \citenamefont {Fu},
  \citenamefont {Shan},\ and\ \citenamefont {Mak}}]{tmd2}%
  \BibitemOpen
  \bibfield  {author} {\bibinfo {author} {\bibfnamefont {T.}~\bibnamefont
  {Li}}, \bibinfo {author} {\bibfnamefont {S.}~\bibnamefont {Jiang}}, \bibinfo
  {author} {\bibfnamefont {L.}~\bibnamefont {Li}}, \bibinfo {author}
  {\bibfnamefont {Y.}~\bibnamefont {Zhang}}, \bibinfo {author} {\bibfnamefont
  {K.}~\bibnamefont {Kang}}, \bibinfo {author} {\bibfnamefont {J.}~\bibnamefont
  {Zhu}}, \bibinfo {author} {\bibfnamefont {K.}~\bibnamefont {Watanabe}},
  \bibinfo {author} {\bibfnamefont {T.}~\bibnamefont {Taniguchi}}, \bibinfo
  {author} {\bibfnamefont {D.}~\bibnamefont {Chowdhury}}, \bibinfo {author}
  {\bibfnamefont {L.}~\bibnamefont {Fu}}, \bibinfo {author} {\bibfnamefont
  {J.}~\bibnamefont {Shan}}, \ and\ \bibinfo {author} {\bibfnamefont {K.~F.}\
  \bibnamefont {Mak}},\ }\href {\doibase 10.1038/s41586-021-03853-0} {\bibfield
   {journal} {\bibinfo  {journal} {Nature}\ }\textbf {\bibinfo {volume}
  {597}},\ \bibinfo {pages} {350—354} (\bibinfo {year}
  {2021}{\natexlab{a}})}\BibitemShut {NoStop}%
\bibitem [{\citenamefont {Jin}\ \emph {et~al.}(2021)\citenamefont {Jin},
  \citenamefont {Tao}, \citenamefont {Li}, \citenamefont {Xu}, \citenamefont
  {Tang}, \citenamefont {Zhu}, \citenamefont {Liu}, \citenamefont {Watanabe},
  \citenamefont {Taniguchi}, \citenamefont {Hone}, \citenamefont {Fu},
  \citenamefont {Shan},\ and\ \citenamefont {Mak}}]{tmd3}%
  \BibitemOpen
  \bibfield  {author} {\bibinfo {author} {\bibfnamefont {C.}~\bibnamefont
  {Jin}}, \bibinfo {author} {\bibfnamefont {Z.}~\bibnamefont {Tao}}, \bibinfo
  {author} {\bibfnamefont {T.}~\bibnamefont {Li}}, \bibinfo {author}
  {\bibfnamefont {Y.}~\bibnamefont {Xu}}, \bibinfo {author} {\bibfnamefont
  {Y.}~\bibnamefont {Tang}}, \bibinfo {author} {\bibfnamefont {J.}~\bibnamefont
  {Zhu}}, \bibinfo {author} {\bibfnamefont {S.}~\bibnamefont {Liu}}, \bibinfo
  {author} {\bibfnamefont {K.}~\bibnamefont {Watanabe}}, \bibinfo {author}
  {\bibfnamefont {T.}~\bibnamefont {Taniguchi}}, \bibinfo {author}
  {\bibfnamefont {J.~C.}\ \bibnamefont {Hone}}, \bibinfo {author}
  {\bibfnamefont {L.}~\bibnamefont {Fu}}, \bibinfo {author} {\bibfnamefont
  {J.}~\bibnamefont {Shan}}, \ and\ \bibinfo {author} {\bibfnamefont {K.~F.}\
  \bibnamefont {Mak}},\ }\href {\doibase 10.1038/s41563-021-00959-8} {\bibfield
   {journal} {\bibinfo  {journal} {Nature Materials}\ }\textbf {\bibinfo
  {volume} {20}},\ \bibinfo {pages} {940} (\bibinfo {year} {2021})}\BibitemShut
  {NoStop}%
\bibitem [{\citenamefont {Ma}\ \emph {et~al.}(2021)\citenamefont {Ma},
  \citenamefont {Nguyen}, \citenamefont {Wang}, \citenamefont {Zeng},
  \citenamefont {Watanabe}, \citenamefont {Taniguchi}, \citenamefont
  {MacDonald}, \citenamefont {Mak},\ and\ \citenamefont {Shan}}]{tmd4}%
  \BibitemOpen
  \bibfield  {author} {\bibinfo {author} {\bibfnamefont {L.}~\bibnamefont
  {Ma}}, \bibinfo {author} {\bibfnamefont {P.~X.}\ \bibnamefont {Nguyen}},
  \bibinfo {author} {\bibfnamefont {Z.}~\bibnamefont {Wang}}, \bibinfo {author}
  {\bibfnamefont {Y.}~\bibnamefont {Zeng}}, \bibinfo {author} {\bibfnamefont
  {K.}~\bibnamefont {Watanabe}}, \bibinfo {author} {\bibfnamefont
  {T.}~\bibnamefont {Taniguchi}}, \bibinfo {author} {\bibfnamefont {A.~H.}\
  \bibnamefont {MacDonald}}, \bibinfo {author} {\bibfnamefont {K.~F.}\
  \bibnamefont {Mak}}, \ and\ \bibinfo {author} {\bibfnamefont
  {J.}~\bibnamefont {Shan}},\ }\href {\doibase 10.1038/s41586-021-03947-9}
  {\bibfield  {journal} {\bibinfo  {journal} {Nature}\ }\textbf {\bibinfo
  {volume} {598}},\ \bibinfo {pages} {585} (\bibinfo {year}
  {2021})}\BibitemShut {NoStop}%
\bibitem [{\citenamefont {Li}\ \emph {et~al.}(2021{\natexlab{b}})\citenamefont
  {Li}, \citenamefont {Jiang}, \citenamefont {Shen}, \citenamefont {Zhang},
  \citenamefont {Li}, \citenamefont {Tao}, \citenamefont {Devakul},
  \citenamefont {Watanabe}, \citenamefont {Taniguchi}, \citenamefont {Fu},
  \citenamefont {Shan},\ and\ \citenamefont {Mak}}]{tmd5}%
  \BibitemOpen
  \bibfield  {author} {\bibinfo {author} {\bibfnamefont {T.}~\bibnamefont
  {Li}}, \bibinfo {author} {\bibfnamefont {S.}~\bibnamefont {Jiang}}, \bibinfo
  {author} {\bibfnamefont {B.}~\bibnamefont {Shen}}, \bibinfo {author}
  {\bibfnamefont {Y.}~\bibnamefont {Zhang}}, \bibinfo {author} {\bibfnamefont
  {L.}~\bibnamefont {Li}}, \bibinfo {author} {\bibfnamefont {Z.}~\bibnamefont
  {Tao}}, \bibinfo {author} {\bibfnamefont {T.}~\bibnamefont {Devakul}},
  \bibinfo {author} {\bibfnamefont {K.}~\bibnamefont {Watanabe}}, \bibinfo
  {author} {\bibfnamefont {T.}~\bibnamefont {Taniguchi}}, \bibinfo {author}
  {\bibfnamefont {L.}~\bibnamefont {Fu}}, \bibinfo {author} {\bibfnamefont
  {J.}~\bibnamefont {Shan}}, \ and\ \bibinfo {author} {\bibfnamefont {K.~F.}\
  \bibnamefont {Mak}},\ }\href {\doibase 10.1038/s41586-021-04171-1} {\bibfield
   {journal} {\bibinfo  {journal} {Nature}\ }\textbf {\bibinfo {volume}
  {600}},\ \bibinfo {pages} {641} (\bibinfo {year}
  {2021}{\natexlab{b}})}\BibitemShut {NoStop}%
\bibitem [{\citenamefont {Hong}\ \emph {et~al.}(2014)\citenamefont {Hong},
  \citenamefont {Kim}, \citenamefont {Shi}, \citenamefont {Zhang},
  \citenamefont {Jin}, \citenamefont {Sun}, \citenamefont {Tongay},
  \citenamefont {Wu}, \citenamefont {Zhang},\ and\ \citenamefont {Wang}}]{pl1}%
  \BibitemOpen
  \bibfield  {author} {\bibinfo {author} {\bibfnamefont {X.}~\bibnamefont
  {Hong}}, \bibinfo {author} {\bibfnamefont {J.}~\bibnamefont {Kim}}, \bibinfo
  {author} {\bibfnamefont {S.-F.}\ \bibnamefont {Shi}}, \bibinfo {author}
  {\bibfnamefont {Y.}~\bibnamefont {Zhang}}, \bibinfo {author} {\bibfnamefont
  {C.}~\bibnamefont {Jin}}, \bibinfo {author} {\bibfnamefont {Y.}~\bibnamefont
  {Sun}}, \bibinfo {author} {\bibfnamefont {S.}~\bibnamefont {Tongay}},
  \bibinfo {author} {\bibfnamefont {J.}~\bibnamefont {Wu}}, \bibinfo {author}
  {\bibfnamefont {Y.}~\bibnamefont {Zhang}}, \ and\ \bibinfo {author}
  {\bibfnamefont {F.}~\bibnamefont {Wang}},\ }\href {\doibase
  10.1038/nnano.2014.167} {\bibfield  {journal} {\bibinfo  {journal} {Nature
  Nanotechnology}\ }\textbf {\bibinfo {volume} {9}},\ \bibinfo {pages} {682}
  (\bibinfo {year} {2014})}\BibitemShut {NoStop}%
\bibitem [{\citenamefont {Peimyoo}\ \emph {et~al.}(2013)\citenamefont
  {Peimyoo}, \citenamefont {Shang}, \citenamefont {Cong}, \citenamefont {Shen},
  \citenamefont {Wu}, \citenamefont {Yeow},\ and\ \citenamefont {Yu}}]{pl2}%
  \BibitemOpen
  \bibfield  {author} {\bibinfo {author} {\bibfnamefont {N.}~\bibnamefont
  {Peimyoo}}, \bibinfo {author} {\bibfnamefont {J.}~\bibnamefont {Shang}},
  \bibinfo {author} {\bibfnamefont {C.}~\bibnamefont {Cong}}, \bibinfo {author}
  {\bibfnamefont {X.}~\bibnamefont {Shen}}, \bibinfo {author} {\bibfnamefont
  {X.}~\bibnamefont {Wu}}, \bibinfo {author} {\bibfnamefont {E.~K.~L.}\
  \bibnamefont {Yeow}}, \ and\ \bibinfo {author} {\bibfnamefont
  {T.}~\bibnamefont {Yu}},\ }\href {\doibase 10.1021/nn4046002} {\bibfield
  {journal} {\bibinfo  {journal} {{ACS} Nano}\ }\textbf {\bibinfo {volume}
  {7}},\ \bibinfo {pages} {10985} (\bibinfo {year} {2013})}\BibitemShut
  {NoStop}%
\bibitem [{\citenamefont {Duan}\ \emph {et~al.}(2014)\citenamefont {Duan},
  \citenamefont {Wang}, \citenamefont {Shaw}, \citenamefont {Cheng},
  \citenamefont {Chen}, \citenamefont {Li}, \citenamefont {Wu}, \citenamefont
  {Tang}, \citenamefont {Zhang}, \citenamefont {Pan}, \citenamefont {Jiang},
  \citenamefont {Yu}, \citenamefont {Huang},\ and\ \citenamefont
  {Duan}}]{ws2app1}%
  \BibitemOpen
  \bibfield  {author} {\bibinfo {author} {\bibfnamefont {X.}~\bibnamefont
  {Duan}}, \bibinfo {author} {\bibfnamefont {C.}~\bibnamefont {Wang}}, \bibinfo
  {author} {\bibfnamefont {J.~C.}\ \bibnamefont {Shaw}}, \bibinfo {author}
  {\bibfnamefont {R.}~\bibnamefont {Cheng}}, \bibinfo {author} {\bibfnamefont
  {Y.}~\bibnamefont {Chen}}, \bibinfo {author} {\bibfnamefont {H.}~\bibnamefont
  {Li}}, \bibinfo {author} {\bibfnamefont {X.}~\bibnamefont {Wu}}, \bibinfo
  {author} {\bibfnamefont {Y.}~\bibnamefont {Tang}}, \bibinfo {author}
  {\bibfnamefont {Q.}~\bibnamefont {Zhang}}, \bibinfo {author} {\bibfnamefont
  {A.}~\bibnamefont {Pan}}, \bibinfo {author} {\bibfnamefont {J.}~\bibnamefont
  {Jiang}}, \bibinfo {author} {\bibfnamefont {R.}~\bibnamefont {Yu}}, \bibinfo
  {author} {\bibfnamefont {Y.}~\bibnamefont {Huang}}, \ and\ \bibinfo {author}
  {\bibfnamefont {X.}~\bibnamefont {Duan}},\ }\href {\doibase
  10.1038/nnano.2014.222} {\bibfield  {journal} {\bibinfo  {journal} {Nature
  Nanotechnology}\ }\textbf {\bibinfo {volume} {9}},\ \bibinfo {pages} {1024}
  (\bibinfo {year} {2014})}\BibitemShut {NoStop}%
\bibitem [{\citenamefont {Wang}\ \emph
  {et~al.}(2012{\natexlab{b}})\citenamefont {Wang}, \citenamefont
  {Kalantar-Zadeh}, \citenamefont {Kis}, \citenamefont {Coleman},\ and\
  \citenamefont {Strano}}]{ws2app2}%
  \BibitemOpen
  \bibfield  {author} {\bibinfo {author} {\bibfnamefont {Q.~H.}\ \bibnamefont
  {Wang}}, \bibinfo {author} {\bibfnamefont {K.}~\bibnamefont
  {Kalantar-Zadeh}}, \bibinfo {author} {\bibfnamefont {A.}~\bibnamefont {Kis}},
  \bibinfo {author} {\bibfnamefont {J.~N.}\ \bibnamefont {Coleman}}, \ and\
  \bibinfo {author} {\bibfnamefont {M.~S.}\ \bibnamefont {Strano}},\ }\href
  {\doibase 10.1038/nnano.2012.193} {\bibfield  {journal} {\bibinfo  {journal}
  {Nature Nanotechnology}\ }\textbf {\bibinfo {volume} {7}},\ \bibinfo {pages}
  {699} (\bibinfo {year} {2012}{\natexlab{b}})}\BibitemShut {NoStop}%
\bibitem [{\citenamefont {Lee}\ \emph {et~al.}(2016)\citenamefont {Lee},
  \citenamefont {Luong}, \citenamefont {Kim}, \citenamefont {Jin},
  \citenamefont {Kim}, \citenamefont {Yun},\ and\ \citenamefont
  {Lee}}]{ws2app3}%
  \BibitemOpen
  \bibfield  {author} {\bibinfo {author} {\bibfnamefont {H.~S.}\ \bibnamefont
  {Lee}}, \bibinfo {author} {\bibfnamefont {D.~H.}\ \bibnamefont {Luong}},
  \bibinfo {author} {\bibfnamefont {M.~S.}\ \bibnamefont {Kim}}, \bibinfo
  {author} {\bibfnamefont {Y.}~\bibnamefont {Jin}}, \bibinfo {author}
  {\bibfnamefont {H.}~\bibnamefont {Kim}}, \bibinfo {author} {\bibfnamefont
  {S.}~\bibnamefont {Yun}}, \ and\ \bibinfo {author} {\bibfnamefont {Y.~H.}\
  \bibnamefont {Lee}},\ }\href {\doibase 10.1038/ncomms13663} {\bibfield
  {journal} {\bibinfo  {journal} {Nature Communications}\ }\textbf {\bibinfo
  {volume} {7}} (\bibinfo {year} {2016}),\ 10.1038/ncomms13663}\BibitemShut
  {NoStop}%
\bibitem [{\citenamefont {Cho}\ \emph {et~al.}(2015)\citenamefont {Cho},
  \citenamefont {Kim}, \citenamefont {Kim}, \citenamefont {Zhao}, \citenamefont
  {Seok}, \citenamefont {Keum}, \citenamefont {Baik}, \citenamefont {Choe},
  \citenamefont {Chang}, \citenamefont {Suenaga}, \citenamefont {Kim},
  \citenamefont {Lee},\ and\ \citenamefont {Yang}}]{Cho_2015}%
  \BibitemOpen
  \bibfield  {author} {\bibinfo {author} {\bibfnamefont {S.}~\bibnamefont
  {Cho}}, \bibinfo {author} {\bibfnamefont {S.}~\bibnamefont {Kim}}, \bibinfo
  {author} {\bibfnamefont {J.~H.}\ \bibnamefont {Kim}}, \bibinfo {author}
  {\bibfnamefont {J.}~\bibnamefont {Zhao}}, \bibinfo {author} {\bibfnamefont
  {J.}~\bibnamefont {Seok}}, \bibinfo {author} {\bibfnamefont {D.~H.}\
  \bibnamefont {Keum}}, \bibinfo {author} {\bibfnamefont {J.}~\bibnamefont
  {Baik}}, \bibinfo {author} {\bibfnamefont {D.-H.}\ \bibnamefont {Choe}},
  \bibinfo {author} {\bibfnamefont {K.~J.}\ \bibnamefont {Chang}}, \bibinfo
  {author} {\bibfnamefont {K.}~\bibnamefont {Suenaga}}, \bibinfo {author}
  {\bibfnamefont {S.~W.}\ \bibnamefont {Kim}}, \bibinfo {author} {\bibfnamefont
  {Y.~H.}\ \bibnamefont {Lee}}, \ and\ \bibinfo {author} {\bibfnamefont
  {H.}~\bibnamefont {Yang}},\ }\href {\doibase 10.1126/science.aab3175}
  {\bibfield  {journal} {\bibinfo  {journal} {Science}\ }\textbf {\bibinfo
  {volume} {349}},\ \bibinfo {pages} {625} (\bibinfo {year}
  {2015})}\BibitemShut {NoStop}%
\bibitem [{\citenamefont {Wang}\ \emph {et~al.}(2019)\citenamefont {Wang},
  \citenamefont {Liu}, \citenamefont {Zhu}, \citenamefont {You}, \citenamefont
  {Bian}, \citenamefont {Zhang}, \citenamefont {Feng},\ and\ \citenamefont
  {Jiang}}]{Wang_2019}%
  \BibitemOpen
  \bibfield  {author} {\bibinfo {author} {\bibfnamefont {Z.}~\bibnamefont
  {Wang}}, \bibinfo {author} {\bibfnamefont {X.}~\bibnamefont {Liu}}, \bibinfo
  {author} {\bibfnamefont {J.}~\bibnamefont {Zhu}}, \bibinfo {author}
  {\bibfnamefont {S.}~\bibnamefont {You}}, \bibinfo {author} {\bibfnamefont
  {K.}~\bibnamefont {Bian}}, \bibinfo {author} {\bibfnamefont {G.}~\bibnamefont
  {Zhang}}, \bibinfo {author} {\bibfnamefont {J.}~\bibnamefont {Feng}}, \ and\
  \bibinfo {author} {\bibfnamefont {Y.}~\bibnamefont {Jiang}},\ }\href
  {\doibase https://doi.org/10.1016/j.scib.2019.10.004} {\bibfield  {journal}
  {\bibinfo  {journal} {Science Bulletin}\ }\textbf {\bibinfo {volume} {64}},\
  \bibinfo {pages} {1750} (\bibinfo {year} {2019})}\BibitemShut {NoStop}%
\bibitem [{\citenamefont {Zhou}\ \emph {et~al.}(2017)\citenamefont {Zhou},
  \citenamefont {Scuri}, \citenamefont {Wild}, \citenamefont {High},
  \citenamefont {Dibos}, \citenamefont {Jauregui}, \citenamefont {Shu},
  \citenamefont {De~Greve}, \citenamefont {Pistunova}, \citenamefont {Joe},
  \citenamefont {Taniguchi}, \citenamefont {Watanabe}, \citenamefont {Kim},
  \citenamefont {Lukin},\ and\ \citenamefont {Park}}]{dark-excitons}%
  \BibitemOpen
  \bibfield  {author} {\bibinfo {author} {\bibfnamefont {Y.}~\bibnamefont
  {Zhou}}, \bibinfo {author} {\bibfnamefont {G.}~\bibnamefont {Scuri}},
  \bibinfo {author} {\bibfnamefont {D.~S.}\ \bibnamefont {Wild}}, \bibinfo
  {author} {\bibfnamefont {A.~A.}\ \bibnamefont {High}}, \bibinfo {author}
  {\bibfnamefont {A.}~\bibnamefont {Dibos}}, \bibinfo {author} {\bibfnamefont
  {L.~A.}\ \bibnamefont {Jauregui}}, \bibinfo {author} {\bibfnamefont
  {C.}~\bibnamefont {Shu}}, \bibinfo {author} {\bibfnamefont {K.}~\bibnamefont
  {De~Greve}}, \bibinfo {author} {\bibfnamefont {K.}~\bibnamefont {Pistunova}},
  \bibinfo {author} {\bibfnamefont {A.~Y.}\ \bibnamefont {Joe}}, \bibinfo
  {author} {\bibfnamefont {T.}~\bibnamefont {Taniguchi}}, \bibinfo {author}
  {\bibfnamefont {K.}~\bibnamefont {Watanabe}}, \bibinfo {author}
  {\bibfnamefont {P.}~\bibnamefont {Kim}}, \bibinfo {author} {\bibfnamefont
  {M.~D.}\ \bibnamefont {Lukin}}, \ and\ \bibinfo {author} {\bibfnamefont
  {H.}~\bibnamefont {Park}},\ }\href {\doibase 10.1038/nnano.2017.106}
  {\bibfield  {journal} {\bibinfo  {journal} {Nature Nanotechnology}\ }\textbf
  {\bibinfo {volume} {12}},\ \bibinfo {pages} {856{\textendash}860} (\bibinfo
  {year} {2017})}\BibitemShut {NoStop}%
\bibitem [{\citenamefont {Attaccalite}\ \emph {et~al.}(2011)\citenamefont
  {Attaccalite}, \citenamefont {Bockstedte}, \citenamefont {Marini},
  \citenamefont {Rubio},\ and\ \citenamefont {Wirtz}}]{Attaccale}%
  \BibitemOpen
  \bibfield  {author} {\bibinfo {author} {\bibfnamefont {C.}~\bibnamefont
  {Attaccalite}}, \bibinfo {author} {\bibfnamefont {M.}~\bibnamefont
  {Bockstedte}}, \bibinfo {author} {\bibfnamefont {A.}~\bibnamefont {Marini}},
  \bibinfo {author} {\bibfnamefont {A.}~\bibnamefont {Rubio}}, \ and\ \bibinfo
  {author} {\bibfnamefont {L.}~\bibnamefont {Wirtz}},\ }\href {\doibase
  10.1103/PhysRevB.83.144115} {\bibfield  {journal} {\bibinfo  {journal} {Phys.
  Rev. B}\ }\textbf {\bibinfo {volume} {83}},\ \bibinfo {pages} {144115}
  (\bibinfo {year} {2011})}\BibitemShut {NoStop}%
\bibitem [{\citenamefont {Wang}\ \emph {et~al.}(2021)\citenamefont {Wang},
  \citenamefont {Wang}, \citenamefont {Ma}, \citenamefont {Chen}, \citenamefont
  {Jiang}, \citenamefont {Chen}, \citenamefont {Xie}, \citenamefont {Li},\ and\
  \citenamefont {Wang}}]{nr1}%
  \BibitemOpen
  \bibfield  {author} {\bibinfo {author} {\bibfnamefont {H.}~\bibnamefont
  {Wang}}, \bibinfo {author} {\bibfnamefont {H.~S.}\ \bibnamefont {Wang}},
  \bibinfo {author} {\bibfnamefont {C.}~\bibnamefont {Ma}}, \bibinfo {author}
  {\bibfnamefont {L.}~\bibnamefont {Chen}}, \bibinfo {author} {\bibfnamefont
  {C.}~\bibnamefont {Jiang}}, \bibinfo {author} {\bibfnamefont
  {C.}~\bibnamefont {Chen}}, \bibinfo {author} {\bibfnamefont {X.}~\bibnamefont
  {Xie}}, \bibinfo {author} {\bibfnamefont {A.-P.}\ \bibnamefont {Li}}, \ and\
  \bibinfo {author} {\bibfnamefont {X.}~\bibnamefont {Wang}},\ }\href {\doibase
  10.1038/s42254-021-00370-x} {\bibfield  {journal} {\bibinfo  {journal}
  {Nature Reviews Physics}\ }\textbf {\bibinfo {volume} {3}},\ \bibinfo {pages}
  {791} (\bibinfo {year} {2021})}\BibitemShut {NoStop}%
\bibitem [{\citenamefont {Magda}\ \emph {et~al.}(2014)\citenamefont {Magda},
  \citenamefont {Jin}, \citenamefont {Hagym{\'{a}}si}, \citenamefont
  {Vancs{\'{o}}}, \citenamefont {Osv{\'{a}}th}, \citenamefont {Nemes-Incze},
  \citenamefont {Hwang}, \citenamefont {Bir{\'{o}}},\ and\ \citenamefont
  {Tapaszt{\'{o}}}}]{nr2}%
  \BibitemOpen
  \bibfield  {author} {\bibinfo {author} {\bibfnamefont {G.~Z.}\ \bibnamefont
  {Magda}}, \bibinfo {author} {\bibfnamefont {X.}~\bibnamefont {Jin}}, \bibinfo
  {author} {\bibfnamefont {I.}~\bibnamefont {Hagym{\'{a}}si}}, \bibinfo
  {author} {\bibfnamefont {P.}~\bibnamefont {Vancs{\'{o}}}}, \bibinfo {author}
  {\bibfnamefont {Z.}~\bibnamefont {Osv{\'{a}}th}}, \bibinfo {author}
  {\bibfnamefont {P.}~\bibnamefont {Nemes-Incze}}, \bibinfo {author}
  {\bibfnamefont {C.}~\bibnamefont {Hwang}}, \bibinfo {author} {\bibfnamefont
  {L.~P.}\ \bibnamefont {Bir{\'{o}}}}, \ and\ \bibinfo {author} {\bibfnamefont
  {L.}~\bibnamefont {Tapaszt{\'{o}}}},\ }\href {\doibase 10.1038/nature13831}
  {\bibfield  {journal} {\bibinfo  {journal} {Nature}\ }\textbf {\bibinfo
  {volume} {514}},\ \bibinfo {pages} {608} (\bibinfo {year}
  {2014})}\BibitemShut {NoStop}%
\bibitem [{\citenamefont {Pan}\ and\ \citenamefont {Zhang}(2012)}]{nr3}%
  \BibitemOpen
  \bibfield  {author} {\bibinfo {author} {\bibfnamefont {H.}~\bibnamefont
  {Pan}}\ and\ \bibinfo {author} {\bibfnamefont {Y.-W.}\ \bibnamefont
  {Zhang}},\ }\href {\doibase 10.1039/C2JM15906F} {\bibfield  {journal}
  {\bibinfo  {journal} {J. Mater. Chem.}\ }\textbf {\bibinfo {volume} {22}},\
  \bibinfo {pages} {7280} (\bibinfo {year} {2012})}\BibitemShut {NoStop}%
\bibitem [{\citenamefont {Tang}\ \emph
  {et~al.}(2022{\natexlab{a}})\citenamefont {Tang}, \citenamefont {Neupane},
  \citenamefont {Neupane}, \citenamefont {Ruan}, \citenamefont {Nepal},\ and\
  \citenamefont {Ruzsinszky}}]{bending2us}%
  \BibitemOpen
  \bibfield  {author} {\bibinfo {author} {\bibfnamefont {H.}~\bibnamefont
  {Tang}}, \bibinfo {author} {\bibfnamefont {B.}~\bibnamefont {Neupane}},
  \bibinfo {author} {\bibfnamefont {S.}~\bibnamefont {Neupane}}, \bibinfo
  {author} {\bibfnamefont {S.}~\bibnamefont {Ruan}}, \bibinfo {author}
  {\bibfnamefont {N.~K.}\ \bibnamefont {Nepal}}, \ and\ \bibinfo {author}
  {\bibfnamefont {A.}~\bibnamefont {Ruzsinszky}},\ }\href {\doibase
  10.1038/s41598-022-06741-3} {\bibfield  {journal} {\bibinfo  {journal}
  {Scientific Reports}\ }\textbf {\bibinfo {volume} {12}},\ \bibinfo {pages}
  {3008} (\bibinfo {year} {2022}{\natexlab{a}})}\BibitemShut {NoStop}%
\bibitem [{\citenamefont {Neupane}\ \emph {et~al.}(2022)\citenamefont
  {Neupane}, \citenamefont {Tang}, \citenamefont {Nepal},\ and\ \citenamefont
  {Ruzsinszky}}]{bending1bn}%
  \BibitemOpen
  \bibfield  {author} {\bibinfo {author} {\bibfnamefont {B.}~\bibnamefont
  {Neupane}}, \bibinfo {author} {\bibfnamefont {H.}~\bibnamefont {Tang}},
  \bibinfo {author} {\bibfnamefont {N.~K.}\ \bibnamefont {Nepal}}, \ and\
  \bibinfo {author} {\bibfnamefont {A.}~\bibnamefont {Ruzsinszky}},\ }\href
  {\doibase 10.1103/PhysRevMaterials.6.014010} {\bibfield  {journal} {\bibinfo
  {journal} {Phys. Rev. Mater.}\ }\textbf {\bibinfo {volume} {6}},\ \bibinfo
  {pages} {014010} (\bibinfo {year} {2022})}\BibitemShut {NoStop}%
\bibitem [{\citenamefont {Tang}\ \emph
  {et~al.}(2022{\natexlab{b}})\citenamefont {Tang}, \citenamefont {Neupane},
  \citenamefont {Yin}, \citenamefont {Breslin},\ and\ \citenamefont
  {Ruzsinszky}}]{spin-anisotropy}%
  \BibitemOpen
  \bibfield  {author} {\bibinfo {author} {\bibfnamefont {H.}~\bibnamefont
  {Tang}}, \bibinfo {author} {\bibfnamefont {S.}~\bibnamefont {Neupane}},
  \bibinfo {author} {\bibfnamefont {L.}~\bibnamefont {Yin}}, \bibinfo {author}
  {\bibfnamefont {J.~M.}\ \bibnamefont {Breslin}}, \ and\ \bibinfo {author}
  {\bibfnamefont {A.}~\bibnamefont {Ruzsinszky}},\ }\href@noop {} {\enquote
  {\bibinfo {title} {Spin-polarization anisotropy included by mechanical
  bending in tungsten diselenide nanoribbons and tunable excitonic states},}\ }
  (\bibinfo {year} {2022}{\natexlab{b}}),\ \Eprint
  {http://arxiv.org/abs/2211.00258} {arXiv:2211.00258 [cond-mat.mtrl-sci]}
  \BibitemShut {NoStop}%
\bibitem [{\citenamefont {Perdew}\ \emph {et~al.}(1996)\citenamefont {Perdew},
  \citenamefont {Burke},\ and\ \citenamefont {Ernzerhof}}]{PBE}%
  \BibitemOpen
  \bibfield  {author} {\bibinfo {author} {\bibfnamefont {J.~P.}\ \bibnamefont
  {Perdew}}, \bibinfo {author} {\bibfnamefont {K.}~\bibnamefont {Burke}}, \
  and\ \bibinfo {author} {\bibfnamefont {M.}~\bibnamefont {Ernzerhof}},\ }\href
  {\doibase 10.1103/PhysRevLett.77.3865} {\bibfield  {journal} {\bibinfo
  {journal} {Phys. Rev. Lett.}\ }\textbf {\bibinfo {volume} {77}},\ \bibinfo
  {pages} {3865} (\bibinfo {year} {1996})}\BibitemShut {NoStop}%
\bibitem [{\citenamefont {Furness}\ \emph {et~al.}(2020)\citenamefont
  {Furness}, \citenamefont {Kaplan}, \citenamefont {Ning}, \citenamefont
  {Perdew},\ and\ \citenamefont {Sun}}]{r2scan}%
  \BibitemOpen
  \bibfield  {author} {\bibinfo {author} {\bibfnamefont {J.~W.}\ \bibnamefont
  {Furness}}, \bibinfo {author} {\bibfnamefont {A.~D.}\ \bibnamefont {Kaplan}},
  \bibinfo {author} {\bibfnamefont {J.}~\bibnamefont {Ning}}, \bibinfo {author}
  {\bibfnamefont {J.~P.}\ \bibnamefont {Perdew}}, \ and\ \bibinfo {author}
  {\bibfnamefont {J.}~\bibnamefont {Sun}},\ }\href {\doibase
  10.1021/acs.jpclett.0c02405} {\bibfield  {journal} {\bibinfo  {journal} {The
  Journal of Physical Chemistry Letters}\ }\textbf {\bibinfo {volume} {11}},\
  \bibinfo {pages} {8208} (\bibinfo {year} {2020})}\BibitemShut {NoStop}%
\bibitem [{\citenamefont {Neupane}\ \emph {et~al.}(2021)\citenamefont
  {Neupane}, \citenamefont {Tang}, \citenamefont {Nepal}, \citenamefont
  {Adhikari},\ and\ \citenamefont {Ruzsinszky}}]{mTASk}%
  \BibitemOpen
  \bibfield  {author} {\bibinfo {author} {\bibfnamefont {B.}~\bibnamefont
  {Neupane}}, \bibinfo {author} {\bibfnamefont {H.}~\bibnamefont {Tang}},
  \bibinfo {author} {\bibfnamefont {N.~K.}\ \bibnamefont {Nepal}}, \bibinfo
  {author} {\bibfnamefont {S.}~\bibnamefont {Adhikari}}, \ and\ \bibinfo
  {author} {\bibfnamefont {A.}~\bibnamefont {Ruzsinszky}},\ }\href {\doibase
  10.1103/PhysRevMaterials.5.063803} {\bibfield  {journal} {\bibinfo  {journal}
  {Phys. Rev. Mater.}\ }\textbf {\bibinfo {volume} {5}},\ \bibinfo {pages}
  {063803} (\bibinfo {year} {2021})}\BibitemShut {NoStop}%
\bibitem [{\citenamefont {Heyd}\ \emph {et~al.}(2003)\citenamefont {Heyd},
  \citenamefont {Scuseria},\ and\ \citenamefont {Ernzerhof}}]{HSE06}%
  \BibitemOpen
  \bibfield  {author} {\bibinfo {author} {\bibfnamefont {J.}~\bibnamefont
  {Heyd}}, \bibinfo {author} {\bibfnamefont {G.~E.}\ \bibnamefont {Scuseria}},
  \ and\ \bibinfo {author} {\bibfnamefont {M.}~\bibnamefont {Ernzerhof}},\
  }\href {\doibase 10.1063/1.1564060} {\bibfield  {journal} {\bibinfo
  {journal} {The Journal of Chemical Physics}\ }\textbf {\bibinfo {volume}
  {118}},\ \bibinfo {pages} {8207} (\bibinfo {year} {2003})},\ \Eprint
  {http://arxiv.org/abs/https://pubs.aip.org/aip/jcp/article-pdf/118/18/8207/10847843/8207\_1\_online.pdf}
  {https://pubs.aip.org/aip/jcp/article-pdf/118/18/8207/10847843/8207\_1\_online.pdf}
  \BibitemShut {NoStop}%
\bibitem [{\citenamefont {Hedin}(1965)}]{LH65}%
  \BibitemOpen
  \bibfield  {author} {\bibinfo {author} {\bibfnamefont {L.}~\bibnamefont
  {Hedin}},\ }\href {\doibase 10.1103/PhysRev.139.A796} {\bibfield  {journal}
  {\bibinfo  {journal} {Phys. Rev.}\ }\textbf {\bibinfo {volume} {139}},\
  \bibinfo {pages} {A796} (\bibinfo {year} {1965})}\BibitemShut {NoStop}%
\bibitem [{\citenamefont {Hedin}\ and\ \citenamefont {Lundqvist}(1969)}]{LH69}%
  \BibitemOpen
  \bibfield  {author} {\bibinfo {author} {\bibfnamefont {L.}~\bibnamefont
  {Hedin}}\ and\ \bibinfo {author} {\bibfnamefont {S.}~\bibnamefont
  {Lundqvist}},\ }{\selectlanguage {English}\enquote {\bibinfo {title} {Effects
  of electron-electron and electron-phonon interactions on the one-electron
  states of solids},}\ }in\ \href {\doibase 10.1016/S0081-1947(08)60615-3}
  {{\selectlanguage {English}\emph {\bibinfo {booktitle} {Solid State
  Physics}}}},\ Vol.~\bibinfo {volume} {23},\ \bibinfo {editor} {edited by\
  \bibinfo {editor} {\bibfnamefont {F.}~\bibnamefont {Seitz}}, \bibinfo
  {editor} {\bibfnamefont {D.}~\bibnamefont {Turnbull}}, \ and\ \bibinfo
  {editor} {\bibfnamefont {H.}~\bibnamefont {Ehrenreich}}}\ (\bibinfo
  {publisher} {Academic Press},\ \bibinfo {address} {United States},\ \bibinfo
  {year} {1969})\ pp.\ \bibinfo {pages} {1--181},\ \bibinfo {note}
  {doi:10.1016/S0081-1947(08)60615-3}\BibitemShut {NoStop}%
\bibitem [{\citenamefont {Hybertsen}\ and\ \citenamefont
  {Louie}(1985)}]{HMLS85}%
  \BibitemOpen
  \bibfield  {author} {\bibinfo {author} {\bibfnamefont {M.~S.}\ \bibnamefont
  {Hybertsen}}\ and\ \bibinfo {author} {\bibfnamefont {S.~G.}\ \bibnamefont
  {Louie}},\ }\href {\doibase 10.1103/PhysRevLett.55.1418} {\bibfield
  {journal} {\bibinfo  {journal} {Phys. Rev. Lett.}\ }\textbf {\bibinfo
  {volume} {55}},\ \bibinfo {pages} {1418} (\bibinfo {year}
  {1985})}\BibitemShut {NoStop}%
\bibitem [{\citenamefont {Aryasetiawan}\ and\ \citenamefont
  {Gunnarsson}(1998)}]{AG98}%
  \BibitemOpen
  \bibfield  {author} {\bibinfo {author} {\bibfnamefont {F.}~\bibnamefont
  {Aryasetiawan}}\ and\ \bibinfo {author} {\bibfnamefont {O.}~\bibnamefont
  {Gunnarsson}},\ }\href {\doibase 10.1088/0034-4885/61/3/002} {\bibfield
  {journal} {\bibinfo  {journal} {Reports on Progress in Physics}\ }\textbf
  {\bibinfo {volume} {61}},\ \bibinfo {pages} {237} (\bibinfo {year}
  {1998})}\BibitemShut {NoStop}%
\bibitem [{\citenamefont {Hohenberg}\ and\ \citenamefont {Kohn}(1964)}]{HK64}%
  \BibitemOpen
  \bibfield  {author} {\bibinfo {author} {\bibfnamefont {P.}~\bibnamefont
  {Hohenberg}}\ and\ \bibinfo {author} {\bibfnamefont {W.}~\bibnamefont
  {Kohn}},\ }\href {\doibase 10.1103/PhysRev.136.B864} {\bibfield  {journal}
  {\bibinfo  {journal} {Phys. Rev.}\ }\textbf {\bibinfo {volume} {136}},\
  \bibinfo {pages} {B864} (\bibinfo {year} {1964})}\BibitemShut {NoStop}%
\bibitem [{\citenamefont {Kohn}\ and\ \citenamefont {Sham}(1965)}]{KS65}%
  \BibitemOpen
  \bibfield  {author} {\bibinfo {author} {\bibfnamefont {W.}~\bibnamefont
  {Kohn}}\ and\ \bibinfo {author} {\bibfnamefont {L.~J.}\ \bibnamefont
  {Sham}},\ }\href {\doibase 10.1103/PhysRev.140.A1133} {\bibfield  {journal}
  {\bibinfo  {journal} {Phys. Rev.}\ }\textbf {\bibinfo {volume} {140}},\
  \bibinfo {pages} {A1133} (\bibinfo {year} {1965})}\BibitemShut {NoStop}%
\bibitem [{\citenamefont {Kresse}\ and\ \citenamefont
  {Furthm\"uller}(1996)}]{VASP}%
  \BibitemOpen
  \bibfield  {author} {\bibinfo {author} {\bibfnamefont {G.}~\bibnamefont
  {Kresse}}\ and\ \bibinfo {author} {\bibfnamefont {J.}~\bibnamefont
  {Furthm\"uller}},\ }\href {\doibase 10.1103/PhysRevB.54.11169} {\bibfield
  {journal} {\bibinfo  {journal} {Phys. Rev. B}\ }\textbf {\bibinfo {volume}
  {54}},\ \bibinfo {pages} {11169} (\bibinfo {year} {1996})}\BibitemShut
  {NoStop}%
\bibitem [{\citenamefont {Bl\"ochl}(1994)}]{pp1}%
  \BibitemOpen
  \bibfield  {author} {\bibinfo {author} {\bibfnamefont {P.~E.}\ \bibnamefont
  {Bl\"ochl}},\ }\href {\doibase 10.1103/PhysRevB.50.17953} {\bibfield
  {journal} {\bibinfo  {journal} {Phys. Rev. B}\ }\textbf {\bibinfo {volume}
  {50}},\ \bibinfo {pages} {17953} (\bibinfo {year} {1994})}\BibitemShut
  {NoStop}%
\bibitem [{\citenamefont {Kresse}\ and\ \citenamefont {Joubert}(1999)}]{pp2}%
  \BibitemOpen
  \bibfield  {author} {\bibinfo {author} {\bibfnamefont {G.}~\bibnamefont
  {Kresse}}\ and\ \bibinfo {author} {\bibfnamefont {D.}~\bibnamefont
  {Joubert}},\ }\href {\doibase 10.1103/PhysRevB.59.1758} {\bibfield  {journal}
  {\bibinfo  {journal} {Phys. Rev. B}\ }\textbf {\bibinfo {volume} {59}},\
  \bibinfo {pages} {1758} (\bibinfo {year} {1999})}\BibitemShut {NoStop}%
\bibitem [{\citenamefont {Onida}\ \emph {et~al.}(2002)\citenamefont {Onida},
  \citenamefont {Reining},\ and\ \citenamefont {Rubio}}]{Onida02}%
  \BibitemOpen
  \bibfield  {author} {\bibinfo {author} {\bibfnamefont {G.}~\bibnamefont
  {Onida}}, \bibinfo {author} {\bibfnamefont {L.}~\bibnamefont {Reining}}, \
  and\ \bibinfo {author} {\bibfnamefont {A.}~\bibnamefont {Rubio}},\ }\href
  {\doibase 10.1103/RevModPhys.74.601} {\bibfield  {journal} {\bibinfo
  {journal} {Rev. Mod. Phys.}\ }\textbf {\bibinfo {volume} {74}},\ \bibinfo
  {pages} {601} (\bibinfo {year} {2002})}\BibitemShut {NoStop}%
\bibitem [{\citenamefont {Albrecht}\ \emph {et~al.}(1998)\citenamefont
  {Albrecht}, \citenamefont {Reining}, \citenamefont {Del~Sole},\ and\
  \citenamefont {Onida}}]{ASR98}%
  \BibitemOpen
  \bibfield  {author} {\bibinfo {author} {\bibfnamefont {S.}~\bibnamefont
  {Albrecht}}, \bibinfo {author} {\bibfnamefont {L.}~\bibnamefont {Reining}},
  \bibinfo {author} {\bibfnamefont {R.}~\bibnamefont {Del~Sole}}, \ and\
  \bibinfo {author} {\bibfnamefont {G.}~\bibnamefont {Onida}},\ }\href
  {\doibase 10.1103/PhysRevLett.80.4510} {\bibfield  {journal} {\bibinfo
  {journal} {Phys. Rev. Lett.}\ }\textbf {\bibinfo {volume} {80}},\ \bibinfo
  {pages} {4510} (\bibinfo {year} {1998})}\BibitemShut {NoStop}%
\bibitem [{\citenamefont {Benedict}\ \emph {et~al.}(1998)\citenamefont
  {Benedict}, \citenamefont {Shirley},\ and\ \citenamefont {Bohn}}]{BLS98}%
  \BibitemOpen
  \bibfield  {author} {\bibinfo {author} {\bibfnamefont {L.~X.}\ \bibnamefont
  {Benedict}}, \bibinfo {author} {\bibfnamefont {E.~L.}\ \bibnamefont
  {Shirley}}, \ and\ \bibinfo {author} {\bibfnamefont {R.~B.}\ \bibnamefont
  {Bohn}},\ }\href {\doibase 10.1103/PhysRevLett.80.4514} {\bibfield  {journal}
  {\bibinfo  {journal} {Phys. Rev. Lett.}\ }\textbf {\bibinfo {volume} {80}},\
  \bibinfo {pages} {4514} (\bibinfo {year} {1998})}\BibitemShut {NoStop}%
\bibitem [{\citenamefont {Rohlfing}\ and\ \citenamefont {Louie}(1998)}]{RML98}%
  \BibitemOpen
  \bibfield  {author} {\bibinfo {author} {\bibfnamefont {M.}~\bibnamefont
  {Rohlfing}}\ and\ \bibinfo {author} {\bibfnamefont {S.~G.}\ \bibnamefont
  {Louie}},\ }\href {\doibase 10.1103/PhysRevLett.80.3320} {\bibfield
  {journal} {\bibinfo  {journal} {Phys. Rev. Lett.}\ }\textbf {\bibinfo
  {volume} {80}},\ \bibinfo {pages} {3320} (\bibinfo {year}
  {1998})}\BibitemShut {NoStop}%
\bibitem [{\citenamefont {Hanke}\ and\ \citenamefont {Sham}(1980)}]{HSL80}%
  \BibitemOpen
  \bibfield  {author} {\bibinfo {author} {\bibfnamefont {W.}~\bibnamefont
  {Hanke}}\ and\ \bibinfo {author} {\bibfnamefont {L.~J.}\ \bibnamefont
  {Sham}},\ }\href {\doibase 10.1103/PhysRevB.21.4656} {\bibfield  {journal}
  {\bibinfo  {journal} {Phys. Rev. B}\ }\textbf {\bibinfo {volume} {21}},\
  \bibinfo {pages} {4656} (\bibinfo {year} {1980})}\BibitemShut {NoStop}%
\bibitem [{\citenamefont {Deslippe}\ \emph {et~al.}(2012)\citenamefont
  {Deslippe}, \citenamefont {Samsonidze}, \citenamefont {Strubbe},
  \citenamefont {Jain}, \citenamefont {Cohen},\ and\ \citenamefont
  {Louie}}]{BGW}%
  \BibitemOpen
  \bibfield  {author} {\bibinfo {author} {\bibfnamefont {J.}~\bibnamefont
  {Deslippe}}, \bibinfo {author} {\bibfnamefont {G.}~\bibnamefont
  {Samsonidze}}, \bibinfo {author} {\bibfnamefont {D.~A.}\ \bibnamefont
  {Strubbe}}, \bibinfo {author} {\bibfnamefont {M.}~\bibnamefont {Jain}},
  \bibinfo {author} {\bibfnamefont {M.~L.}\ \bibnamefont {Cohen}}, \ and\
  \bibinfo {author} {\bibfnamefont {S.~G.}\ \bibnamefont {Louie}},\ }\href
  {\doibase https://doi.org/10.1016/j.cpc.2011.12.006} {\bibfield  {journal}
  {\bibinfo  {journal} {Computer Physics Communications}\ }\textbf {\bibinfo
  {volume} {183}},\ \bibinfo {pages} {1269} (\bibinfo {year}
  {2012})}\BibitemShut {NoStop}%
\bibitem [{\citenamefont {Giannozzi}\ \emph {et~al.}(2009)\citenamefont
  {Giannozzi}, \citenamefont {Baroni}, \citenamefont {Bonini}, \citenamefont
  {Calandra}, \citenamefont {Car}, \citenamefont {Cavazzoni}, \citenamefont
  {Ceresoli}, \citenamefont {Chiarotti}, \citenamefont {Cococcioni},
  \citenamefont {Dabo}, \citenamefont {Corso}, \citenamefont {de~Gironcoli},
  \citenamefont {Fabris}, \citenamefont {Fratesi}, \citenamefont {Gebauer},
  \citenamefont {Gerstmann}, \citenamefont {Gougoussis}, \citenamefont
  {Kokalj}, \citenamefont {Lazzeri}, \citenamefont {Martin-Samos},
  \citenamefont {Marzari}, \citenamefont {Mauri}, \citenamefont {Mazzarello},
  \citenamefont {Paolini}, \citenamefont {Pasquarello}, \citenamefont
  {Paulatto}, \citenamefont {Sbraccia}, \citenamefont {Scandolo}, \citenamefont
  {Sclauzero}, \citenamefont {Seitsonen}, \citenamefont {Smogunov},
  \citenamefont {Umari},\ and\ \citenamefont {Wentzcovitch}}]{QE}%
  \BibitemOpen
  \bibfield  {author} {\bibinfo {author} {\bibfnamefont {P.}~\bibnamefont
  {Giannozzi}}, \bibinfo {author} {\bibfnamefont {S.}~\bibnamefont {Baroni}},
  \bibinfo {author} {\bibfnamefont {N.}~\bibnamefont {Bonini}}, \bibinfo
  {author} {\bibfnamefont {M.}~\bibnamefont {Calandra}}, \bibinfo {author}
  {\bibfnamefont {R.}~\bibnamefont {Car}}, \bibinfo {author} {\bibfnamefont
  {C.}~\bibnamefont {Cavazzoni}}, \bibinfo {author} {\bibfnamefont
  {D.}~\bibnamefont {Ceresoli}}, \bibinfo {author} {\bibfnamefont {G.~L.}\
  \bibnamefont {Chiarotti}}, \bibinfo {author} {\bibfnamefont {M.}~\bibnamefont
  {Cococcioni}}, \bibinfo {author} {\bibfnamefont {I.}~\bibnamefont {Dabo}},
  \bibinfo {author} {\bibfnamefont {A.~D.}\ \bibnamefont {Corso}}, \bibinfo
  {author} {\bibfnamefont {S.}~\bibnamefont {de~Gironcoli}}, \bibinfo {author}
  {\bibfnamefont {S.}~\bibnamefont {Fabris}}, \bibinfo {author} {\bibfnamefont
  {G.}~\bibnamefont {Fratesi}}, \bibinfo {author} {\bibfnamefont
  {R.}~\bibnamefont {Gebauer}}, \bibinfo {author} {\bibfnamefont
  {U.}~\bibnamefont {Gerstmann}}, \bibinfo {author} {\bibfnamefont
  {C.}~\bibnamefont {Gougoussis}}, \bibinfo {author} {\bibfnamefont
  {A.}~\bibnamefont {Kokalj}}, \bibinfo {author} {\bibfnamefont
  {M.}~\bibnamefont {Lazzeri}}, \bibinfo {author} {\bibfnamefont
  {L.}~\bibnamefont {Martin-Samos}}, \bibinfo {author} {\bibfnamefont
  {N.}~\bibnamefont {Marzari}}, \bibinfo {author} {\bibfnamefont
  {F.}~\bibnamefont {Mauri}}, \bibinfo {author} {\bibfnamefont
  {R.}~\bibnamefont {Mazzarello}}, \bibinfo {author} {\bibfnamefont
  {S.}~\bibnamefont {Paolini}}, \bibinfo {author} {\bibfnamefont
  {A.}~\bibnamefont {Pasquarello}}, \bibinfo {author} {\bibfnamefont
  {L.}~\bibnamefont {Paulatto}}, \bibinfo {author} {\bibfnamefont
  {C.}~\bibnamefont {Sbraccia}}, \bibinfo {author} {\bibfnamefont
  {S.}~\bibnamefont {Scandolo}}, \bibinfo {author} {\bibfnamefont
  {G.}~\bibnamefont {Sclauzero}}, \bibinfo {author} {\bibfnamefont {A.~P.}\
  \bibnamefont {Seitsonen}}, \bibinfo {author} {\bibfnamefont {A.}~\bibnamefont
  {Smogunov}}, \bibinfo {author} {\bibfnamefont {P.}~\bibnamefont {Umari}}, \
  and\ \bibinfo {author} {\bibfnamefont {R.~M.}\ \bibnamefont {Wentzcovitch}},\
  }\href {\doibase 10.1088/0953-8984/21/39/395502} {\bibfield  {journal}
  {\bibinfo  {journal} {Journal of Physics: Condensed Matter}\ }\textbf
  {\bibinfo {volume} {21}},\ \bibinfo {pages} {395502} (\bibinfo {year}
  {2009})}\BibitemShut {NoStop}%
\bibitem [{\citenamefont {{Cho}}\ \emph {et~al.}(2015)\citenamefont {{Cho}},
  \citenamefont {{Kim}}, \citenamefont {{Kim}}, \citenamefont {{Zhao}},
  \citenamefont {{Seok}}, \citenamefont {{Keum}}, \citenamefont {{Baik}},
  \citenamefont {{Choe}}, \citenamefont {{Chang}}, \citenamefont {{Suenaga}},
  \citenamefont {{Kim}}, \citenamefont {{Lee}},\ and\ \citenamefont
  {{Yang}}}]{laser-driven}%
  \BibitemOpen
  \bibfield  {author} {\bibinfo {author} {\bibfnamefont {S.}~\bibnamefont
  {{Cho}}}, \bibinfo {author} {\bibfnamefont {S.}~\bibnamefont {{Kim}}},
  \bibinfo {author} {\bibfnamefont {J.~H.}\ \bibnamefont {{Kim}}}, \bibinfo
  {author} {\bibfnamefont {J.}~\bibnamefont {{Zhao}}}, \bibinfo {author}
  {\bibfnamefont {J.}~\bibnamefont {{Seok}}}, \bibinfo {author} {\bibfnamefont
  {D.~H.}\ \bibnamefont {{Keum}}}, \bibinfo {author} {\bibfnamefont
  {J.}~\bibnamefont {{Baik}}}, \bibinfo {author} {\bibfnamefont {D.-H.}\
  \bibnamefont {{Choe}}}, \bibinfo {author} {\bibfnamefont {K.~J.}\
  \bibnamefont {{Chang}}}, \bibinfo {author} {\bibfnamefont {K.}~\bibnamefont
  {{Suenaga}}}, \bibinfo {author} {\bibfnamefont {S.~W.}\ \bibnamefont
  {{Kim}}}, \bibinfo {author} {\bibfnamefont {Y.~H.}\ \bibnamefont {{Lee}}}, \
  and\ \bibinfo {author} {\bibfnamefont {H.}~\bibnamefont {{Yang}}},\ }\href
  {\doibase 10.1126/science.aab3175} {\bibfield  {journal} {\bibinfo  {journal}
  {Science}\ }\textbf {\bibinfo {volume} {349}},\ \bibinfo {pages} {625}
  (\bibinfo {year} {2015})}\BibitemShut {NoStop}%
\bibitem [{\citenamefont {{Perdew}}\ and\ \citenamefont
  {{Schmidt}}(2001)}]{Jacob-ladder}%
  \BibitemOpen
  \bibfield  {author} {\bibinfo {author} {\bibfnamefont {J.~P.}\ \bibnamefont
  {{Perdew}}}\ and\ \bibinfo {author} {\bibfnamefont {K.}~\bibnamefont
  {{Schmidt}}},\ }in\ \href {\doibase 10.1063/1.1390175} {\emph {\bibinfo
  {booktitle} {Density Functional Theory and its Application to Materials}}},\
  \bibinfo {series} {American Institute of Physics Conference Series}, Vol.\
  \bibinfo {volume} {577}\ (\bibinfo {year} {2001})\ pp.\ \bibinfo {pages}
  {1--20}\BibitemShut {NoStop}%
\bibitem [{\citenamefont {Yang}\ \emph {et~al.}(2016)\citenamefont {Yang},
  \citenamefont {Peng}, \citenamefont {Sun},\ and\ \citenamefont
  {Perdew}}]{gKS}%
  \BibitemOpen
  \bibfield  {author} {\bibinfo {author} {\bibfnamefont {Z.-h.}\ \bibnamefont
  {Yang}}, \bibinfo {author} {\bibfnamefont {H.}~\bibnamefont {Peng}}, \bibinfo
  {author} {\bibfnamefont {J.}~\bibnamefont {Sun}}, \ and\ \bibinfo {author}
  {\bibfnamefont {J.~P.}\ \bibnamefont {Perdew}},\ }\href {\doibase
  10.1103/PhysRevB.93.205205} {\bibfield  {journal} {\bibinfo  {journal} {Phys.
  Rev. B}\ }\textbf {\bibinfo {volume} {93}},\ \bibinfo {pages} {205205}
  (\bibinfo {year} {2016})}\BibitemShut {NoStop}%
\bibitem [{\citenamefont {Perdew}\ \emph {et~al.}(2017)\citenamefont {Perdew},
  \citenamefont {Yang}, \citenamefont {Burke}, \citenamefont {Yang},
  \citenamefont {Gross}, \citenamefont {Scheffler}, \citenamefont {Scuseria},
  \citenamefont {Henderson}, \citenamefont {Zhang}, \citenamefont {Ruzsinszky},
  \citenamefont {Peng}, \citenamefont {Sun}, \citenamefont {Trushin},\ and\
  \citenamefont {Görling}}]{scan-bandgap}%
  \BibitemOpen
  \bibfield  {author} {\bibinfo {author} {\bibfnamefont {J.~P.}\ \bibnamefont
  {Perdew}}, \bibinfo {author} {\bibfnamefont {W.}~\bibnamefont {Yang}},
  \bibinfo {author} {\bibfnamefont {K.}~\bibnamefont {Burke}}, \bibinfo
  {author} {\bibfnamefont {Z.}~\bibnamefont {Yang}}, \bibinfo {author}
  {\bibfnamefont {E.~K.~U.}\ \bibnamefont {Gross}}, \bibinfo {author}
  {\bibfnamefont {M.}~\bibnamefont {Scheffler}}, \bibinfo {author}
  {\bibfnamefont {G.~E.}\ \bibnamefont {Scuseria}}, \bibinfo {author}
  {\bibfnamefont {T.~M.}\ \bibnamefont {Henderson}}, \bibinfo {author}
  {\bibfnamefont {I.~Y.}\ \bibnamefont {Zhang}}, \bibinfo {author}
  {\bibfnamefont {A.}~\bibnamefont {Ruzsinszky}}, \bibinfo {author}
  {\bibfnamefont {H.}~\bibnamefont {Peng}}, \bibinfo {author} {\bibfnamefont
  {J.}~\bibnamefont {Sun}}, \bibinfo {author} {\bibfnamefont {E.}~\bibnamefont
  {Trushin}}, \ and\ \bibinfo {author} {\bibfnamefont {A.}~\bibnamefont
  {Görling}},\ }\href {\doibase 10.1073/pnas.1621352114} {\bibfield  {journal}
  {\bibinfo  {journal} {Proceedings of the National Academy of Sciences}\
  }\textbf {\bibinfo {volume} {114}},\ \bibinfo {pages} {2801} (\bibinfo {year}
  {2017})},\ \Eprint
  {http://arxiv.org/abs/https://www.pnas.org/doi/pdf/10.1073/pnas.1621352114}
  {https://www.pnas.org/doi/pdf/10.1073/pnas.1621352114} \BibitemShut {NoStop}%
\bibitem [{\citenamefont {Kothakonda}\ \emph {et~al.}(2023)\citenamefont
  {Kothakonda}, \citenamefont {Kaplan}, \citenamefont {Isaacs}, \citenamefont
  {Bartel}, \citenamefont {Furness}, \citenamefont {Ning}, \citenamefont
  {Wolverton}, \citenamefont {Perdew},\ and\ \citenamefont
  {Sun}}]{r2scan-bandgap}%
  \BibitemOpen
  \bibfield  {author} {\bibinfo {author} {\bibfnamefont {M.}~\bibnamefont
  {Kothakonda}}, \bibinfo {author} {\bibfnamefont {A.~D.}\ \bibnamefont
  {Kaplan}}, \bibinfo {author} {\bibfnamefont {E.~B.}\ \bibnamefont {Isaacs}},
  \bibinfo {author} {\bibfnamefont {C.~J.}\ \bibnamefont {Bartel}}, \bibinfo
  {author} {\bibfnamefont {J.~W.}\ \bibnamefont {Furness}}, \bibinfo {author}
  {\bibfnamefont {J.}~\bibnamefont {Ning}}, \bibinfo {author} {\bibfnamefont
  {C.}~\bibnamefont {Wolverton}}, \bibinfo {author} {\bibfnamefont {J.~P.}\
  \bibnamefont {Perdew}}, \ and\ \bibinfo {author} {\bibfnamefont
  {J.}~\bibnamefont {Sun}},\ }\href {\doibase 10.1021/acsmaterialsau.2c00059}
  {\bibfield  {journal} {\bibinfo  {journal} {ACS Materials Au}\ }\textbf
  {\bibinfo {volume} {3}},\ \bibinfo {pages} {102} (\bibinfo {year}
  {2023})}\BibitemShut {NoStop}%
\bibitem [{\citenamefont {Aschebrock}\ and\ \citenamefont
  {K\"ummel}(2019)}]{task}%
  \BibitemOpen
  \bibfield  {author} {\bibinfo {author} {\bibfnamefont {T.}~\bibnamefont
  {Aschebrock}}\ and\ \bibinfo {author} {\bibfnamefont {S.}~\bibnamefont
  {K\"ummel}},\ }\href {\doibase 10.1103/PhysRevResearch.1.033082} {\bibfield
  {journal} {\bibinfo  {journal} {Phys. Rev. Res.}\ }\textbf {\bibinfo {volume}
  {1}},\ \bibinfo {pages} {033082} (\bibinfo {year} {2019})}\BibitemShut
  {NoStop}%
\bibitem [{\citenamefont {Palummo}\ \emph {et~al.}(2015)\citenamefont
  {Palummo}, \citenamefont {Bernardi},\ and\ \citenamefont
  {Grossman}}]{radlife}%
  \BibitemOpen
  \bibfield  {author} {\bibinfo {author} {\bibfnamefont {M.}~\bibnamefont
  {Palummo}}, \bibinfo {author} {\bibfnamefont {M.}~\bibnamefont {Bernardi}}, \
  and\ \bibinfo {author} {\bibfnamefont {J.~C.}\ \bibnamefont {Grossman}},\
  }\href {\doibase 10.1021/nl503799t} {\bibfield  {journal} {\bibinfo
  {journal} {Nano Letters}\ }\textbf {\bibinfo {volume} {15}},\ \bibinfo
  {pages} {2794} (\bibinfo {year} {2015})}\BibitemShut {NoStop}%
\bibitem [{\citenamefont {Shi}\ \emph {et~al.}(2013)\citenamefont {Shi},
  \citenamefont {Yan}, \citenamefont {Bertolazzi}, \citenamefont {Brivio},
  \citenamefont {Gao}, \citenamefont {Kis}, \citenamefont {Jena}, \citenamefont
  {Xing},\ and\ \citenamefont {Huang}}]{exciton-dynamics}%
  \BibitemOpen
  \bibfield  {author} {\bibinfo {author} {\bibfnamefont {H.}~\bibnamefont
  {Shi}}, \bibinfo {author} {\bibfnamefont {R.}~\bibnamefont {Yan}}, \bibinfo
  {author} {\bibfnamefont {S.}~\bibnamefont {Bertolazzi}}, \bibinfo {author}
  {\bibfnamefont {J.}~\bibnamefont {Brivio}}, \bibinfo {author} {\bibfnamefont
  {B.}~\bibnamefont {Gao}}, \bibinfo {author} {\bibfnamefont {A.}~\bibnamefont
  {Kis}}, \bibinfo {author} {\bibfnamefont {D.}~\bibnamefont {Jena}}, \bibinfo
  {author} {\bibfnamefont {H.~G.}\ \bibnamefont {Xing}}, \ and\ \bibinfo
  {author} {\bibfnamefont {L.}~\bibnamefont {Huang}},\ }\href {\doibase
  10.1021/nn303973r} {\bibfield  {journal} {\bibinfo  {journal} {{ACS} Nano}\
  }\textbf {\bibinfo {volume} {7}},\ \bibinfo {pages} {1072} (\bibinfo {year}
  {2013})}\BibitemShut {NoStop}%
\end{thebibliography}%
\end{document}